\documentclass[prc,aps,float,nofootinbib,superscriptaddress,byrevtex,twocolumn]{revtex4-1}
\usepackage{amssymb}
\usepackage{amsmath}
\usepackage{amsfonts}
\usepackage{epsfig}
\usepackage{bbm}
\usepackage{comment}

\usepackage{multirow}
\usepackage{longtable}
\usepackage{array}
\usepackage{subfigure}
\usepackage{graphics}

\newcommand\bbone{\ensuremath{\mathbbm{1}}}

\usepackage[usenames,dvipsnames]{color}

\allowdisplaybreaks
\begin{document}
\title{Symmetry broken and restored coupled-cluster theory\\
I. Rotational symmetry and angular momentum}
\author{T. Duguet}
\email{thomas.duguet@cea.fr} 
\affiliation{CEA-Saclay DSM/Irfu/SPhN, F-91191 Gif sur Yvette Cedex, France} 
\affiliation{National Superconducting Cyclotron Laboratory and Department of Physics and Astronomy, Michigan State University, East Lansing, MI 48824, USA}


%
\date{\today}
%
%
\begin{abstract}
We extend coupled-cluster theory performed on top of a Slater determinant breaking rotational symmetry to allow for the exact restoration of the angular momentum at any truncation order. The main objective relates to the description of near-degenerate finite quantum systems with an open-shell character. As such, the newly developed many-body formalism offers a wealth of potential applications and further extensions dedicated to the ab initio description of, e.g., doubly open-shell atomic nuclei and molecule dissociation. The formalism, which encompasses both single-reference coupled cluster theory and projected Hartree-Fock theory as particular cases, permits the computation of usual sets of connected diagrams while consistently incorporating static correlations through the highly non-perturbative restoration of rotational symmetry. Interestingly, the yrast spectroscopy of the system, i.e. the lowest energy associated with each angular momentum, is accessed within a single calculation. A key difficulty presently overcome relates to the necessity to handle generalized energy {\it and} norm kernels for which naturally terminating coupled-cluster expansions could be eventually obtained. The present work focuses on $SU(2)$ but can be extended to any (locally) compact Lie group and to discrete groups, such as most point groups. In particular, the formalism will be soon generalized to $U(1)$ symmetry associated with particle number conservation. This is relevant to Bogoliubov coupled cluster theory that was recently applied to singly open-shell nuclei.  
\end{abstract}
\maketitle

%

\section{Introduction}
\label{introduction}

High-quality ab-initio many-body methods have been revived or developed in the last ten years to go beyond the p-shell nuclei that were and are still addressed with success on the basis of, e.g., Faddeev-Yakubowski~\cite{friar88,glockle93,nogga97}, Green function Monte-Carlo~\cite{pudliner97,wiringa98,wiringa00} or the no-core shell model~\cite{navratil98a,navratil98b,Quaglioni:2007qe,Navratil:2009ut} methods. A decisive step was the re-introduction of coupled cluster (CC) techniques to nuclear theory~\cite{Dean:2003vc,Kowalski:2003hp,Wloch:2005za,Wloch:2005qq,Gour:2005dm,Gour:2008zz,Hagen:2010gd,Jansen:2011gb,Binder:2012mk} after a long period of intense development in quantum chemistry~\cite{shavitt09a}. Alongside, self-consistent Dyson-Green's function (SC{\it Dy}GF) theory~\cite{Barbieri:2000pg,Barbieri:2001gt,Dickhoff:2004xx,Barbieri:2004tp,Waldecker:2011by,Cipollone:2013zma} and in-medium similarity renormalization group (IMSRG) techniques~\cite{Tsukiyama:2010rj,Hergert:2012nb} have provided quantitatively analogous results, opening new paths to mid-mass nuclei. As a matter of fact, essentially converged calculations have been recently achieved on the basis of realistic two- and three-nucleon interactions~\cite{Hagen:2007ew,Binder:2013oea,Hergert:2013uja,Carbone:2013eqa,Cipollone:2013zma,Soma:2013vca,Soma:2013xha,Binder:2013xaa} up to mass $A\sim 130$. Although impressive, these developments were limited until recently to doubly closed-shell nuclei plus those accessed via the addition and the removal of 1 or 2 nucleons. These systems eventually represent a very limited fraction of the nuclei relevant to problems of current interest and planned to be studied at the upcoming generation of nuclear radioactive ion beam facilities. 

The extension to near-degenerate or genuinely open-shell systems constitutes a major difficulty as it requires to expand the many-body solution around a near-degenerate reference state. Doing so necessarily complicates the formalism and increases the computational cost. A first route appropriate to a set of particular cases consists of employing high-order non-perturbative, e.g. CC, methods~\cite{noga87a,olsen96a,piecuch05a,piecuch09a} based on a single symmetry-restricted reference state appropriate to closed- or open-shell~\cite{roothaan60a,glaesemann10a,tsuchimochi10a} systems and a wave operator that is possibly spin restricted~\cite{paldus77a,nakatsuji78a,adams79a,rittby88a} or even spin adapted~\cite{heckert06a}. A second way to overcome the near degeneracy of the reference state relies on the development of multi-reference (MR) methods based on, e.g., many-body perturbation theory (MBPT) and CC theory~\cite{bartlett07a,shavitt09a}. This option has been heavily pursued in quantum chemistry over the last thirty years. In the nuclear context, a multi-reference IMSRG technique has been developed recently to address (singly) open-shell nuclei~\cite{Hergert:2013uja} whereas CC-based~\cite{Jansen:2014qxa} and IMSRG-based~\cite{Bogner:2014baa} configuration interaction methods have been proposed even more recently. 

There exists an alternative route based on the spontaneous symmetry breaking of the reference state. In the nuclear context, this relates essentially to the breaking of $U(1)$ gauge symmetry associated with particle-number conservation as a way to account for superfluidity in singly open-shell nuclei. One must add the breaking of $SU(2)$ rotational symmetry associated with angular momentum conservation to capture quadrupole correlations in doubly open-shell nuclei. The key benefit of breaking a symmetry relates to commuting the degeneracy of the reference state with respect to particle-hole excitations into a degeneracy with respect to transformations of the associated symmetry group, i.e. it produces a (pseudo) Goldstone mode in the manifold of "deformed" {\it closed-shell} product states connected via symmetry transformations. This trade-off allows one to take a good first shot at near-degenerate systems on the basis of a {\it single-reference} (SR) method, while postponing the handling of the pseudo Goldstone mode to a later stage. This idea has been recently exploited in the nuclear context to address singly open-shell nuclei by allowing for the breaking of $U(1)$ gauge symmetry associated with particle-number conservation. As for Green's function techniques, this was achieved via the first realistic application of self-consistent Gorkov-Green's function (SC{\it Go}GF) theory to finite nuclei~\cite{soma11a,Soma:2012zd,Barbieri:2012rd,Soma:2013vca,Soma:2013xha}. As for CC techniques, this relates to the even more recent development and implementation of the so-called Bogoliubov coupled-cluster theory~\cite{StolarczykMonkhorst,signoracci13a,Henderson:2014vka}. Doubly-open shell systems can already be addressed by performing standard SCGF and CC calculations on the basis of a deformed Slater determinant, i.e. using a reference state that breaks $SU(2)$ symmetry associated with angular momentum conservation. An even better approach to doubly open shell nuclei to be developed in the future consists of breaking both $U(1)$ and $SU(2)$ symmetries at the same time in CC and SCGF calculations.

However, the description of open-shell systems is achieved in this way at the price of losing good symmetry quantum numbers. This corresponds to describing the system via a wave packet that spans several irreducible representations of the symmetry group rather than via the proper eigenstate. Correspondingly, the degeneracy associated with the pseudo Goldstone mode must be resolved as the symmetry breaking is in fact fictitious in finite systems. Indeed, quantum fluctuations eventually lift the near degeneracy because of the finiteness of the associated inertia. See, e.g., Refs.~\cite{ui83a,yannouleas07a,Papenbrock:2013cra} for more detailed discussions. It is thus mandatory to restore good symmetry quantum numbers, which does not only revise the energy (dramatically in certain situations) but also allows the proper handling of transition operators characterized by symmetry selection rules.

In principle, symmetry-unrestricted methods automatically restore good symmetries in the limit of exact calculations. In practice, however, it is not clear to what extent this is the case with presently tractable many-body truncations. While one may monitor the restoration (residual breaking) of the symmetry by computing the expectation value and the variance of the members of the Lie algebra~\cite{bartlett83a,cole87a,stanton94a,hagen07c}, one cannot however evaluate the remaining contamination of the energy. It is a compelling question that can only be answered by developing proper symmetry-restoration tools within which a full account of static correlations can be achieved. Projection techniques are convenient instruments to restore symmetries exactly at the mean-field, e.g. Hartree-Fock (HF), level and they are used extensively in both nuclear physics~\cite{ring80a,Duguet:2013dga} and quantum chemistry~\cite{jimenez12a}.  However, such techniques cannot be straightforwardly merged with ab-initio methods performed on top of a symmetry-breaking reference state to go beyond symmetry-projected Hartree-Fock. Still, the insertion of projectors at second-order in perturbation theory was investigated at some point in nuclear physics~\cite{peierls73a,atalay73a,atalay74a,atalay75a,atalay78a} and quantum chemistry~\cite{schlegel86a,schlegel88a,knowles88a} but not pursued since. In the latter case in particular, the main goal was to improve on the obvious defects left in potential energy surfaces of bond breaking computed from symmetry-unrestricted Hartree-Fock and symmetry-projected Hartree-Fock theories\footnote{The typical situation is that (i) symmetry-restricted Hartree Fock calculations do not give the correct dissociation limit, (ii) symmetry-unrestricted Hartree-Fock calculations do provide the right dissociation limit but are highly incorrect at intermediate distances, (iii) symmetry-projected Hartree-Fock displays a discontinuity at the point where the symmetry breaks spontaneously. Going to higher orders, the situation is that (i) symmetry-restricted MBPT breaks down as the bond is stretched, (ii) low order symmetry-unrestricted MBPT calculations converge to the right dissociation limit but are still incorrect at intermediate distances with a very slow convergence of the expansion, (iii) low order symmetry-projected MBPT calculations do improve on projected Hartree Fock but do not recover the full configuration interaction (CI) results. As for non-perturbative symmetry restricted or unrestricted CC calculations, they systematically improve over perturbative calculations but they do not correct for their failures entirely. Eventually, one is lacking today a consistent symmetry-restored CC theory that would approach full CI results over the whole dissociation path.}. Those methods relied on L\"{o}wdin's representation of the spin projector~\cite{lowdin55c}, often approximating it to only remove the next highest spin. Thus, those methods were often only approximately restoring broken spin symmetry, did not offer any transparent understanding of the connected/disconnected nature of the expansion and were eventually limited to low-order perturbative expansions. In the end, one is still missing today  satisfying extensions of symmetry-unrestricted  SCGF or CC methods within which symmetries are {\it exactly} restored without spoiling the most wanted features of those approaches, e.g. non-perturbative resummation of dynamical correlations, size extensivity and a “gentle” computation cost. 

It is the objective of the present paper to formulate a generalization of symmetry-unrestricted CC theory that properly incorporates the exact restoration of the broken symmetry at {\it any} truncation order. While grasping dynamical correlations through a CC diagrammatic based on a symmetry-breaking reference state, one wishes to incorporate static correlations via the explicit restoration of the symmetry. The proposed method provides not only access to the ground state but also to the yrast spectroscopy, i.e. to the lowest energy associated with each irreducible representation of the symmetry group. The approach is meant to be valid for any symmetry that can be broken (spontaneously or enforcing it) by the reference state and to be applicable to any system independently of its closed-shell, near-degenerate or open-shell character. While the present paper entirely focuses on $SU(2)$ rotational symmetry, extensions to other (locally) compact Lie group and to discrete groups, such as most point groups are possible. As a matter of fact, the method will be generalized in a forthcoming publication to the restoration of (good) particle number in connection with the recently proposed Bogoliubov coupled cluster theory~\cite{signoracci13a,Henderson:2014vka}. 

The many-body method designed to accomplish this task is naturally of {\it multi-reference} character. However, its MR nature is different from any of those at play in MR-CC methods developed in quantum chemistry~\cite{bartlett07a}. Indeed, reference states are not obtained from one another via particle-hole excitations but via highly non-perturbative (collective) symmetry transformations. Additionally, the method leads in practice to solving a finite number of single-reference-like CC calculations. This work builds on a first step taken in Ref.~\cite{duguet03a} that was however not entirely satisfactory as it was limited to perturbation theory and did not clarify the handling of disconnected diagrams.

The paper is organized as follows. Section~\ref{definitions} provides the ingredients necessary to set up the formalism while Sec.~\ref{normkernel} elaborates on the general principles of the approach, independently of the approximation method eventually employed. In Sec.~\ref{sectionMBPT}, a many-body perturbation theory is developed and acts as the foundation for the coupled-cluster approach.  Section~\ref{CCtheory} introduces the coupled-cluster scheme itself. It is shown how generalized energy and norm kernels can be computed from {\it naturally terminating} coupled-cluster expansions. The way to recover standard SR-CC theory on the one hand and projected Hartree-Fock theory on the other hand is illustrated. Eventually, the algorithm the owner of a symmetry-unrestricted single-reference coupled-cluster code must follow to implement the symmetry restoration step is provided. The body of the paper is restricted to discussing the overall scheme, limiting technical details to the minimum. Diagrammatic rules, analytic derivations and proofs are provided in an extended set of appendices.

\section{Basic ingredients}
\label{definitions}

We now introduce necessary ingredients to make the paper self-contained. Although pedestrian, this section displays definitions and identities that are crucial to the building of the formalism later on.

\subsection{Hamiltonian}

Let the Hamiltonian $H=T+V$ of the system be of the form\footnote{The formalism can be extended to a Hamiltonian containing three- and higher-body forces without running into any fundamental problem. Also, one subtracts the center of mass kinetic energy to the Hamiltonian in actual calculations of finite nuclei. As far as the present work is concerned, this simply leads to a redefinition of the one- and two-body matrix elements $t_{\alpha\beta} $ and $v_{\alpha\beta\gamma\delta}$ of the Hamiltonian without changing any aspect of the many-body formalism that follows.}
\begin{eqnarray}
\label{e:ham}
H &=& \sum _{\alpha\beta} t_{\alpha\beta} \, c^{\dagger}_{\alpha} c_{\beta} + \frac{1}{2} \sum _{\alpha\beta\gamma\delta} v_{\alpha\beta\gamma\delta} \, c^{\dagger}_{\alpha} c^{\dagger}_{\beta} c_{\delta} c_{\gamma} \, ,
\end{eqnarray} 
where (direct-product) matrix elements of the kinetic energy and of the two-body interaction are defined, respectively, through 
\begin{subequations}
\begin{align}
t_{\alpha\beta} & \equiv \langle 1: \alpha | T | 1: \beta \rangle \, , \\
v_{\alpha\beta\gamma\delta} & \equiv \langle 1: \alpha; 2: \beta | V | 1: \gamma ; 2: \delta \rangle \, ,
\end{align}
\end{subequations}
such that antisymmetrized matrix elements of the latter are given by $\bar{v}_{\alpha\beta\gamma\delta} \equiv v_{\alpha\beta\gamma\delta} - v_{\alpha\beta\delta\gamma}$.

\subsection{$SU(2)$ symmetry group}

We consider the non-abelian compact Lie group $SU(2)\equiv \{R(\Omega), \Omega \in  D_{SU(2)}\}$ associated with the rotation of a A-body fermion system with integer or half-integer angular momentum. The group is parametrized by three Euler angles $\Omega \equiv (\alpha,\beta,\gamma)$ whose domain of definition is 
\begin{equation}
D_{SU(2)} \equiv D_{\alpha} \times D_{\beta} \times D_{\gamma}  = [0,4\pi] \times [0,\pi] \times [0,2\pi] \, . 
\end{equation}
As $SU(2)$ is considered to be a symmetry group of $H$, the commutation relations
\begin{eqnarray}
\left[H,R(\Omega)\right]&=& \left[T,R(\Omega)\right] = \left[V,R(\Omega)\right] =0 \, , \label{commutation}
\end{eqnarray}
hold for $\Omega \in  D_{SU(2)}$.

We utilize the unitary representation of $SU(2)$ on the Fock space given by  
\begin{equation}
R(\Omega)  =e^{-\frac{i}{\hbar}\alpha J_{z}}e^{-\frac{i}{\hbar}\beta J_{y}}e^{-\frac{i}{\hbar}\gamma J_{z}} \, ,
\end{equation}
where the three components of the angular momentum vector $\vec{J} = \sum_{n=1}^{\text{A}} \vec{j}(n)$ take the second-quantized form
\begin{equation}
J_{i} = \sum_{\alpha\beta} (j_{i})_{\alpha\beta}  \, c^{\dagger}_{\alpha} c_{\beta} \, ,
\end{equation}
with $i=x,y,z$ and $(j_{i})_{\alpha\beta} \equiv \langle 1: \alpha | j_{i} | 1: \beta \rangle$. Those one-body operators make up the Lie algebra
\begin{equation}
[J_{i},J_{j}]=\epsilon_{ijk} i\hbar \, J_{k} \, , \label{Lieidentity}
\end{equation}
where $\epsilon_{ijk}$ denotes the Levi-Civita tensor. The Casimir operator of the group built from the infinitesimal generators through a non-degenerate invariant bilinear form is the total angular momentum
\begin{equation}
J^{2}  \equiv \sum_{i=x,y,z} J^2_{i} \, ,
\end{equation}
which is the sum of a one-body and a two-body term, respectively defined as
\begin{subequations}
\label{J21and2}
\begin{eqnarray}
J^{2}_{(1)} &\equiv& \sum_{n=1}^{\text{A}} j^2(n)  \nonumber \\
&=& \sum_{\alpha\beta} j^2_{\alpha\beta}  \, c^{\dagger}_{\alpha} c_{\beta} \, , \\
J^{2}_{(2)} &\equiv& \sum_{n\neq n'=1}^{\text{A}}  \vec{j}(n)\cdot \vec{j}(n')  \nonumber \\
&=& \frac{1}{2} \sum _{\alpha\beta\gamma\delta} j\!j_{\alpha\beta\gamma\delta} \, c^{\dagger}_{\alpha} c^{\dagger}_{\beta} c_{\delta} c_{\gamma}  \, ,
\end{eqnarray}
\end{subequations}
with (direct-product) matrix elements given by
\begin{subequations}
\begin{eqnarray}
j^2_{\alpha\beta}  & \equiv& \langle 1: \alpha | j^2 | 1: \beta \rangle \nonumber \\
&=& \sum_{i=x,y,z} \langle 1: \alpha | j^2_{i} | 1: \beta \rangle \, , \\
j\!j_{\alpha\beta\gamma\delta} &\equiv& \langle 1: \alpha; 2: \beta | j\!j | 1: \gamma ; 2: \delta \rangle \nonumber \\
&=&  2 \sum_{i=x,y,z} \langle 1 :  \alpha | j_{i} | 1: \gamma \rangle  \, \langle 2: \beta | j_{i} | 2: \delta \rangle  \, ,
\end{eqnarray}
\end{subequations}
from which antisymmetrized matrix elements are obtained though $\overline{\jmath \jmath}_{\alpha\beta\gamma\delta} \equiv j\!j_{\alpha\beta\gamma\delta} - j\!j_{\alpha\beta\delta\gamma}$.

Matrix elements of the irreducible representations (IRREPs) of $SU(2)$ are given by the so-called Wigner $D$-functions~\cite{varshalovich88a}
\begin{equation}
\langle \Psi^{JM}_{\mu} |  R(\Omega)  |\Psi^{J'K}_{\mu'} \rangle \equiv \delta_{\mu\mu'} \delta_{JJ'} D_{MK}^{J}(\Omega) \, ,
\end{equation}
where $| \Psi^{JM}_{\mu} \rangle$ is an eigenstate of $J^2$ and $J_{z}$
\begin{subequations}
\label{eigenequationJ}
\begin{eqnarray}
J^2 | \Psi^{JM}_{\mu} \rangle &=&  J(J+1)\hbar^2 | \Psi^{JM}_{\mu} \rangle \,\,\, , \\
J_{z} | \Psi^{JM}_{\mu} \rangle &=&  M\hbar | \Psi^{JM}_{\mu} \rangle \,\,\, .
\end{eqnarray}
\end{subequations}
with $2J \in \mathbb{N}$, $2M\in \mathbb{Z}$, $J-M\in \mathbb{N}$ and $-J\leq M \leq +J$. The $(2J\!+\!1)$-dimensional IRREPs are labeled by $J$ and are spanned by the $\{| \Psi^{JM}_{\mu} \rangle\}$ for fixed $J$ and $\mu$. By virtue of Eq.~\ref{commutation}, $\mu=0, 1, 2\ldots$ orders the eigenenergies for fixed $(J,M)$ according to
\begin{eqnarray}
H | \Psi^{J M}_{\mu} \rangle &=&  E^{J}_{\mu} \, | \Psi^{J M}_{\mu} \rangle \,\,\, , \label{schroed}
\end{eqnarray}
knowing that $E^{J}_{\mu}$ is independent of $M$. Wigner $D$-functions can be expressed as $D_{MK}^{J}(\Omega)\equiv e^{-iM\alpha} \, d_{MK}^{J}(\beta) \, e^{-iK\gamma}$, where the so-called reduced Wigner $d$-functions are real and defined through $d_{MK}^{J}(\beta) = \langle \Psi^{JM}_{\mu} |  R(0,\beta,0)  |\Psi^{JK}_{\mu} \rangle$.

The volume of the group is
\begin{eqnarray*}
v_{SU(2)} &\equiv& \int_{D_{SU(2)}} \hspace{-0.7cm} d\Omega \\
&\equiv& \int_{0}^{4\pi}\!d\alpha\!\int_{0}^{\pi}\! d\beta\sin\beta\!\int_{0}^{2\pi}\!d\gamma \\
&=& 16\pi^2 \,\,\, ,
\end{eqnarray*}
such that the orthogonality of Wigner $D$-functions reads
\begin{equation}
\int_{D_{SU(2)}} \hspace{-0.7cm} d\Omega \, D_{MK}^{J \, \ast}(\Omega) \, D_{M'K'}^{J'}(\Omega)  =\frac{16\pi^{2}}{2J+1}\delta_{JJ'}\delta_{MM'}\delta_{KK'} \, . \label{orthogonality}
\end{equation}%

An irreducible tensor operator $T^{J}_{K}$ of rank $J$ and a state $| \Psi^{J K}_{\mu} \rangle$ transform under rotation according to
\begin{subequations}
\label{eq:ten:def}
\begin{eqnarray}
R(\Omega) \, T^{J}_{K} \, R(\Omega)^{-1}  &=& \sum_{M}  T^{J}_{M} \, D^{J}_{MK}(\Omega) \,\,\, ,\label{eq:ten:def1} \\
R(\Omega) \, | \Psi^{J K}_{\mu} \rangle &=& \sum_{M}  | \Psi^{J M}_{\mu} \rangle \, D^{J}_{MK}(\Omega) \,\,\, . \label{eq:ten:def2}
\end{eqnarray}
\end{subequations}


A key feature for the following is that, any function $f(\Omega)$ defined on $D_{SU(2)}$ can be expanded over the IRREPs of the group. Such a decomposition reads
\begin{equation}
f(\Omega) \equiv \sum_{J MK} \, f^{J}_{MK} \, \, D^{J}_{MK}(\Omega) \, , \label{decomposition_general}
\end{equation}
and defines the set of expansion coefficients $\{f^{J}_{MK}\}$.

Of importance later on is the fact that Wigner $D$-functions fulfill three coupled ordinary differential equation (ODE)~\cite{varshalovich88a}
\begin{widetext}
\begin{subequations}
\label{wignerODE}
\begin{eqnarray}
-i \frac{\partial}{\partial \alpha} D_{MK}^{J}(\Omega) &=& -M \, D_{MK}^{J}(\Omega) \, ,\label{wignerODE1} \\
-i \frac{\partial}{\partial \gamma} D_{MK}^{J}(\Omega) &=&  -K \, D_{MK}^{J}(\Omega) \, ,\label{wignerODE2} \\
-\left[\frac{\partial}{\partial \beta}\left(\sin \beta \frac{\partial}{\partial \beta}\right) + \frac{1}{\sin^2\beta} \left(\frac{\partial^2}{\partial \alpha^2} - 2\cos \beta\frac{\partial^2}{\partial \alpha\partial \gamma} +  \frac{\partial^2}{\partial \gamma^2}\right)\right]  D_{MK}^{J}(\Omega) &=& J(J+1) \, D_{MK}^{J}(\Omega) \, . \label{wignerODE3}
\end{eqnarray}
\end{subequations}
\end{widetext}

\subsection{Time-dependent state}
\label{timedependentstate}

The generalized many-body scheme proposed in the present work is conveniently formulated within an imaginary-time framework. Accessing the ground-state information eventually leads to taking the imaginary time to infinity. However, and as will become clear below, setting up an explicit time-dependent formalism also allows one to access yrast states and is thus beneficial. 

We introduce the evolution operator in {\it imaginary} time as\footnote{The time is given in units of MeV$^{-1}$.}
\begin{equation}
{\cal U}(\tau) \equiv e^{-\tau H} \, , \label{evoloperator}
\end{equation}
with $\tau$ real. A key quantity throughout the present study is the time-evolved many-body state
\begin{subequations}
\label{evolstate}
\begin{eqnarray}
| \Psi (\tau) \rangle &\equiv& {\cal U}(\tau) | \Phi \rangle \\
&=& \sum_{\mu J M} e^{-\tau E^{J}_{\mu}} \,  |  \Psi^{J M}_{\mu} \rangle \, \langle \Psi^{J M}_{\mu} |  \Phi \rangle  \, ,
\end{eqnarray}
\end{subequations}
where we have inserted a completeness relationship on the A-fermion Hilbert space under the form
\begin{equation}
\bbone =  \sum_{\mu J M}   |  \Psi^{J M}_{\mu} \rangle \, \langle \Psi^{J M}_{\mu} | \, . \label{completeness}
\end{equation}
In Eq.~\ref{evolstate}, $| \Phi \rangle$ is an arbitrary reference state belonging to the A-fermion Hilbert space. It is straightforward to demonstrate that $| \Psi (\tau) \rangle$ satisfies the time-dependent Schroedinger equation
\begin{equation}
H \, | \Psi (\tau) \rangle = -\partial_{\tau} | \Psi (\tau) \rangle  \, . \label{schroedinger}
\end{equation}

\subsection{Large and infinite time limits}
\label{timelimits}

Below, we will be interested in first looking at the {\it large} $\tau$ limit of various quantities before eventually taking their {\it infinite} time limit. Although we utilize the same mathematical symbol ($\lim\limits_{\tau \to \infty}$) in both cases for simplicity, the reader must not be confused by the fact that there remains a residual $\tau$ dependence in the first case, which typically disappears by considering ratios before actually the time to infinity. The large $\tau$ limit is essentially defined as $\tau \gg \Delta E^{-1}$, where $\Delta E$ is the energy difference between the ground state and the first excited state. Depending on the system, the latter can be the first excited state in the IRREP of the ground state or the lowest state of another IRREP.

\subsection{Ground state}
\label{groundstateSeC}

Taking the large $\tau$ limit provides the A-body ground state of $H$ under the form
\begin{subequations}
\label{groundstate}
\begin{eqnarray}
| \Psi^{J_0}_{0} \rangle &\equiv& \lim\limits_{\tau \to \infty} | \Psi (\tau) \rangle \\
&=& e^{-\tau E^{J_0}_0}  \sum_{M}  |  \Psi^{J_0 M}_{0} \rangle \, \langle \Psi^{J_0 M}_{0} |  \Phi \rangle  \, ,
\end{eqnarray}
\end{subequations}
where one supposes that the IRREP $J_0$ of the ground state is not necessarily the trivial one, i.e. one allows for the possibility of a degenerate ground state. As will become clear below, the many-body scheme developed in the present work relies on choosing $| \Phi \rangle$ as the ground state of an unperturbed Hamiltonian $H_0$ that breaks $SU(2)$ symmetry. As such, $| \Phi \rangle$ mixes several IRREPS and is thus likely to contain a component belonging to the one of the actual ground state. This is necessary for $| \Psi^{J_0}_{0} \rangle$ to actually correspond to the ground state of $H$. If $| \Phi \rangle$ were to be orthogonal to the true ground state, $| \Psi^{J_0}_{0} \rangle$ would provide access to the lowest eigenstate not orthogonal to $| \Phi \rangle$. Eventually, one obtains that
\begin{eqnarray}
H | \Psi^{J_0}_{0} \rangle &=&  E^{J_0}_0 \, | \Psi^{J_0}_{0} \rangle \,\,\, . \label{schroedevolstate}
\end{eqnarray}

\subsection{Off-diagonal kernels}
\label{transitionkernels}

Starting from the above definitions, we now introduce a set of off-diagonal, i.e. $\Omega$-dependent, time-dependent kernels
\begin{subequations}
\label{defkernels}
\begin{eqnarray}
N(\tau,\Omega) &\equiv& \langle \Psi (\tau)  | \bbone | \Phi(\Omega) \rangle  \, , \label{defnormkernel} \\
H(\tau,\Omega) &\equiv& \langle \Psi (\tau) | H | \Phi(\Omega) \rangle  \, , \label{defenergykernel} \\
J_i(\tau,\Omega) &\equiv& \langle \Psi (\tau)  | J_i | \Phi(\Omega) \rangle  \, , \label{defJzkernel} \\
J^2(\tau,\Omega) &\equiv& \langle \Psi (\tau) | J^2 | \Phi(\Omega) \rangle  \, , \label{defJ2kernel} 
\end{eqnarray}
\end{subequations}
where $| \Phi(\Omega) \rangle \equiv R(\Omega) | \Phi \rangle$ denotes the {\it rotated} reference state. Equation~\ref{defkernels} defines the norm, energy, angular momentum projections and total angular momentum kernels, respectively. The energy and total angular momentum kernels can be further split into their one- and two-body components according to
\begin{subequations}
\label{defkernels1and2body}
\begin{eqnarray}
H(\tau,\Omega) &=& T(\tau,\Omega)+V(\tau,\Omega) \, , \label{defHkernel} \\
J^2(\tau,\Omega) &=& J^2_{(1)}(\tau,\Omega)+J^2_{(2)}(\tau,\Omega) \, . \label{defJ2oneandtwokernel}
\end{eqnarray}
\end{subequations}
In the following, we refer to a generic operator as $O$ and to its off-diagonal kernel as 
\begin{eqnarray}
O(\tau,\Omega) &\equiv& \langle \Psi (\tau)  | O | \Phi(\Omega) \rangle  \, .
\end{eqnarray}
Additionally, use will often be made of the {\it reduced} kernel defined through
\begin{equation}
{\cal O}(\tau,\Omega) \equiv \frac{O(\tau,\Omega)}{N(\tau,0)} \, , \label{reducedkernels}
\end{equation}
which corresponds, for $O=\bbone$, to working with {\it intermediate normalization} at $\Omega=0$, i.e. ${\cal N}(\tau,0) \equiv 1$ for all $\tau$.

\section{Master equations}
\label{normkernel}

This section presents a set of master equations providing the basis for the newly proposed many-body method, i.e. they constitute {\it exact} equations of reference on top of which the actual expansion scheme will be designed in the remaining of the paper.

\subsection{Expanded kernels over the IRREPs}
\label{kernelexpansion}

Inserting twice Eq.~\ref{completeness} into Eqs.~\ref{defkernels} while making use of Eqs.~\ref{eigenequationJ}, \ref{schroed} and~\ref{eq:ten:def2}, one obtains
\begin{subequations}
\label{expandedkernels}
\begin{eqnarray}
N(\tau,\Omega)  &=&  \sum_{\mu J} e^{-\tau E^{J}_{\mu}} \sum_{MK}  N^{J \mu}_{MK} \, D^{J}_{MK}(\Omega)  \, , \label{expandedkernels1} \\
H(\tau,\Omega) &=&  \sum_{\mu J} e^{-\tau E^{J}_{\mu}} \, E^{J}_{\mu}  \sum_{MK}   N^{J \mu}_{MK} \, D^{J}_{MK}(\Omega) \, , \label{expandedkernels2} \\
J_z(\tau,\Omega) &=&  \sum_{\mu J} e^{-\tau E^{J}_{\mu}}  \sum_{MK}  M\hbar \,  N^{J \mu}_{MK} \, D^{J}_{MK}(\Omega) \, , \label{expandedkernels3} \\
J^2(\tau,\Omega) &=&  \sum_{\mu J} e^{-\tau E^{J}_{\mu}} \, J(J+1)\hbar^2  \sum_{MK}   N^{J \mu}_{MK} \, D^{J}_{MK}(\Omega) \, , \label{expandedkernels4}
\end{eqnarray}
\end{subequations}
with $N^{J \mu}_{MK}\equiv \langle \Phi |  \Psi^{J M}_{\mu} \rangle \, \langle \Psi^{J K}_{\mu} |  \Phi \rangle$.

\subsection{Ground-state energy}
\label{energy}

Defining the large $\tau$ limit of a kernel via
\begin{eqnarray}
O(\Omega) &\equiv & \lim\limits_{\tau \to \infty} O(\tau,\Omega)  \, , \label{limitoperator}
\end{eqnarray}
we obtain
\begin{subequations}
\label{limitkernels}
\begin{eqnarray}
N(\Omega) &=&  e^{-\tau E^{J_0}_0} \sum_{MK} N^{J_0 0}_{MK} \, D^{J_0}_{MK}(\Omega)  \, , \label{limitnorm} \\
H(\Omega) &=& e^{-\tau E^{J_0}_0} \, E^{J_0}_0  \sum_{MK} N^{J_0 0}_{MK} \, D^{J_0}_{MK}(\Omega)  \, , \label{limitenergy} \\
J_z(\Omega) &=& e^{-\tau E^{J_0}_0}  \sum_{MK} M\hbar \, N^{J_0 0}_{MK} \, D^{J_0}_{MK}(\Omega)  \, , \label{limitJz} \\
J^2(\Omega) &=& e^{-\tau E^{J_0}_0} \, J_0(J_0+1)\hbar^2  \sum_{MK} N^{J_0 0}_{MK} \, D^{J_0}_{MK}(\Omega)  \, , \label{limitJ2} 
\end{eqnarray}
\end{subequations}
where the residual time dependence typically disappears by eventually employing reduced kernels as defined in Eq.~\ref{reducedkernels}. Equation~\ref{limitkernels} leads, in agreement with Eq.~\ref{schroedevolstate}, to
\begin{eqnarray}
H(\Omega) &=&  E^{J_0}_0 \, N(\Omega) \,\, , \label{kernelequation}
\end{eqnarray}
or equivalently with intermediate normalization to
\begin{eqnarray}
{\cal H}(\Omega) &=&  E^{J_0}_0 \, {\cal N}(\Omega) \,\, . \label{kernelequationintermediate}
\end{eqnarray}
In Eq.~\ref{limitkernels}, the $\Omega$ dependence originally built into the time-dependent kernels reduces to that of the single IRREP $J_0$ of the ground state. This specific feature, trivially valid for the exact kernels, testifies that the selected eigenstate $| \Psi^{J_0}_{0} \rangle$ of $H$ carries good angular momentum $J_0$. Let us now consider the case of interest where the kernels are approximated in a way that breaks $SU(2)$ symmetry. In this situation, Eqs.~\ref{limitnorm} and~\ref{limitenergy} must be replaced by
\begin{subequations}
\label{limitkernelsapprox}
\begin{eqnarray}
{\cal N}_{\text{approx}}(\Omega) &\equiv&   \sum_{J} \sum_{MK} \textmd{N}^{J}_{MK} \, D^{J}_{MK}(\Omega)  \, , \label{limitnormapprox} \\
{\cal H}_{\text{approx}}(\Omega) &\equiv& \sum_{J}  \sum_{MK} \textmd{E}^{J}_{MK} \, \textmd{N}^{J}_{MK} \, D^{J}_{MK}(\Omega)  \, , \label{limitenergyapprox}
\end{eqnarray}
\end{subequations}
where the remaining sum over $J$ and the dependence of the expansion coefficient of the energy kernel on $M$ and $K$ signal the breaking of the symmetry induced by the approximation. Note that Eq.~\ref{limitkernelsapprox} always exists as the expansion over the IRREPs of $SU(2)$ of a function $f(\Omega)$ defined on ${\cal D}_{SU(2)}$; i.e. by virtue of Eq.~\ref{decomposition_general}.

Except for going back to an exact computation of the kernels, such that all the expansion coefficients but the physical one become zero in Eq.~\ref{limitkernelsapprox}, taking the straight ratio ${\cal H}_{\text{approx}}(\Omega)/{\cal N}_{\text{approx}}(\Omega)$ does not provide an approximate energy that is in one-to-one correspondence with the physical IRREP $J_0$. However, one can take advantage of the $\Omega$ dependence built into ${\cal N}_{\text{approx}}(\Omega)$ and ${\cal H}_{\text{approx}}(\Omega)$ to extract the one component associated with that physical IRREP.  Indeed, by virtue of the orthogonality of the IRREPs (Eq.~\ref{orthogonality}), the approximation to $E^{J_0}_0$ can be extracted as  
\begin{eqnarray}
E^{J_0}_0 &=& \frac{\sum_{MK} f^{J_0\ast}_M \, f^{J_0}_K  \int_{D_{SU(2)}} \! d\Omega \, D^{J_0 \, \ast}_{MK}(\Omega) \, {\cal H}_{\text{approx}}(\Omega)}{\sum_{MK} f^{J_0\ast}_M \, f^{J_0}_K  \int_{D_{SU(2)}} \! d\Omega \, D^{J_0 \, \ast}_{MK}(\Omega) \, {\cal N}_{\text{approx}}(\Omega)} \nonumber \\
&=& \frac{\sum_{MK} f^{J_0\ast}_M \, f^{J_0}_K  \, \textmd{E}^{J_0}_{MK} \, \textmd{N}^{J_0}_{MK} }{\sum_{MK} f^{J_0\ast}_M \, f^{J_0}_K \, \textmd{N}^{J_0}_{MK}  }\, , \label{projected_energy}
\end{eqnarray}
where the sum over $(M,K)$ mixes the components of the targeted IRREP to remove a nonphysical dependence on the orientation of the deformed reference state. The coefficients of the mixing $f^{J_0}_K$ are generally unknown and are typically determined utilizing the fact that the ground-state energy is a variational minimum. This eventually leads to solving a Hill-Wheeler-Griffin equation~\cite{Hill53,ring80a,bender03b} for the lowest eigenvalue
\begin{equation}
\sum_{K=-J}^{+J}  \left(\textmd{E}^{J_0}_{MK} -  E^{J_0}_0 \right)\textmd{N}^{J_0}_{MK} \, f^{J_0}_K = 0 \, . \label{HWG}
\end{equation}
Of course, in the exact limit or when the approximation scheme respects the symmetry, Eq.~\ref{projected_energy} extracts $E^{J_0}_0$ trivially. As a matter of fact, the integration over the domain of $SU(2)$ becomes superfluous in this case such that one can simply take the straight ratio of the energy and the norm kernels in Eq.~\ref{kernelequationintermediate} to access $E^{J_0}_0$. 

\subsection{Comparison with standard approaches}
\label{standardapproaches}

Applying standard symmetry-unrestricted MBPT or CC theory amounts to expanding {\it diagonal} $(\Omega=0)$ kernels around a symmetry-breaking reference state  $| \Phi \rangle$. This is sufficient in the limit of exact calculations given that summing all diagrams does restore the symmetry by definition. As discussed above, approximate kernels however mix components associated with {\it different} IRREPs of $SU(2)$ and thus contain spurious contaminations from the symmetry viewpoint. The difficulty resides in the fact that the $\Omega$ dependence is absent in standard approaches, i.e. given that $D^{J}_{MK}(0)=\delta_{MK}$ for all $J$ Eq.~\ref{limitkernelsapprox} is replaced by
\begin{subequations}
\label{limitkernelsapprox0}
\begin{eqnarray}
{\cal N}_{\text{approx}}(0) &=&   \sum_{J} \sum_{M} \textmd{N}^{J}_{MM}  \, , \label{limitnormapprox0} \\
{\cal H}_{\text{approx}}(0) &=& \sum_{J} \sum_{M} \textmd{E}^{J}_{MM} \,  \textmd{N}^{J}_{MM} \, , \label{limitenergyapprox0}
\end{eqnarray}
\end{subequations}
from which the coefficient associated with the physical IRREP cannot be extracted. Accordingly, the key feature of the proposed approach is to utilize {\it off-diagonal} kernels incorporating, from the outset, the effect of the rotation $R(\Omega)$. The associated $\Omega$ dependence leaves a fingerprint of the artificial symmetry breaking built into approximated kernels. Eventually, this fingerprint can be exploited to extract the physical component of interest through Eq.~\ref{projected_energy}, i.e. to remove symmetry contaminants. 


\subsection{Yrast spectroscopy}
\label{yrast}

Now that the benefit of performing the integral over the domain of $SU(2)$ has been highlighted for the ground-state energy, let us step back to Eq.~\ref{expandedkernels} and slightly modify the procedure to access the lowest eigenenergy $E^J_0$ associated with {\it each} IRREP, i.e. to access the yrast spectroscopy\footnote{More exactly, one can access the energy of the lowest eigenstate of each IRREP {\it not orthogonal} to $| \Phi \rangle$.}. To do so, we invert the order in which the limit $\tau \to \infty$ and the integral over the domain of the group are performed. We thus extract the expansion coefficient associated with an arbitrary IRREP $J$ of interest
\begin{subequations}
\label{integratedkernels}
\begin{eqnarray}
n^{J}_{MK}(\tau) &\equiv &  \frac{2J+1}{16\pi^{2}}  \int_{D_{SU(2)}} \!\! d\Omega \, D^{J \, \ast}_{MK}(\Omega) \, {\cal N}(\tau,\Omega) \nonumber \\
&=&  \sum_{\mu} e^{-\tau E^{J}_{\mu}} \, N^{J \mu}_{MK}/N(\tau,0)  \, , \label{integratednorm} \\
h^{J}_{MK}(\tau)n^{J }_{MK}(\tau) &\equiv & \frac{2J+1}{16\pi^{2}}  \int_{D_{SU(2)}} \!\! d\Omega \, D^{J \, \ast}_{MK}(\Omega) \,   {\cal H}(\tau,\Omega)  \nonumber \\ 
&=& \sum_{\mu} e^{-\tau E^{J}_{\mu}} \, E^{J}_{\mu}  \, N^{J \mu}_{MK}/N(\tau,0)    \, . \label{integratedenergy}
\end{eqnarray}
\end{subequations}
and the take the limit $\tau \to \infty$ to access the lowest eigenenergy $E^{J}_{0}$ through
\begin{eqnarray}
E^{J}_{0} &=& \lim\limits_{\tau \to \infty} \frac{\sum_{MK} f^{J\ast}_M \, f^{J}_K  \, h^{J}_{MK}(\tau)n^{J }_{MK}(\tau)}{\sum_{MK} f^{J\ast}_M \, f^{J}_K  \, n^{J }_{MK}(\tau)} . \label{yrast_projected_energy}
\end{eqnarray}
This above analysis is based on the exact kernels respecting the symmetries and requires the extraction of the IRREP of interest prior to taking the large time limit. As explained above, the large time limit of approximate kernels based on a symmetry breaking reference state still mixes the IRREPS of $SU(2)$. This can be used at our advantage to actually extract yrast states with various $J$ from the infinite time kernels, i.e. Eq.\ref{yrast_projected_energy} is eventually replaced by
\begin{eqnarray}
E^{J}_0 &=& \frac{\sum_{MK} f^{J\ast}_M \, f^{J}_K  \int_{D_{SU(2)}} \! d\Omega \, D^{J \, \ast}_{MK}(\Omega) \, {\cal H}_{\text{approx}}(\Omega)}{\sum_{MK} f^{J\ast}_M \, f^{J}_K  \int_{D_{SU(2)}} \! d\Omega \, D^{J \, \ast}_{MK}(\Omega) \, {\cal N}_{\text{approx}}(\Omega)} \, . \label{yrast_projected_energy2nd}
\end{eqnarray}

Everything exposed so far is valid independently of the many-body method employed to approximate the off-diagonal energy and norm kernels. The remainder of the paper is devoted to the computation of ${\cal N}(\tau,\Omega)$ and ${\cal H}(\tau,\Omega)$ on the basis of MBPT and CC techniques. Once this is achieved, the yrast spectroscopy of near-degenerate systems can be extracted through Eqs.~\ref{integratedkernels}-\ref{yrast_projected_energy}.

\section{Perturbation theory}
\label{sectionMBPT}

Coupled-cluster theory usually starts from a similarity-transformed Hamiltonian or from an energy functional. The Baker-Campbell-Hausdorff identity applied to the similarity-transformed Hamiltonian on the basis of the standard Wick theorem~\cite{shavitt09a} provides a direct access to the naturally terminating expansion of the diagonal energy kernel. In the present case, such a property cannot be obtained directly for the more general off-diagonal energy kernel at play. The off-diagonal Wick theorem~\cite{balian69a} that can be used to expand the kernel, which does not apply to {\it operators} but only to their {\it fully contracted part}, i.e. to the matrix element $\langle \Phi | O | \Phi (\Omega) \rangle$, does not permit to obtain straightforwardly the standard connected structure of the kernel. This situation makes necessary to first develop the perturbation theory of off-diagonal energy and norm kernels. With the perturbation theory at hand, we will be in position to elaborate the coupled cluster scheme in Sec.~\ref{CCtheory} and obtain the same connected structure as for the traditional diagonal energy kernel.

\subsection{Unperturbed system}
\label{chap:slater}

\begin{figure}[t!]
\begin{center}
\includegraphics[clip=,width=0.18\textwidth,angle=0]{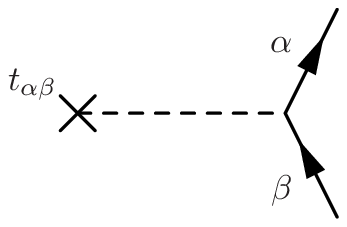}\\ \vspace{0.4cm}
\includegraphics[clip=,width=0.18\textwidth,angle=0]{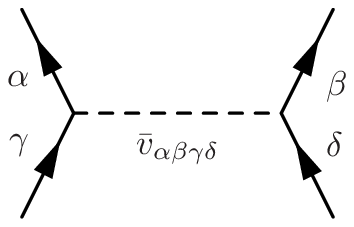}\\ \vspace{0.4cm}
\includegraphics[clip=,width=0.18\textwidth,angle=0]{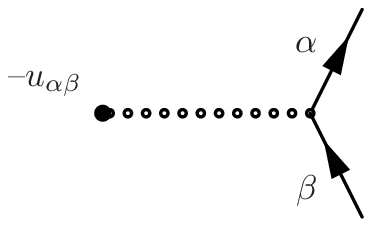}\\ \vspace{0.4cm}
\includegraphics[clip=,width=0.18\textwidth,angle=0]{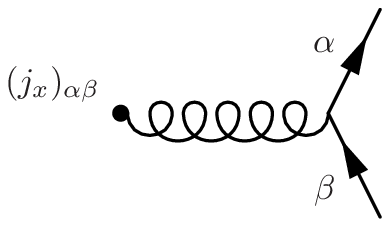}\\ \vspace{0.4cm}
\includegraphics[clip=,width=0.18\textwidth,angle=0]{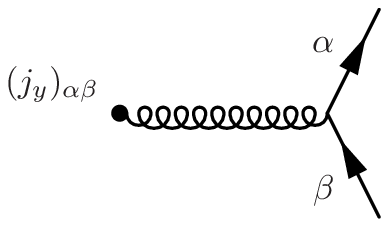}\\ \vspace{0.4cm}
\includegraphics[clip=,width=0.18\textwidth,angle=0]{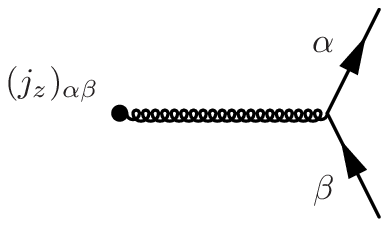}\\ \vspace{0.4cm}
\includegraphics[clip=,width=0.18\textwidth,angle=0]{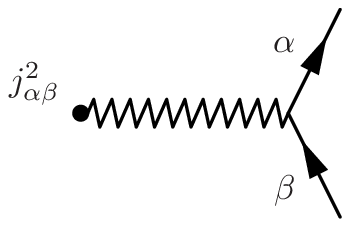}\\ \vspace{0.4cm}
\includegraphics[clip=,width=0.18\textwidth,angle=0]{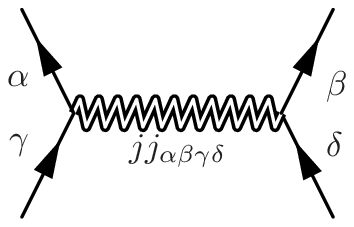}
\end{center}
\caption{
\label{variousvertices}
From top to bottom: diagrammatic representation of the operators $T$, $V$, $-U$, $J_x$, $J_y$, $J_z$, $J^2_{(1)}$ and $J^2_{(2)}$ in the Schroedinger representation.}
\end{figure}

The Hamiltonian is split into a one-body part $H_{0}$ and a residual two-body part $H_1$
\begin{equation}
\label{split1}
H \equiv H_{0} + H_{1} \, ,
\end{equation} 
such that $H_{0}\equiv T+U$ and $H_{1}\equiv V-U$, where $U$ is a one-body operator that remains to be specified. Having introduced $U$,  Fig.~\ref{variousvertices} displays for later use the diagrammatic representation of the various operators of interest in the Schroedinger representation.

While $H$, $T$ and $V$ commute with the transformations of $SU(2)$, we are interested in the case where $H_0$, and thus $H_1$, do {\it not} commute with $R(\Omega)$, i.e.
\begin{subequations}
\label{commutators}
\begin{eqnarray}
\left[H_0,R(\Omega)\right]&\neq& 0 \, , \label{commutators2} \\
\left[H_1,R(\Omega)\right]&\neq& 0 \, . \label{commutators3} 
\end{eqnarray}
\end{subequations}
For a given number of interacting Fermions, the key is to choose $H_0$ with a low-enough symmetry for its ground state $| \Phi \rangle$ to be non-degenerate with respect to particle-hole excitations\footnote{For example, $H_{0}$ can be taken as the the sum of the zero and one-body part obtained by normal-ordering $H$ with respect to the Slater determinant $|  \Phi \rangle $ minimizing the expectation value of $H$ under the possible breaking of $SU(2)$ symmetry, i.e. solving deformed Hartree-Fock equations. Correspondingly, $H_1$ is the (symmetry-breaking) normal-ordered two-body part of $V$. In this context, one can let the reference state break $SU(2)$ symmetry spontaneously or force it to do so via the addition of an appropriate Lagrange constraint.}. The product state $|  \Phi \rangle$ is {\it deformed} and is thus not an eigenstate of $J^2$; i.e. it spans several IRREPs of $SU(2)$\footnote{Although we do not consider it at this point, we will later specify the set of equations to the particular case where  $[H_0,J_z]=0$ and $J_z |  \Phi \rangle=0$, i.e. to the case where the reference state remains axially symmetric.}.

The operator $H_{0}$ can be written in diagonal form in terms of its deformed one-body eigenstates
\begin{equation}
H_{0}\equiv \sum_{\alpha}(e_{\alpha}\!-\!\mu) a_{\alpha}^{\dagger} a_{\alpha} \label{hzero} \, ,
\end{equation}
where the chemical potential $\mu$ is introduced for convenience. Associated creation and annihilation operators read in the interaction representation
\begin{subequations}
\label{aalphatau}
\begin{eqnarray}
a_{\alpha}^{\dagger}\left(  \tau\right)  &\equiv& e^{+\tau H_{0}} \, a_{\alpha}^{\dagger} \, e^{-\tau
H_{0}}=e^{+\tau\left(e_{\alpha}-\mu\right)  } \, a_{\alpha}^{\dagger} \, , \label{aalphatau1}\\
a_{\alpha} \left(  \tau\right)  &\equiv& e^{+\tau H_{0}} \, a_{\alpha} \, e^{-\tau H_{0}}=e^{-\tau\left(
e_{\alpha}-\mu\right)  } \, a_{\alpha} \label{aalphatau2} \, .
\end{eqnarray}
\end{subequations}
As mentioned above, the deformed Slater determinant $|  \Phi \rangle$ necessarily possesses a {\it closed-shell} character, i.e. there exists a finite energy gap between the fully occupied shells below the Fermi energy (chemical potential) and the unoccupied levels above. Thus, $|  \Phi \rangle$ is defined by $A$ occupied (hole) orbitals labeled by $(i,j,k,l\ldots)$
\begin{subequations}
\begin{eqnarray}
|  \Phi \rangle &\equiv& \prod_{i=1}^{A}a_{i}^{\dagger} \, | 0 \rangle \, ,  \label{slater} \\
H_{0}\, |  \Phi \rangle &=& \varepsilon_{0} \, |  \Phi \rangle \, , \\
\varepsilon_{0} &=& \sum_{i=1}^{A} (e_{i}-\mu)  \label{phi} \, ,
\end{eqnarray}
\end{subequations}
while the unoccupied (particle) orbitals are labeled by $(a,b,c,d\ldots)$. As already visible from Eq.~\ref{hzero}, states labeled $(\alpha,\beta,\gamma,\delta\ldots)$ relate from there on to any single-particle eigenstate of $H_0$. The chemical potential is chosen to lie in the energy gap separating particle and hole orbitals, i.e.
\begin{equation}
e_{a}>\mu\;\; \text{and}\;\;\;e_{i}<\mu\label{eph} \, .
\end{equation}
Excited eigenstates of $H_0$ are obtained as particle-hole excitations of $| \Phi \rangle$ 
\begin{eqnarray}
| \Phi^{ab\ldots}_{ij\ldots} \rangle &\equiv& A^{ab\ldots}_{ij\ldots}  |  \Phi \rangle \, , 
\end{eqnarray}
where
\begin{eqnarray}
A^{ab\ldots}_{ij\ldots} &\equiv& a^{\dagger}_{a} \, a_{i} \, a^{\dagger}_{b} \, a_{j} \ldots  \, ,  \label{phexcitation}
\end{eqnarray}
with the eigenenergy
\begin{subequations}
\begin{eqnarray}
H_{0}\,| \Phi^{ab\ldots}_{ij\ldots} \rangle &=& (\varepsilon_0 +\varepsilon^{ab\ldots}_{ij\ldots}) \, | \Phi^{ab\ldots}_{ij\ldots} \rangle \, , \\
\varepsilon^{ab\ldots}_{ij\ldots} &=& e_a\!+\! e_b\! +\!\ldots\! -\! e_i\! -\! e_j\! -\! \ldots \label{phiexcit} \, .
\end{eqnarray}
\end{subequations}

\subsection{Rotated reference state}
\label{chap:rotatedbasis}

Given the Slater determinant $|  \Phi \rangle$, we define its rotated partner
\begin{equation}
|  \Phi (\Omega) \rangle \equiv R(\Omega) |  \Phi \rangle = \prod_{i=1}^{A} a_{\bar{\imath}}^{\dagger} \, |  0 \rangle \, .
\end{equation}
where rotated orbitals are defined through
\begin{subequations}
\begin{eqnarray}
| \bar{\alpha} \rangle &\equiv& R(\Omega)  | \alpha \rangle = \sum_{\beta} R_{\beta\alpha}(\Omega)  | \beta \rangle \, , \\
 a_{\bar{\alpha}}^{\dagger} &\equiv& R(\Omega) \, a^{\dagger}_{\alpha} \, R^{\dagger}(\Omega) = \sum_{\beta} R_{\beta\alpha}(\Omega)  a_{\beta}^{\dagger} \, ,
\end{eqnarray}
\end{subequations}
with $R_{\alpha\beta}(\Omega) \equiv \langle \alpha | R(\Omega) | \beta \rangle $ the unitary transformation matrix connecting the rotated basis to the unrotated one. The Slater determinant $| \Phi(\Omega) \rangle$ is the ground-state of the rotated Hamiltonian $H_0(\Omega)\equiv R(\Omega) H_0 R^{\dagger}(\Omega)$ with the $\Omega$-independent eigenvalue $\varepsilon_0$. This feature characterizes the fact that, while the deformed unperturbed ground-state is non-degenerate with respect to particle-hole excitations, there exists a degeneracy, i.e. a zero mode, in the manifold of its rotated partners.

As proven in, e.g., Ref.~\cite{blaizot86}, the overlap between $|  \Phi \rangle$ and $|  \Phi (\Omega) \rangle$ can be expressed as
\begin{eqnarray}
\langle \Phi | \Phi(\Omega) \rangle &=& \text{det} M(\Omega) \, , \label{kernel}
\end{eqnarray}
where $M_{ij}(\Omega)$ is the $A\times A$ reduction of $R_{\alpha\beta}(\Omega)$ to the subspace of hole states of $| \Phi \rangle$. 

\subsection{Unperturbed off-diagonal density matrix}
\label{transitiondens}

We now introduce the unperturbed one-body off-diagonal density matrix $\rho(\Omega)$  defined through its matrix elements in the eigenbasis of $H_0$
\begin{eqnarray}
\rho_{\alpha\beta}(\Omega) &\equiv& \langle \alpha | \rho(\Omega) | \beta \rangle \equiv \frac{\langle \Phi | a_{\beta}^{\dagger} a_{\alpha}  | \Phi(\Omega) \rangle}{\langle \Phi | \Phi(\Omega) \rangle}  \, . \label{onebodyrhotrans}
\end{eqnarray}
It is possible to write it as~\cite{blaizot86}
\begin{equation}
\rho(\Omega)  = \sum_{ij=1}^{A} |  \bar{\imath} \rangle \, M_{ij}^{-1}(\Omega) \, \langle j |  \label{rhohh} \, ,
\end{equation}
such that it acquires the form
\begin{eqnarray}
\rho(\Omega) &\equiv& \left(
\begin{array} {cc}
\rho^{hh}(\Omega) & \rho^{hp}(\Omega) \\
\rho^{ph}(\Omega) & \rho^{pp}(\Omega) 
\end{array}
\right) \nonumber \\
&=& \left(
\begin{array} {cc}
\bbone^{hh} & 0   \\
0 & 0
\end{array}
\right) + \left(
\begin{array} {cc}
0 & 0  \\
R(\Omega)M^{-1}(\Omega) & 0
\end{array}
\right) \nonumber \\
&\equiv& \rho + \rho^{ph}(\Omega) \, , \label{transdens}
\end{eqnarray}
where $\bbone^{hh}$ is the identity operator on the hole subspace of the one-body Hilbert space. In Eq.~\ref{transdens}, $\rho\equiv\rho(0)$ is nothing but the diagonal one-body density matrix associated with the unrotated reference state $| \Phi \rangle$. The $\Omega$-dependent part $\rho^{ph}(\Omega)$, which only connects particle kets to hole bras, vanishes for $\Omega=0$, i.e. $\rho^{ph}(0)=0$. The above partitioning of the off-diagonal density matrix can be summarized by writing its matrix elements under the form 
\begin{eqnarray}
\rho_{\alpha\beta}(\Omega) &\equiv& n_{\alpha} \, \delta_{\alpha\beta} +  (1-n_{\alpha}) \, n_{\beta} \, \rho^{ph}_{\alpha\beta}(\Omega) \, , \label{truc} 
\end{eqnarray}
where $n_{i}=1$ for hole states and $n_{a}=0$ for particle states. Making particle and hole indices explicit, one obtains equivalently
\begin{subequations}
\label{contractionsrho}
\begin{eqnarray}
\rho_{a b}(\Omega) &=& 0 \, , \label{contractionsrho1} \\
\rho_{i b}(\Omega) &=& 0 \, , \label{contractionsrho2} \\
\rho_{ij}(\Omega) &=& \delta_{ij} \, , \label{contractionsrho3} \\
\rho_{aj}(\Omega) &=& \rho^{ph}_{aj}(\Omega) \, . \label{contractionsrho4}
\end{eqnarray}
\end{subequations}
Similarly, one introduces
\begin{eqnarray}
(\bbone-\rho(\Omega))_{\alpha\beta} &\equiv&  \frac{\langle \Phi |  a_{\alpha} a_{\beta}^{\dagger}  | \Phi(\Omega) \rangle}{\langle \Phi | \Phi(\Omega) \rangle}  \, , \label{1minusrho}
\end{eqnarray}
whose properties can be summarized through
\begin{subequations}
\label{contractions1moinsrho}
\begin{eqnarray}
(\bbone-\rho(\Omega))_{ab} &=& \delta_{ab} \, , \label{contractions1moinsrho1} \\
(\bbone-\rho(\Omega))_{ib} &=& 0 \, , \label{contractions1moinsrho2} \\
(\bbone-\rho(\Omega))_{ij} &=& 0 \, , \label{contractions1moinsrho3} \\
(\bbone-\rho(\Omega))_{aj} &=& -\rho^{ph}_{aj}(\Omega) \, . \label{contractions1moinsrho4}
\end{eqnarray}
\end{subequations}

\subsection{Unperturbed off-diagonal propagator}
\label{prop}

\begin{figure}[t!]
\begin{center}
\includegraphics[clip=,width=0.05\textwidth,angle=0]{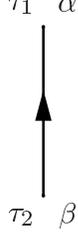}\\
\end{center}
\caption{
\label{prop1}
Diagrammatic representation of the unperturbed off-diagonal one-body propagator $G^{0}_{\alpha\beta}(\tau_1,\tau_2 ; \Omega)$.}
\end{figure}

The unperturbed off-diagonal one-body propagator $G^{0}(\Omega)$ is defined through its matrix elements in the eigenbasis of $H_0$
\begin{eqnarray}
\langle \alpha \tau_1 | G^{0}(\Omega) | \beta \tau_2 \rangle &\equiv& \frac{\langle \Phi |  \textmd{T}[a_{\alpha}(\tau_1) a_{\beta}^{\dagger}(\tau_2)] | \Phi(\Omega) \rangle}{\langle \Phi | \Phi(\Omega) \rangle}  \nonumber \\
&\equiv & G^{0}_{\alpha\beta}(\tau_1,\tau_2 ; \Omega)  \, , \label{onebodyproptrans}
\end{eqnarray}
where $\textmd{T}$ denotes the time ordering operator. Combining Eqs.~\ref{aalphatau} and \ref{truc} together with the anti commutation of creation and annihilation operators, one rewrites the propagator as
\begin{widetext}
\begin{eqnarray}
G^{0}_{\alpha\beta}(\tau_1,\tau_2 ; \Omega) &=&  e^{-\tau_{1}\left(e_{\alpha}-\mu\right)} \, e^{\tau_{2}\left(e_{\beta}-\mu\right)} \{
\theta\left(  \tau_{1}-\tau_{2}\right) \, (\bbone-\rho(\Omega))_{\alpha\beta} -\theta\left(  \tau_{2}-\tau
_{1}\right) \rho_{\alpha\beta}(\Omega)\} \nonumber \\
&\equiv& G^{0}_{\alpha\beta}(\tau_1,\tau_2) + G^{0,ph}_{\alpha\beta}(\tau_1,\tau_2 ; \Omega) \, , \label{decompoprop}
\end{eqnarray}
where $\theta\left(  \tau\right)$ is the Heaviside function and where
\begin{subequations}
\label{twocontribprop}
\begin{eqnarray}
G^{0}_{\alpha\beta}(\tau_1,\tau_2) &\equiv& + e^{-\left(\tau_{1}-\tau_{2}\right)\left(e_{\alpha}-\mu\right)} \, \{ \theta\left(  \tau_{1}-\tau_{2}\right)   \left(1-n_{\alpha}\right)  -\theta\left(\tau_{2}-\tau_{1}\right)  n_{\alpha} \} \, \delta_{\alpha\beta} \, , \label{twocontribprop1} \\
G^{0,ph}_{\alpha\beta}(\tau_1,\tau_2 ; \Omega) &\equiv& - e^{-\tau_{1}\left(e_{\alpha}-\mu\right)} \, e^{\tau_{2}\left(e_{\beta}-\mu\right)} (1-n_{\alpha}) \, n_{\beta} \, \rho^{ph}_{\alpha\beta}(\Omega) \, . \label{twocontribpro2}
\end{eqnarray}
\end{subequations}
\end{widetext}
The propagator $G^{0}_{\alpha\beta}(\tau_1,\tau_2 ; \Omega)$ is displayed diagrammatically in Fig.~\ref{prop1} where its $\Omega$ dependence is left implicit. 
The $\Omega$-independent part $G^{0}=G^{0}(0)$ is nothing but the standard unperturbed propagator associated with $| \Phi \rangle$. Correspondingly, the $\Omega$-dependent (purely off-diagonal) part $G^{0,ph}(\Omega)$, which connects particle bras to hole kets, vanishes for $\Omega=0$, i.e. $G^{0,ph}(0)=0$. Whereas $G^{0}$ depends only on the difference of its time arguments, it is not the case of $G^{0,ph}(\Omega)$. In both contributions, one further notices that the coefficients in front of the (positive) time variables are necessarily strictly negative. The equal time propagator must be treated separately. Because it only arises from the contraction of two operators belonging to the same vertex, it must be related to the part of  $G^{0}_{\alpha\beta}(\tau_1,\tau_2 ; \Omega)$ that displays the creation and annihilation operators in the order $a_{\beta}^{\dagger}(\tau_2)a_{\alpha}(\tau_1)$. Given the definition of $G^{0}_{\alpha\beta}(\tau_1,\tau_2 ; \Omega)$, this corresponds to defining the equal-time propagator according to 
\begin{eqnarray}
G^{0}_{\alpha\beta}(\tau,\tau ; \Omega) &\equiv& -\frac{\langle \Phi |  a_{\beta}^{\dagger}(\tau) a_{\alpha}(\tau) | \Phi(\Omega) \rangle}{\langle \Phi | \Phi(\Omega) \rangle}   \nonumber \\
&=& -e^{-\tau\left(  e_{\alpha}-e_{\beta}\right)  } \, \rho_{\alpha\beta}(\Omega) \nonumber \\
&\equiv& G^{0}_{\alpha\beta}(\tau,\tau) + G^{0,ph}_{\alpha\beta}(\tau,\tau ; \Omega) \, , \label{decompopropsametime}
\end{eqnarray}
where
\begin{subequations}
\label{twocontribpropequaltime}
\begin{eqnarray}
G^{0}_{\alpha\beta}(\tau_,\tau)\! &\equiv&\! -n_{\alpha} \, \delta_{\alpha\beta}  \, , \label{twocontribpropequaltime1} \\
G^{0,ph}_{\alpha\beta}(\tau,\tau ; \Omega)\! &\equiv&\! -e^{-\tau\left(e_{\alpha}-e_{\beta}\right)  } (1-n_{\alpha})  n_{\beta} \rho^{ph}_{\alpha\beta}(\Omega)  \, . \label{twocontribpropequaltime2}
\end{eqnarray}
\end{subequations}
The diagonal equal-time propagator $G^{0}_{\alpha\beta}(\tau_,\tau)$ is independent of $\tau$ whereas $G^{0,ph}_{\alpha\beta}(\tau,\tau ; \Omega)$ does depend on time.

\subsection{Expansion of the evolution operator}

As recalled in App.~\ref{perturbativeannexe}, the "evolution operator" can be expanded in powers of $H_{1}$ under the form
\begin{eqnarray}
{\cal U}(\tau) &=& e^{-\tau H_{0}} \, \textmd{T}e^{-\int_{0}^{\tau}dt H_{1}\left(t\right) \, ,
} \label{evol1}%
\end{eqnarray}
where
\begin{equation}
H_{1}\left( \tau\right)  \equiv e^{\tau H_{0}}H_{1}e^{-\tau H_{0}} \, ,
\end{equation}
defines the perturbation in the interaction representation. 

\subsection{Norm kernel}
\label{normkernel2}

\subsubsection{Expansion}

Expressing $V$ and $-U$ (and thus $H_1$) in the eigenbasis of $H_0$, one obtains from Eq.~\ref{evol1}
\begin{widetext}
\begin{eqnarray}
N(\tau,\Omega) &=& \langle \Phi | e^{-\tau H_{0}} \, \textmd{T}e^{-\int_{0}^{\tau}dt H_{1}\left(t\right)} | \Phi(\Omega) \rangle \nonumber  \\
&=& e^{-\tau\varepsilon_{0}} \langle \Phi |\Big\{  1-\int_{0}^{\tau}d\tau_1 H_{1}\left(  \tau_1\right)  +\frac{1}{2!}\int_{0}^{\tau}d\tau_{1}d\tau
_{2}\textmd{T}\left[  H_{1}\left(  \tau_{1}\right)  H_{1}\left(  \tau_{2}\right)  \right]
+... \Big\}|  \Phi(\Omega) \rangle \nonumber\\
&=& e^{-\tau\varepsilon_{0}}\Big\{  \sum_{k=0}^{\infty}\frac{\left(  -\right)
^{k}}{k!}\int_{0}^{\tau}d\tau_{1}\ldots d\tau_{k}\,\frac{1}{2}\sum_{\alpha_1\beta_1\gamma_1\delta_1}v_{\alpha_1\beta_1\gamma_1\delta_1} \ldots \frac{1}{2}\sum_{\alpha_k\beta_k\gamma_k\delta_k}v_{\alpha_k\beta_k\gamma_k\delta_k}  \nonumber\\
&& \hspace{3cm} \times \langle \Phi |  \textmd{T}\left[  a_{\alpha_1}^{\dagger}\left(  \tau_{1}\right)
a_{\beta_1}^{\dagger}\left(  \tau_{1}\right)  a_{\delta_1}\left(  \tau_{1}\right)
a_{\gamma_1}\left(  \tau_{1}\right) \ldots a_{\alpha_k}^{\dagger}\left(  \tau_{k}\right)
a_{\beta_k}^{\dagger}\left(  \tau_{k}\right)  a_{\delta_k}\left(  \tau_{k}\right)
a_{\gamma_k}\left(  \tau_{k}\right)\right]  |  \Phi(\Omega) \rangle \nonumber \\
&& \hspace{1cm} +  \sum_{k=0}^{\infty}\frac{\left(  -\right)
^{k}}{k!}\int_{0}^{\tau}d\tau_{1}\ldots d\tau_{k} \sum_{\alpha\beta}(-u_{\alpha_1\beta_1})\ldots  \sum_{\alpha_k\beta_k}(-u_{\alpha_k\beta
_k})\langle \Phi |  \textmd{T}\left[  a_{\alpha_1}^{\dagger}\left(  \tau_{1}\right)
a_{\beta_1}\left(  \tau_{1}\right) \ldots  a_{\alpha_k}^{\dagger}\left(  \tau_{k}\right)
a_{\beta_k}\left(  \tau_{k}\right)\right]  |  \Phi(\Omega) \rangle \nonumber\\
&& + \, \text{all cross terms involving both} \, \, V \, \, \text{and} \, \, -\!U \, \Big\} . \label{expansionnormkernel}
\end{eqnarray}
\end{widetext}
The off-diagonal matrix elements of products of time-dependent field operators appearing in Eq.~\ref{expansionnormkernel} can be
expressed~\cite{balian69a} as the sum of all possible systems of products of elementary contractions $G^{0}_{\alpha\beta}(\tau_1,\tau_2 ; \Omega)$ (Eq.~\ref{onebodyproptrans}), eventually multiplied by the unperturbed norm kernel $\langle \Phi | \Phi(\Omega) \rangle$ (Eq.~\ref{kernel}). Consequently, it is possible to represent $N(\tau,\Omega)$ diagrammatically following standard~\cite{blaizot86} techniques usually applied~\cite{bloch58a} to the diagonal norm kernel $N(\tau,0)$. Details of the diagrammatic approach are given in App.~\ref{diagrams}.

The above considerations rely on a generalized Wick theorem for matrix elements of products of field operators between {\it different} (non-orthogonal) left and right vacua, i.e. presently $\langle \Phi |$ and $| \Phi(\Omega) \rangle$. This constitutes an efficient way to deal exactly with the presence of the rotation operator $R(\Omega)$. The off-diagonal Wick theorem~\cite{balian69a} only holds for {\it matrix elements} of products of operators, i.e. no extension of the standard Wick theorem holds for the operators themselves and no analogue of normal ordering can be used in the present context.

\subsubsection{Exponentiation of connected diagrams}

Diagrams representing the off-diagonal norm kernel are vacuum-to-vacuum diagrams, i.e. diagrams with no incoming or outgoing external lines. In general, a diagram consists of disconnected parts which are joined neither by vertices nor by propagators. Consider a diagram contributing to Eq.~\ref{expansionnormkernel} and consisting of $n_{1}$ identical unlabeled connected parts $\Gamma_{1}(\tau,\Omega)$, of $n_{2}$ identical unlabeled connected parts $\Gamma_{2}(\tau,\Omega)$, and so on. Using for simplicity the same symbol to designate both the diagram and its contribution, the whole diagram gives
\begin{equation}
\Gamma(\tau,\Omega) = \frac{\left[\Gamma_{1}(\tau,\Omega)\right]^{n_{1}}}{n_{1}!}\frac{\left[  \Gamma_{2}(\tau,\Omega)\right]^{n_{2}}}{n_{2}!}...
\end{equation}
The factor $n_{1}!$ is the symmetry factor due to the exchange of time labels among the $n_{1}$ identical diagrams $\Gamma_{1}$ (see App.~\ref{diagrams}). It follows that the sum of all vacuum-to-vacuum diagrams is equal to the exponential of the sum of {\it connected} vacuum-to-vacuum diagrams
\begin{eqnarray}
\sum_{\Gamma} \Gamma(\tau,\Omega) &=& \sum_{n_{1}n_{2}...}\frac{\left[  \Gamma_{1}(\tau,\Omega)\right]^{n_{1}}}{n_{1}!}\frac{\left[\Gamma_{2}(\tau,\Omega)\right]^{n_{2}}}{n_{2}!}... \nonumber \\
&=& e^{\Gamma_{1}(\tau,\Omega)+\Gamma_{2}(\tau,\Omega)+...}  \label{condiag} \, .
\end{eqnarray}
Consequently, the norm can be written as
\begin{equation}
N(\tau,\Omega) =   e^{-\tau \varepsilon_{0} + n(\tau,\Omega)}  \, \langle \Phi |  \Phi(\Omega) \rangle \label{wexp3} \, ,
\end{equation}
where $n(\tau,\Omega)\equiv\sum^{\infty}_{n=1}n^{(n)}(\tau,\Omega)$, with $n^{(n)}(\tau,\Omega)$ the sum of all $\Omega$-dependent connected vacuum-to-vacuum diagrams of order $n$.

\subsubsection{Computing diagrams}
\label{computingdiagramsNc}

\begin{figure}[t!]
\begin{center}
\includegraphics[clip=,width=0.22\textwidth,angle=0]{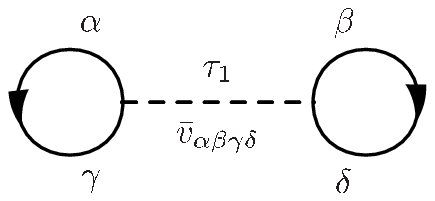}\\ \vspace{0.4cm}
\includegraphics[clip=,width=0.22\textwidth,angle=0]{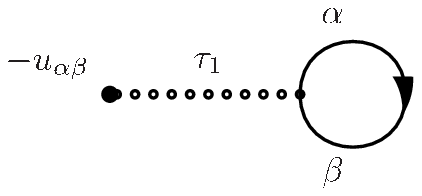}\\
\end{center}
\caption{
\label{diagramsNc}
First-order Feynman diagrams contributing to $n(\tau,\Omega)$.}
\end{figure}

\begin{figure}[t!]
\begin{center}
\includegraphics[clip=,width=0.18\textwidth,angle=0]{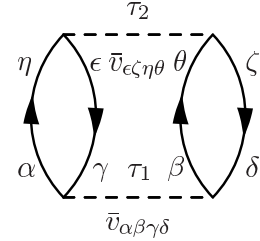}\\ \vspace{0.4cm}
\includegraphics[clip=,width=0.18\textwidth,angle=0]{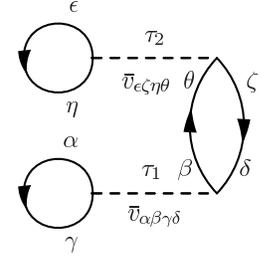}\\ \vspace{0.4cm}
\includegraphics[clip=,width=0.18\textwidth,angle=0]{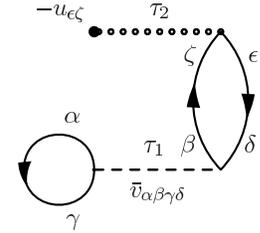}\\ \vspace{0.4cm}
\includegraphics[clip=,width=0.18\textwidth,angle=0]{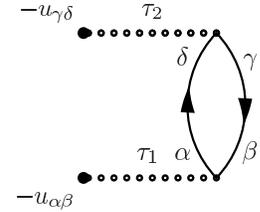}\\ 
\end{center}
\caption{
\label{diagramsNc2}
Second-order Feynman diagrams contributing to $n(\tau,\Omega)$.}
\end{figure}

First- and second-order diagrams contributing to $n(\tau,\Omega)$ are displayed in Figs.~\ref{diagramsNc} and~\ref{diagramsNc2}, respectively, where a propagator line denotes $G^{0}(\Omega)$. The actual calculation of the diagrams is performed in detail in  App.~\ref{diagrams}. For illustration, the starting expression of the first diagram appearing in Fig.~\ref{diagramsNc} reads as
\begin{widetext}
\begin{equation}
n^{(1)}_{V}(\tau,\Omega)=-\frac{1}{2} \sum_{\alpha\beta\gamma\delta}\int_{0}^{\tau}d\tau_1 \, \bar{v}_{\alpha\beta\gamma\delta} \, G^{0}_{\gamma\alpha}(\tau_1,\tau_1 ; \Omega) \, G^{0}_{\delta\beta}(\tau_1,\tau_1 ; \Omega) \label{gamharbody} \, ,
\end{equation}
while the starting expression of the first diagram appearing in Fig.~\ref{diagramsNc2} is
\begin{eqnarray}
n^{(2)}_{V1}(\tau,\Omega) &=& \frac{1}{8}\sum_{\alpha\beta\gamma\delta}\sum_{\epsilon\zeta\eta\theta}\int_{0}^{\tau}\!\!\!d\tau_{1}\!
\int_{0}^{\tau}\!\!\!d\tau_{2}\; \bar{v}_{\alpha\beta\gamma\delta} \, \bar{v}_{\epsilon\zeta\eta\theta} \, G^{0}_{\gamma\epsilon}(\tau_1,\tau_2 ; \Omega) \, G^{0}_{\eta\alpha}(\tau_2,\tau_1 ; \Omega) \, G^{0}_{\delta\zeta}(\tau_1,\tau_2 ; \Omega) \, G^{0}_{\theta\beta}(\tau_2,\tau_1 ; \Omega) \, . \label{total2ndorderdiagbody}
\end{eqnarray}
\end{widetext}

\subsubsection{Dependence on $\tau$ and $\Omega$}
\label{structure1}

Diagrams $n^{(n)}\left(\tau,\Omega\right)$ can always be split according to
\begin{equation}
n^{(n)}\left(\tau,\Omega\right) \equiv n^{(n)}\left(\tau\right)  + \aleph^{(n)}\left(\tau,\Omega\right) \, , \label{split1a}
\end{equation}
where $n^{(n)}\left(\tau\right) \equiv n^{(n)}\left(\tau,0\right)$ is the sum of vacuum-to-vacuum connected diagrams arising in standard, i.e. diagonal, MBPT~\cite{bloch58a}. For a given diagram $n^{(n)}\left(\tau,\Omega\right)$ the two terms on the right hand-side of Eq.~\ref{split1a} are obtained by splitting each propagator $G^{0}(\Omega)$ according to Eqs.~\ref{decompoprop}-\ref{twocontribprop}. More specifically, $n^{(n)}\left(\tau\right)$ is obtained by replacing {\it all} propagators $G^{0}(\Omega)$ by {\it diagonal} ones $G^{0}$. Correspondingly, $\aleph^{(n)}\left(\tau,\Omega\right)$ sums the contributions generated by taking {\it at least} one propagator to be $G^{0,ph}(\Omega)$. As a consequence of Eq.~\ref{split1a}, Eq.~\ref{wexp3} becomes
\begin{eqnarray}
N(\tau,\Omega) &\equiv& N(\tau,0) \, e^{\aleph (\tau,\Omega)}  \, \langle \Phi |  \Phi(\Omega) \rangle  \, , \label{splittedN} 
\end{eqnarray}
where 
\begin{eqnarray}
N(\tau,0) &=& e^{-\tau\varepsilon_{0}+n\left(\tau\right)}   \label{betalimtruc1} \, , 
\end{eqnarray}
and  $\aleph(\tau,0)=0$.  In view of Eq.~\ref{limitnorm}, one is interested in the large $\tau$ limit
\begin{subequations}
\label{betalim}
\begin{eqnarray}
\underset{\tau\rightarrow\infty}{\lim}n\left(\tau\right)  &\equiv& -\tau \Delta E^{J_0}_0 + \ln\left[\sum_{M}  |\langle \Phi |  \Psi^{J_0 M}_{0} \rangle|^2\right] \label{betalim1} \, , \\
 \underset{\tau\rightarrow\infty}{\lim} \aleph\left(\tau,\Omega\right) &\equiv& \aleph\left(\Omega\right) \label{betalim2} \, ,
\end{eqnarray}
\end{subequations}
where the correction to the unperturbed ground-state energy is given by 
\begin{eqnarray}
\Delta E^{J_0}_0 &\equiv& E^{J_0}_0-\varepsilon_0 \nonumber \\
&=& \langle \Phi | H_{1} \sum_{k=1}^{\infty} \left(\frac{1}{\varepsilon_0-H_0} H_{1} \right)^{k-1} | \Phi \rangle_{c} \, , \label{frombloch}%
\end{eqnarray}
and is nothing but the usual Goldstone's formula~\cite{goldstone57a} computed relative to the deformed reference state $| \Phi \rangle$. This expansion of the ground-state energy does not constitute the solution to the problem of present interest but is anyway recovered as a byproduct. Relation~\ref{betalim1} was demonstrated in Ref.~\cite{bloch58a} and proves that, in the large $\tau$ limit, the $\Omega$-independent part $n\left(\tau\right)$ is  made of a term independent of $\tau$ plus a term linear in $\tau$. Contrarily, Eq.~\ref{betalim2} states that the $\Omega$-dependent counterpart $\aleph\left(\tau,\Omega\right)$ is independent of $\tau$, i.e. it converges to a finite value when $\tau$ goes to infinity. These characteristic behaviors at large imaginary time are proven for any arbitrary order in App.~\ref{structure2}.

In Eq.~\ref{betalim1}, the contribution that does not depend on $\tau$ provides the overlap between the unrotated unperturbed state and the correlated ground-state. This overlap is not equal to $1$, which underlines that the expansion of $N(\tau,\Omega)$ does not rely on intermediate normalization at $\Omega=0$. Equation~\ref{betalim2} only contains a term independent of $\tau$ because the dependence on $\Omega$ of the norm kernel does not modify $\Delta E^{J_0}_0$ but simply amends the overlap between  $| \Phi(\Omega) \rangle$ and the eigenstate selected in the large $\tau$ limit.

\subsection{Energy kernel}
\label{energykernel}

\subsubsection{Expansion}

Proceeding similarly to $N(\tau,\Omega)$, and taking the energy as a particular example, one obtains the perturbative expansion of an operator kernel as
\begin{widetext}
\begin{eqnarray}
H(\tau,\Omega) &=& \langle \Phi | e^{-\tau H_{0}} \, \textmd{T}e^{-\int_{0}^{\tau}dt H_{1}\left(t\right)} (T+V) | \Phi(\Omega) \rangle \nonumber  \\
&=& e^{-\tau\varepsilon_{0}} \langle \Phi | T(0)-\int_{0}^{\tau}d\tau_1 \textmd{T}\left[H_{1}\left(  \tau_1\right) T(0)\right]  +\frac{1}{2!}\int_{0}^{\tau}d\tau_{1}d\tau
_{2}\textmd{T}\left[  H_{1}\left(  \tau_{1}\right)  H_{1}\left(  \tau_{2}\right) T(0)  \right]
+... |  \Phi(\Omega) \rangle \nonumber\\
&=& e^{-\tau\varepsilon_{0}}\Big\{  \sum_{p=0}^{\infty}\frac{\left(  -\right)
^{p}}{p!}\int_{0}^{\tau}d\tau_{1}\ldots d\tau_{p}\,\frac{1}{2}\sum_{\alpha_1\beta_1\gamma_1\delta_1}v_{\alpha_1\beta_1\gamma_1\delta_1} \ldots \frac{1}{2}\!\!\!\sum_{\alpha_p\beta_p\gamma_p\delta_p}v_{\alpha_p\beta_p\gamma_p\delta_p} \sum_{\epsilon\zeta} t_{\epsilon\zeta}  \nonumber\\
&& \hspace{1.5cm} \times \langle \Phi |  \textmd{T}\left[  a_{\alpha_1}^{\dagger}\left(  \tau_{1}\right)
a_{\beta_1}^{\dagger}\left(  \tau_{1}\right)  a_{\delta_1}\left(  \tau_{1}\right)
a_{\gamma_1}\left(  \tau_{1}\right) \ldots a_{\alpha_p}^{\dagger}\left(  \tau_{p}\right)
a_{\beta_p}^{\dagger}\left(  \tau_{p}\right)  a_{\delta_p}\left(  \tau_{p}\right)
a_{\gamma_p}\left(  \tau_{p}\right) a_{\epsilon}^{\dagger}\left(0\right)  a_{\zeta}\left(0\right)\right]  |  \Phi(\Omega) \rangle \nonumber \\
&& \hspace{0.8cm}  + \sum_{p=0}^{\infty}\frac{\left(  -\right)
^{p}}{p!}\int_{0}^{\tau}d\tau_{1}\ldots d\tau_{p}\, \sum_{\alpha_1\beta_1}(-u_{\alpha_1\beta_1})\ldots \sum_{\alpha_p\beta_p}(-u_{\alpha_p\beta_p}) \sum_{\epsilon\zeta} t_{\epsilon\zeta}   \nonumber \\
&& \hspace{1.5cm} \times \langle \Phi |  \textmd{T}\left[  a_{\alpha_1}^{\dagger}\left(  \tau_{1}\right)
a_{\beta_1}\left(  \tau_{1}\right) \ldots a_{\alpha_p}^{\dagger}\left(  \tau_{p}\right)
a_{\beta_p}\left(  \tau_{p}\right)  a_{\epsilon}^{\dagger}\left(0\right)  a_{\zeta}\left(0\right)\right]  |  \Phi(\Omega) \rangle \nonumber\\
&& + \, \text{all cross terms involving both} \, \, V \, \, \text{and} \, \, -\!U \,  \nonumber\\
&& + \, \text{the same set of terms obtained when replacing} \, \, T(0) \, \, \text{by} \, \, V(0) \Big\} \,. \label{expansionenergykernel}
\end{eqnarray}
\end{widetext}
We have assigned a time $t=0$ to the time-independent operators $T$ and $V$ stemming from $H$ in the definition of $H(\tau,\Omega)$. This allows us to insert them inside the product of time-ordered operators. Expansion~\ref{expansionenergykernel} differs from the one of $N(\tau,\Omega)$ by the presence of the operator $T$ or $V$ at a fixed time $(t=0)$. Consequences are discussed at length in App.~\ref{diagramsE}. Just as for $N(\tau,\Omega)$, it is possible to express $H(\tau,\Omega)$ diagrammatically
\begin{equation}
H(\tau,\Omega) = e^{-\tau\varepsilon_{0}} \, \sum^{\infty}_{n=0}\left[ T^{(n)}(\tau,\Omega) + V^{(n)}(\tau,\Omega) \right] \langle \Phi |  \Phi(\Omega) \rangle \,  ,
\end{equation}
where $T^{(n)}(\tau,\Omega)$ ($V^{(n)}(\tau,\Omega)$) denotes the sum of all vacuum-to-vacuum diagrams of order $n$ that include the operator $T$ ($V$) at fixed time $t=0$. The convention is that the zero-order diagram $T^{(0)}(\tau,\Omega)$ ($V^{(0)}(\tau,\Omega)$) solely contains the fixed-time operator $T$ ($V$).

\begin{figure}[t!]
\begin{center}
\includegraphics[clip=,width=0.21\textwidth,angle=0]{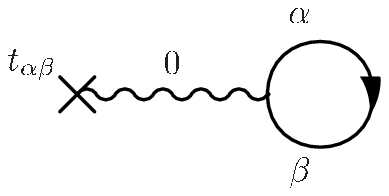}\\ \vspace{0.4cm}
\includegraphics[clip=,width=0.18\textwidth,angle=0]{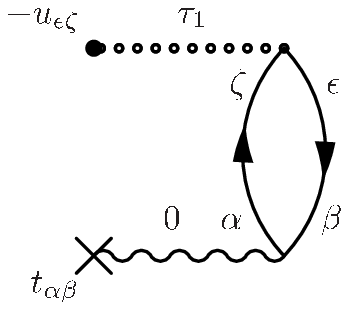}\\ \vspace{0.4cm}
\includegraphics[clip=,width=0.19\textwidth,angle=0]{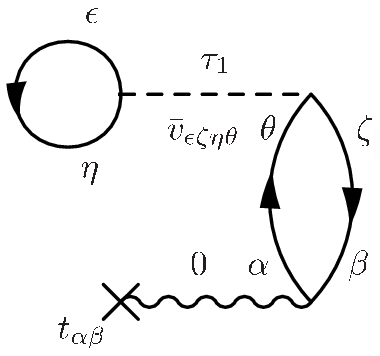}\\
\end{center}
\caption{
\label{diagramsTL}
Zero- and first-order Feynman diagrams contributing to $t(\tau,\Omega)$.}
\end{figure}

\subsubsection{Exponentiation of disconnected diagrams}

Any diagram $T^{(n)}(\tau,\Omega)$ ($V^{(n)}(\tau,\Omega)$) consists of a part that is {\it linked} to the operator $T(0)$ ($V(0)$), i.e. that results from contractions involving the creation and annihilation operators of $T(0)$ ($V(0)$), and parts that are unlinked. Effectively, each vacuum-to-vacuum diagram linked to $T(0)$ ($V(0)$) multiplies the complete set of vacuum-to-vacuum diagrams making up $N(\tau,\Omega)$. As a result, one obtains the factorization
\begin{equation}
H(\tau,\Omega) \equiv h(\tau,\Omega) \, N(\tau,\Omega) \label{linked1} \, ,
\end{equation}
with
\begin{subequations}
\label{linked2}
\begin{eqnarray}
h(\tau,\Omega) &\equiv& t(\tau,\Omega) +  v(\tau,\Omega)  \label{linked2b} \\ 
&\equiv&  \sum^{\infty}_{n=0} \left[t^{(n)}(\tau,\Omega) +  v^{(n)}(\tau,\Omega)\right] \label{linked2a}  \, ,
\end{eqnarray}
\end{subequations}
where $t(\tau,\Omega)$ ($v(\tau,\Omega)$) denotes the sum of all connected vacuum-to-vacuum diagrams of order $n$ {\it linked} to $T(0)$ ($V(0)$). 

The fact that the (reduced) kernel of an operator $O$ (${\cal O}$) factorizes into its linked/connected part $o(\tau,\Omega)$ times the (reduced) norm kernel   $N(\tau,\Omega)$ (${\cal N}(\tau,\Omega)$) is a fundamental result that will be exploited extensively in the remainder of the paper.

\subsubsection{Large $\tau$ limit}

According to Eq.~\ref{limitkernels}, $N(\tau,\Omega)$ and $H(\tau,\Omega)$ carry the same dependence on $\Omega$ in the large $\tau$ limit, which leads to the remarkable result that the complete sum $h(\Omega)$ of {\it all} vacuum-to-vacuum diagrams linked to the fixed-time operator $H(0)$ is independent of $\Omega$ in this limit. This corresponds to the fact that the expansion does fulfill the symmetry in the exact limit independently of whether the expansion is performed around a symmetry conserving or symmetry breaking reference state. In the latter case, however, each individual contribution $h^{(n)}(\Omega)$ or any partial sum of diagrams carries a dependence on $\Omega$. While the dependence of $\aleph\left(\Omega\right)$ on $\Omega$ is genuine, the dependence of $h(\Omega)$ is not and must be dealt with to restore the symmetry.

\begin{figure}[t!]
\begin{center}
\includegraphics[clip=,width=0.22\textwidth,angle=0]{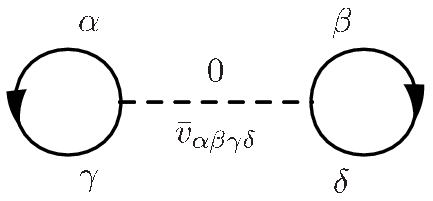}\\ \vspace{0.4cm}
\includegraphics[clip=,width=0.18\textwidth,angle=0]{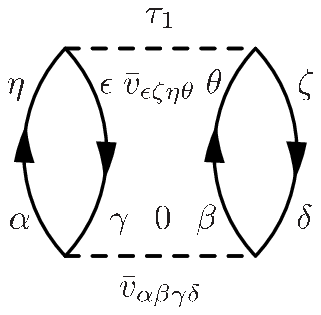}\\ \vspace{0.4cm}
\includegraphics[clip=,width=0.19\textwidth,angle=0]{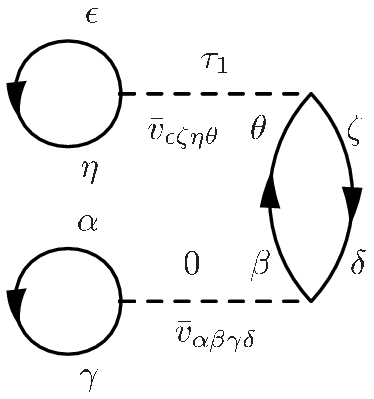}\\ \vspace{0.4cm}
\includegraphics[clip=,width=0.19\textwidth,angle=0]{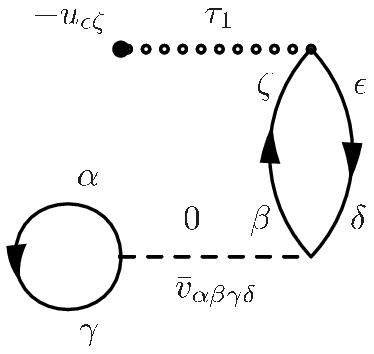}\\
\end{center}
\caption{
\label{diagramsVL}
Zero- and first-order Feynman diagrams contributing to $v(\tau,\Omega)$.}
\end{figure}

\subsubsection{Computing diagrams}
\label{computingdiagramsEL}

Zero- and first-order diagrams contributing to $t(\tau,\Omega)$ and $v(\tau,\Omega)$ are displayed in Figs.~\ref{diagramsTL} and~\ref{diagramsVL}, respectively.  The actual calculation of those diagrams is performed in detail in  App.~\ref{diagramsE}. For illustration, the starting expressions of the  zero order contribution to $v(\tau,\Omega)$, along with the first first-order diagram appearing in Fig.~\ref{diagramsVL} are given by
\begin{eqnarray}
v^{(0)}(\tau,\Omega) &=& \frac{1}{2} \sum_{\alpha\beta\gamma\delta}  \bar{v}_{\alpha\beta\gamma\delta} \, G^{0}_{\gamma\alpha}(0,0 ; \Omega) \, G^{0}_{\delta\beta}(0,0 ; \Omega)  \label{0orderVbody} \, ,
\end{eqnarray}
and
\begin{widetext}
\begin{eqnarray}
v^{(1)}_{V1}(\tau,\Omega) &=& -\frac{1}{4}\sum_{\alpha\beta\gamma\delta}\sum_{\epsilon\zeta\eta\theta}\int_{0}^{\tau}d\tau_{1}%
\; \bar{v}_{\alpha\beta\gamma\delta} \, \bar{v}_{\epsilon\zeta\eta\theta} \, G^{0}_{\gamma\epsilon}(0,\tau_1 ; \Omega) \, G^{0}_{\eta\alpha}(\tau_1,0 ; \Omega) \, G^{0}_{\delta\zeta}(0,\tau_1 ; \Omega) \, G^{0}_{\theta\beta}(\tau_1,0 ; \Omega) \, , \label{total2ndorderdiag2body}
\end{eqnarray}
\end{widetext}
respectively.

\section{Coupled cluster expansion}
\label{CCtheory}

Having their MBPT expansions at hand, we now design the coupled cluster expansions of $H(\tau,\Omega)$ and $N(\tau,\Omega)$.

\subsection{Energy kernel}
\label{energykernelMBPTtoCC}

We first show that the perturbative expansion of the linked/connected kernel $h(\tau,\Omega)$ can be recast in terms of an exponentiated cluster operator whose expansion naturally terminates. 

\subsubsection{From MBPT to cluster operators}

\begin{figure}[t!]
\begin{center}
\includegraphics[clip=,width=0.19\textwidth,angle=0]{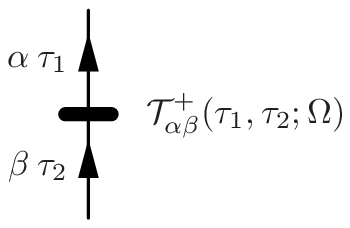}\\ \vspace{0.4cm}
\includegraphics[clip=,width=0.41\textwidth,angle=0]{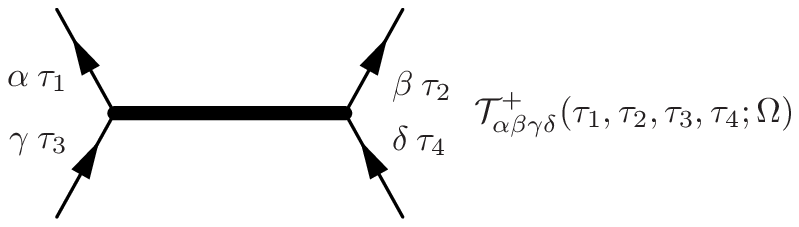}\\
\end{center}
\caption{
\label{TD1B2BCA}
Feynman diagrams representing  one- (first line) and two-body (second line) cluster amplitudes.}
\end{figure}

We introduce the $\tau$- and $\Omega$-dependent n-body cluster operator through
\begin{widetext}
\begin{eqnarray}
{\cal T}^{\dagger}_{n}(\tau,\Omega) &\equiv& \left(\frac{1}{n!}\right)^2 \!\! \sum_{\alpha \ldots \beta\gamma\ldots\delta} \int_{0}^{\tau}\!\prod_{k=1}^{2n}d\tau_{k} \, {\cal T}^{\dagger}_{\alpha\ldots\beta \gamma\ldots\delta}(\tau_{1},\ldots\tau_{n},\tau_{n\!+\!1},\ldots \tau_{2n}; \Omega) \, T[a^{\dagger}_{\alpha}(\tau_1) \ldots a^{\dagger}_{\beta}(\tau_n) \, a_{\delta}(\tau_{2n}) \ldots a_{\gamma}(\tau_{n\!+\!1})] \, , \label{clusteroperators}
\end{eqnarray}
\end{widetext}
where the Feynman amplitude ${\cal T}^{\dagger}_{\alpha_1\ldots\alpha_{n} \alpha_{n\!+\!1}\ldots\alpha_{2n}}(\tau_{1},\ldots\tau_{n},\tau_{n\!+\!1},\ldots \tau_{2n}; \Omega)$
is antisymmetric under the exchange of  $(\alpha_k,\tau_k)$ and $(\alpha_l,\tau_l)$ whenever $(k,l)\in\{1,\ldots n\}^2$ or $\{n\!+\!1,\ldots 2n\}^2$. One- and two-body cluster amplitudes are represented diagrammatically in Fig.~\ref{TD1B2BCA}. For historical reasons, the operators introduced in Eq.~\ref{clusteroperators} reduce to the {\it Hermitian conjugate} of the usual cluster operators when considering the diagonal energy kernel, i.e. $\Omega=0$.

\begin{figure}[t!]
\begin{center}
\includegraphics[clip=,width=0.20\textwidth,angle=0]{figures/MBPTdiag12_v2.eps}\\ 
\includegraphics[clip=,width=0.19\textwidth,angle=0]{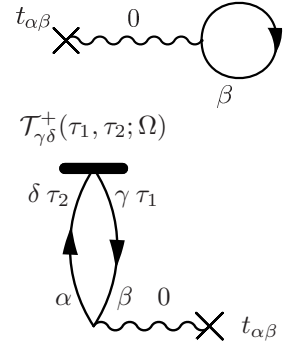}\\ 
\end{center}
\caption{
\label{T1contribtokinetic}
Feynman diagrams representing the two contributions to $t(\tau,\Omega)$.}
\end{figure}

\begin{figure}[t!]
\begin{center}
\includegraphics[clip=,width=0.20\textwidth,angle=0]{figures/MBPTdiag11_v2.eps}\\ \vspace{0.4cm}
\includegraphics[clip=,width=0.19\textwidth,angle=0]{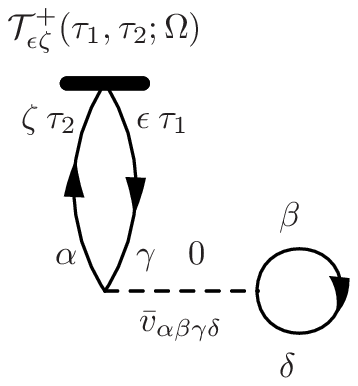}\\ \vspace{0.4cm}
\includegraphics[clip=,width=0.20\textwidth,angle=0]{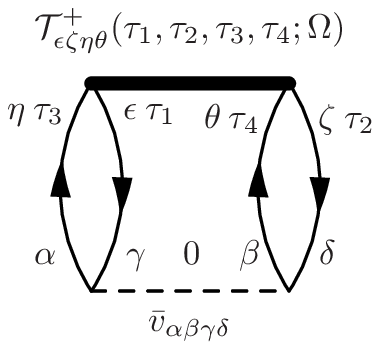}\\ \vspace{0.4cm}
\includegraphics[clip=,width=0.24\textwidth,angle=0]{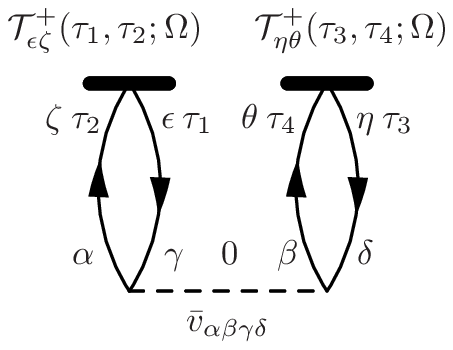}\\
\end{center}
\caption{
\label{variousTcontribtopotential}
Feynman diagrams representing the four contributions to $v(\tau,\Omega)$.}
\end{figure}

As discussed in Sec.~\ref{energykernel}, $t(\tau,\Omega)$ and $v(\tau,\Omega)$ represent the infinite set of connected diagrams linked at time zero to the one-body operator $T$ and to the two-body operator $V$, respectively. By virtue of their linked character, diagrams entering $t(\tau,\Omega)$ necessarily possess the topology of one of the two diagrams represented in Fig.~\ref{T1contribtokinetic}. Similarly, those entering $v(\tau,\Omega)$ necessarily possess the topology of one of the four diagrams represented in Fig.~\ref{variousTcontribtopotential}. In both cases, the first diagram simply isolates the contribution with no propagating leg, i.e the zero-order contribution associated with the matrix elements of $T$ and $V$ between the reference state and its rotated partner.  

All diagrams entering $t(\tau,\Omega)$ beyond zero order are thus captured by the second topology in Fig.~\ref{T1contribtokinetic}. This leads to defining the one-body cluster amplitude ${\cal T}^{\dagger}_{\alpha\beta}(\tau_1,\tau_2 ; \Omega)$ as the {\it complete} sum of connected diagrams generated through perturbation theory with one line entering at an arbitrary time $\tau_2$ and one line exiting at an arbitrary time $\tau_1$. In Fig.~\ref{T1contribtokinetic}, these two lines propagate downwards to contract with $T$ at time zero. Similarly, all diagrams beyond zero order entering $v(\tau,\Omega)$ are captured by the last three topologies in Fig.~\ref{variousTcontribtopotential}. This leads to defining the two-body cluster amplitude ${\cal T}^{\dagger}_{\alpha\beta\gamma\delta}(\tau_1,\tau_2, \tau_3,\tau_4 ; \Omega)$ as the {\it complete} sum of connected diagrams with two lines entering at arbitrary times $\tau_3$ and $ \tau_4$, and two lines exiting at arbitrary times $\tau_1$ and $\tau_2$. In the third diagram of Fig.~\ref{variousTcontribtopotential}, these four lines propagate downwards to contract with $V$ at time zero. This definition trivially extends to higher-body cluster operators. First-order expressions of ${\cal T}^{\dagger}_{\alpha\beta}(\tau_1,\tau_2 ; \Omega)$ and ${\cal T}^{\dagger}_{\alpha\beta\gamma\delta}(\tau_1,\tau_2, \tau_3,\tau_4 ; \Omega)$ are provided in Sec.~\ref{firstorderToperators}.

Thus, the introduction of cluster operators allows one to group the complete set of linked/connected vacuum-to-vacuum diagrams making up $t(\tau,\Omega)$ and $v(\tau,\Omega)$ under the form
\begin{widetext}
\begin{subequations}
\label{CCenergyequation}
\begin{eqnarray}
t(\tau,\Omega) &=& \langle \Phi |  T  +  {\cal T}^{\dagger}_{1}(\tau,\Omega) \, T  |\Phi (\Omega) \rangle_{c} \, \langle \Phi |  \Phi (\Omega) \rangle^{-1} \, , \label{CCenergyequation1}\\
v(\tau,\Omega)  &=& \langle \Phi |  V + {\cal T}^{\dagger}_{1}(\tau,\Omega) \, V + {\cal T}^{\dagger}_{2}(\tau,\Omega) \, V +  \frac{1}{2} {\cal T}^{\dagger\, 2}_{1}(\tau,\Omega) \, V |\Phi (\Omega) \rangle_{c} \, \langle \Phi |  \Phi (\Omega) \rangle^{-1} \, , \label{CCenergyequation2}
\end{eqnarray}
\end{subequations}
\end{widetext}
where the subscript $c$ means that  cluster operators are all linked to $T$ or $V$ through strings of contractions but are not connected together. Contractions between creation and annihilation operators {\it within} a given cluster operator are also excluded in Eq.~\ref{CCenergyequation} (see below). As off-diagonal contractions within a given cluster operator are {\it not} zero a priori, the rule that those contractions are to be excluded when computing contributions to Eq.~\ref{CCenergyequation} must indeed be stated explicitly here. The Hamiltonian being of two-body character, the sum of terms in Eq.~\ref{CCenergyequation} does exhaust exactly the complete set of diagrams generated through perturbation theory. The $1/2$ factor entering the last term of Eq.~\ref{CCenergyequation2} can be justified order by order by considering a contribution ${\cal T}^{\dagger(n)}_{1}(\tau,\Omega)$ extracted from an arbitrary diagram of order $n$ having the topology of the second term of Eq.~\ref{CCenergyequation1} (or of Eq.~\ref{CCenergyequation2}). The corresponding contribution to $v(\tau,\Omega)$ of order $2n$ associated with the last term of Eq.~\ref{CCenergyequation2} acquires a factor $1/2$ because exchanging at once all time labels entering the two identical ${\cal T}^{\dagger (n)}_{1}(\tau,\Omega)$ pieces provides an equivalent diagram. This is nothing but the counting associated with so-called equivalent cluster operators in standard CC theory~\cite{shavitt09a}.

Eventually, one can rewrite Eq.~\ref{CCenergyequation} under the characteristic form
\begin{subequations}
\label{clusterexpansion}
\begin{eqnarray}
h(\tau,\Omega) &=& \frac{\langle \Phi | e^{{\cal T}^{\dagger}(\tau,\Omega)} H  |\Phi (\Omega) \rangle_{c}}{\langle \Phi |  \Phi (\Omega) \rangle} \, , \\
{\cal T}^{\dagger}(\tau,\Omega) &\equiv& \sum_{n=1}^{A} {\cal T}^{\dagger}_{n}(\tau,\Omega) \, ,
\end{eqnarray}
\end{subequations}
given that no cluster operator beyond ${\cal T}^{\dagger}_{1}$ and ${\cal T}^{\dagger}_{2}$ can actually contribute to the linked/connected kernels $t(\tau,\Omega)$ and $v(\tau,\Omega)$. Again, it is understood that (i) cluster operators must all be linked to $H$ and that (ii)  no contraction can occur among cluster operators or within a given cluster operator when expanding back the exponential. Contracting creation and annihilation operators originating from different cluster operators (e.g. from ${\cal T}^{\dagger}_1$ and ${\cal T}^{\dagger}_2$) or within the same cluster operator generate diagrams that are already contained in  a connected cluster of lower rank and would thus lead to double counting. The fact that Eq.~\ref{clusterexpansion} does indeed reduce to Eq.~\ref{CCenergyequation} generalizes to off-diagonal energy kernels the natural termination of the CC expansion of the standard, i.e. diagonal, energy kernel. As mentioned earlier, the termination and the specific connected structure of the resulting terms are usually obtained from the similarity transformed Hamilton operator $\bar{H}\equiv e^{-T}He^{T}$ on the basis of the Baker-Campbell-Hausdorff identity and the standard Wick theorem~\cite{shavitt09a}. In the present case, the long detour through perturbation theory applied to off-diagonal kernels was necessary to obtain the same connected structure as in the standard case, including the possibility to omit from the outset contractions within a cluster operator or among them. 

\subsubsection{Computation of CC diagrams}

The two contributions to Eq.~\ref{CCenergyequation1}, which correspond to the diagrams displayed in Fig.~\ref{T1contribtokinetic}, read explicitly as
\begin{widetext}
\begin{subequations}
\label{Tctimedependent}
\begin{eqnarray}
\frac{\langle \Phi | T  |\Phi (\Omega) \rangle_{c}}{\langle \Phi |  \Phi (\Omega) \rangle} &=& -\sum_{\alpha\beta} t_{\alpha\beta} \, G^{0}_{\beta\alpha}(0,0;\Omega) \, , \\
\frac{\langle \Phi | {\cal T}^{\dagger}_{1}(\tau,\Omega)\, T  |\Phi (\Omega) \rangle_{c}}{\langle \Phi |  \Phi (\Omega) \rangle} &=& -\sum_{\alpha\beta} t_{\alpha\beta} \sum_{\gamma\delta} \int_{0}^{\tau}\!\!\int_{0}^{\tau}\!\! d\tau_{1}d\tau_{2} \, {\cal T}^{\dagger}_{\gamma\delta}(\tau_1,\tau_2;\Omega) \, G^{0}_{\delta\alpha}(\tau_2,0;\Omega) \, G^{0}_{\beta\gamma}(0,\tau_1;\Omega) \, .
\end{eqnarray}
\end{subequations}
while those appearing in Eq.~\ref{CCenergyequation2}, which correspond\footnote{Up to the factor $1/2$ that must multiply Eq.~\ref{Vctimedependent4} to match the corresponding diagram.} to the diagrams displayed in Fig.~\ref{variousTcontribtopotential}, are
\begin{subequations}
\label{Vctimedependent}
\begin{eqnarray}
\frac{\langle \Phi | V  |\Phi (\Omega) \rangle_{c}}{\langle \Phi |  \Phi (\Omega) \rangle} &=& \frac{1}{2}\sum_{\alpha\beta\gamma\delta} \bar{v}_{\alpha\beta\gamma\delta} \, G^{0}_{\gamma\alpha}(0,0;\Omega) \, G^{0}_{\delta\beta}(0,0;\Omega) \, ,\\
\frac{\langle \Phi | {\cal T}^{\dagger}_{1}(\tau,\Omega)\, V  |\Phi (\Omega) \rangle_{c}}{\langle \Phi |  \Phi (\Omega) \rangle} &=& \sum_{\alpha\beta\gamma\delta} \bar{v}_{\alpha\beta\gamma\delta} \, G^{0}_{\gamma\alpha}(0,0;\Omega) \sum_{\zeta\epsilon} \int_{0}^{\tau}\!\!\int_{0}^{\tau}\!\! d\tau_{1}d\tau_{2} \, {\cal T}^{\dagger}_{\epsilon\zeta}(\tau_1,\tau_2;\Omega) \, G^{0}_{\zeta\beta}(\tau_2,0;\Omega) \, G^{0}_{\delta\epsilon}(0,\tau_1;\Omega) \, , \\
\frac{\langle \Phi | {\cal T}^{\dagger}_{2}(\tau,\Omega)\, V  |\Phi (\Omega) \rangle_{c}}{\langle \Phi |  \Phi (\Omega) \rangle} &=&  \frac{1}{4}\sum_{\alpha\beta\gamma\delta} \bar{v}_{\alpha\beta\gamma\delta} \sum_{\zeta\epsilon\eta\theta} \int_{0}^{\tau}\!\!\int_{0}^{\tau}\!\!\int_{0}^{\tau}\!\!\int_{0}^{\tau}\!\! d\tau_{1}d\tau_{2}d\tau_{3}d\tau_{4}\, {\cal T}^{\dagger}_{\epsilon\zeta\eta\theta}(\tau_1,\tau_2,\tau_3,\tau_4;\Omega)  \, G^{0}_{\eta\alpha}(\tau_3,0;\Omega) \nonumber \\
&& \hspace{6cm} \times G^{0}_{\gamma\epsilon}(0,\tau_1;\Omega) \,  G^{0}_{\theta\beta}(\tau_4,0;\Omega) \,  G^{0}_{\delta\zeta}(0,\tau_2;\Omega) \,  , \\
\frac{\langle \Phi | {\cal T}^{\dagger\, 2}_{1}(\tau,\Omega)\, V  |\Phi (\Omega) \rangle_{c}}{\langle \Phi |  \Phi (\Omega) \rangle} &=& \sum_{\alpha\beta\gamma\delta} \bar{v}_{\alpha\beta\gamma\delta} \sum_{\zeta\epsilon} \int_{0}^{\tau}\!\!\int_{0}^{\tau}\!\! d\tau_{1}d\tau_{2} \, {\cal T}^{\dagger}_{\epsilon\zeta}(\tau_1,\tau_2;\Omega) \, G^{0}_{\zeta\alpha}(\tau_2,0;\Omega) \, G^{0}_{\gamma\epsilon}(0,\tau_1;\Omega) \nonumber \\
&& \hspace{1.3cm} \times \sum_{\theta\eta} \int_{0}^{\tau}\!\!\int_{0}^{\tau}\!\! d\tau_{3}d\tau_{4} \, {\cal T}^{\dagger}_{\eta\theta}(\tau_3,\tau_4;\Omega) \, G^{0}_{\theta\beta}(\tau_4,0;\Omega) \, G^{0}_{\delta\eta}(0,\tau_3;\Omega) \, . \label{Vctimedependent4}
\end{eqnarray}
\end{subequations}
\end{widetext}

One- and two-body\footnote{The same operation trivially follows for any n-tuple cluster operator ${\cal T}^{\dagger}_{n}(\tau,\Omega)$.} cluster operators can be rewritten in terms of Goldstone amplitudes and creation/annihilation operators in the Schroedinger representation as
\begin{subequations}
\label{Toperators}
\begin{eqnarray}
{\cal T}^{\dagger}_{1}(\tau,\Omega) &\equiv& \frac{1}{(1!)^2} \sum_{\alpha\beta} {\cal T}^{\dagger}_{\alpha\beta}(\tau,\Omega)  \, a^{\dagger}_{\alpha} \, a_{\beta}  \, , \label{T1} \\ 
{\cal T}^{\dagger}_{2}(\tau,\Omega) &\equiv& \frac{1}{(2!)^2} \sum_{\alpha\beta\gamma\delta} {\cal T}^{\dagger}_{\alpha\beta\gamma\delta}(\tau,\Omega) \, a^{\dagger}_{\alpha} \, a^{\dagger}_{\beta} \, a_{\delta} \, a_{\gamma}  \label{T2} \, ,
\end{eqnarray}
\end{subequations}
where
\begin{widetext}
\begin{subequations}
\label{another}
\begin{eqnarray}
{\cal T}^{\dagger}_{\alpha\beta}(\tau,\Omega) \!\! &=& \!\! \int_{0}^{\tau}\!\!\int_{0}^{\tau}\!\! d\tau_1d\tau_2 {\cal T}^{\dagger}_{\alpha\beta}(\tau_1,\tau_2;\Omega) \, e^{\tau_1 (e_{\alpha}\!-\!\mu)} e^{\tau_2(\mu\!-\!e_{\beta})}  \label{T1matrixelements} \, , \\
{\cal T}^{\dagger}_{\alpha\beta\gamma\delta}(\tau,\Omega) \!\! &=&  \!\! \int_{0}^{\tau}\!\!\int_{0}^{\tau}\!\!\int_{0}^{\tau}\!\!\int_{0}^{\tau}\!\! d\tau_1d\tau_2d\tau_3d\tau_4 {\cal T}^{\dagger}_{\alpha\beta\gamma\delta}(\tau_1,\tau_2,\tau_3,\tau_4;\Omega) \, e^{\tau_1 (e_{\alpha}\!-\!\mu)} e^{\tau_2(e_{\beta}\!-\!\mu)}  e^{\tau_3(\mu\!-\! e_{\gamma})}  e^{\tau_4(\mu\!-\! e_{\delta})} \label{T2matrixelements} \, .
\end{eqnarray}
\end{subequations}
\end{widetext}
Matrix elements of ${\cal T}^{\dagger}_{2}(\tau,\Omega)$ are anti-symmetric under the exchange of pairs of in-going or out-going indices. Expanding the propagators according to Eq.~\ref{decompoprop} and making use of Eq.~\ref{another}, Eqs.~\ref{Tctimedependent} and~\ref{Vctimedependent} become
\begin{widetext}
\begin{subequations}
\label{expressionTC}
\begin{eqnarray}
\frac{\langle \Phi | T  |\Phi (\Omega) \rangle}{\langle \Phi |  \Phi (\Omega) \rangle} &=& + \sum_{i} t_{ii} \\
&& +  \sum_{ia} t_{ia} \, \rho^{ph}_{ai}(\Omega) \, , \nonumber\\
\frac{\langle \Phi | {\cal T}^{\dagger}_{1}(\tau,\Omega)\, T  |\Phi (\Omega) \rangle_{c}}{\langle \Phi |  \Phi (\Omega) \rangle} &=& + \sum_{ia}  {\cal T}^{\dagger}_{ia}(\tau,\Omega) \, t_{ai} \\
&& + \sum_{iab}  {\cal T}^{\dagger}_{ia}(\tau,\Omega) \, t_{ab} \, \rho^{ph}_{bi}(\Omega)  - \sum_{ija} {\cal T}^{\dagger}_{ia}(\tau,\Omega) \, \rho^{ph}_{aj}(\Omega) \, t_{ji}   \nonumber \\
&& - \sum_{ijab}  {\cal T}^{\dagger}_{ia}(\tau,\Omega) \, \rho^{ph}_{aj}(\Omega) \, t_{jb} \, \rho^{ph}_{bi}(\Omega)   \, , \nonumber \\
\frac{\langle \Phi | V  |\Phi (\Omega) \rangle}{\langle \Phi |  \Phi (\Omega) \rangle} &=& + \frac{1}{2}\sum_{ij} \bar{v}_{ijij}  \\
&& + \frac{1}{2} \sum_{ijc} \bar{v}_{ijcj} \, \rho^{ph}_{ci}(\Omega) + \frac{1}{2} \sum_{ijd} \bar{v}_{ijid} \, \rho^{ph}_{dj}(\Omega)  \nonumber \\
&&+ \frac{1}{2} \sum_{ijab} \bar{v}_{ijab} \, \rho^{ph}_{ai}(\Omega) \, \rho^{ph}_{bj}(\Omega)\, , \nonumber\\
\frac{\langle \Phi | {\cal T}^{\dagger}_{1}(\tau,\Omega)\, V  |\Phi (\Omega) \rangle_{c}}{\langle \Phi |  \Phi (\Omega) \rangle} &=& + \sum_{ikd}  {\cal T}^{\dagger}_{kd}(\tau,\Omega) \, \bar{v}_{idik} \\
&& + \sum_{ikcd}  {\cal T}^{\dagger}_{kd}(\tau,\Omega) \, \bar{v}_{idic}  \, \rho^{ph}_{ck}(\Omega) - \sum_{ikld}  {\cal T}^{\dagger}_{kd}(\tau,\Omega)  \, \rho^{ph}_{dl}(\Omega) \, \bar{v}_{ilik} + \sum_{ikad}  {\cal T}^{\dagger}_{kd}(\tau,\Omega) \, \bar{v}_{idak}  \, \rho^{ph}_{ai}(\Omega) \nonumber \\
&& - \sum_{iklcd}  {\cal T}^{\dagger}_{kd}(\tau,\Omega)  \, \rho^{ph}_{dl}(\Omega) \, \bar{v}_{ilic}  \, \rho^{ph}_{ck}(\Omega) \nonumber \\
&& + \sum_{ikacd}  {\cal T}^{\dagger}_{kd}(\tau,\Omega) \, \bar{v}_{idac}  \, \rho^{ph}_{ai}(\Omega)  \, \rho^{ph}_{ck}(\Omega) \nonumber \\
&& - \sum_{iklad}  {\cal T}^{\dagger}_{kd}(\tau,\Omega) \, \rho^{ph}_{dl}(\Omega) \, \bar{v}_{ilak}  \, \rho^{ph}_{ai}(\Omega)  \nonumber \\
&& - \sum_{iklacd}  {\cal T}^{\dagger}_{kd}(\tau,\Omega) \, \rho^{ph}_{dl}(\Omega) \, \bar{v}_{ilac}  \, \rho^{ph}_{ai}(\Omega) \, \rho^{ph}_{ck}(\Omega)  \nonumber \, , \\
\frac{\langle \Phi | {\cal T}^{\dagger}_{2}(\tau,\Omega)\, V  |\Phi (\Omega) \rangle_{c}}{\langle \Phi |  \Phi (\Omega) \rangle} &=& +\frac{1}{4} \sum_{abij}  {\cal T}^{\dagger}_{ijab}(\tau,\Omega) \, \bar{v}_{abij}  \\
&& +\frac{1}{4} \sum_{ijabc}  {\cal T}^{\dagger}_{ijab}(\tau,\Omega) \, \bar{v}_{abic} \, \rho^{ph}_{cj}(\Omega)-\frac{1}{4} \sum_{ijkab}   {\cal T}^{\dagger}_{ijab}(\tau,\Omega) \, \rho^{ph}_{bk}(\Omega) \, \bar{v}_{akij} \nonumber \\
&& +\frac{1}{4} \sum_{ijabc}  {\cal T}^{\dagger}_{ijab}(\tau,\Omega) \, \bar{v}_{abcj} \, \rho^{ph}_{ci}(\Omega)-\frac{1}{4} \sum_{ijkab}   {\cal T}^{\dagger}_{ijab}(\tau,\Omega) \, \rho^{ph}_{ak}(\Omega) \, \bar{v}_{kbij} \nonumber \\
&& -\frac{1}{4} \sum_{ijkabc}   {\cal T}^{\dagger}_{ijab}(\tau,\Omega) \, \rho^{ph}_{bk}(\Omega) \, \bar{v}_{akic} \, \rho^{ph}_{cj}(\Omega)+\frac{1}{4} \sum_{ijabcd}  {\cal T}^{\dagger}_{ijab}(\tau,\Omega) \, \bar{v}_{abcd} \, \rho^{ph}_{ci}(\Omega) \, \rho^{ph}_{dj}(\Omega) \nonumber \\
&& -\frac{1}{4} \sum_{ijkabc}   {\cal T}^{\dagger}_{ijab}(\tau,\Omega) \, \rho^{ph}_{bk}(\Omega) \, \bar{v}_{akcj} \, \rho^{ph}_{ci}(\Omega)-\frac{1}{4} \sum_{ijkabc}  {\cal T}^{\dagger}_{ijab}(\tau,\Omega) \, \rho^{ph}_{ak}(\Omega) \, \bar{v}_{kbic} \, \rho^{ph}_{cj}(\Omega) \nonumber \\
&& +\frac{1}{4} \sum_{ijklab}   {\cal T}^{\dagger}_{ijab}(\tau,\Omega) \, \rho^{ph}_{ak}(\Omega) \, \rho^{ph}_{bl}(\Omega) \,  \bar{v}_{klij}-\frac{1}{4} \sum_{ijkabc}  {\cal T}^{\dagger}_{ijab}(\tau,\Omega) \, \rho^{ph}_{ak}(\Omega) \, \bar{v}_{kbcj} \, \rho^{ph}_{ci}(\Omega) \nonumber \\
&& -\frac{1}{4} \sum_{ijkabcd}  {\cal T}^{\dagger}_{ijab}(\tau,\Omega) \,  \rho^{ph}_{bk}(\Omega) \,  \bar{v}_{akcd} \, \rho^{ph}_{ci}(\Omega) \, \rho^{ph}_{dj}(\Omega) \nonumber \\
&& +\frac{1}{4} \sum_{ijklabc}  {\cal T}^{\dagger}_{ijab}(\tau,\Omega) \, \rho^{ph}_{ak}(\Omega) \, \rho^{ph}_{bl}(\Omega) \,  \bar{v}_{klic} \, \rho^{ph}_{cj}(\Omega)\nonumber \\
&& -\frac{1}{4} \sum_{ijkabcd}  {\cal T}^{\dagger}_{ijab}(\tau,\Omega) \, \rho^{ph}_{ak}(\Omega) \,  \bar{v}_{kbcd} \, \rho^{ph}_{ci}(\Omega) \, \rho^{ph}_{dj}(\Omega) \nonumber \\
&& +\frac{1}{4} \sum_{ijklabc}   {\cal T}^{\dagger}_{ijab}(\tau,\Omega) \, \rho^{ph}_{ak}(\Omega) \, \rho^{ph}_{bl}(\Omega) \,  \bar{v}_{klcj} \, \rho^{ph}_{ci}(\Omega) \nonumber \\
&& +\frac{1}{4} \sum_{ijklabcd}  {\cal T}^{\dagger}_{ijab}(\tau,\Omega) \, \rho^{ph}_{ak}(\Omega) \, \rho^{ph}_{bl}(\Omega) \,  \bar{v}_{klcd} \, \rho^{ph}_{ci}(\Omega) \, \rho^{ph}_{dj}(\Omega) \nonumber \\
\frac{\langle \Phi | {\cal T}^{\dagger\, 2}_{1}(\tau,\Omega)\, V  |\Phi (\Omega) \rangle_{c}}{\langle \Phi |  \Phi (\Omega) \rangle} &=& + \sum_{ikbd} {\cal T}^{\dagger}_{ib}(\tau,\Omega) \, {\cal T}^{\dagger}_{kd}(\tau,\Omega) \, \bar{v}_{bdik} \\
&& + \sum_{ikbdc} {\cal T}^{\dagger}_{ib}(\tau,\Omega) \, {\cal T}^{\dagger}_{kd}(\tau,\Omega) \, \bar{v}_{bdic} \, \rho^{ph}_{ck}(\Omega)  - \sum_{iklbd} {\cal T}^{\dagger}_{ib}(\tau,\Omega) \, {\cal T}^{\dagger}_{kd}(\tau,\Omega) \, \rho^{ph}_{dl}(\Omega) \, \bar{v}_{blik} \nonumber \\
&& + \sum_{ikabd} {\cal T}^{\dagger}_{ib}(\tau,\Omega) \, {\cal T}^{\dagger}_{kd}(\tau,\Omega) \, \bar{v}_{bdak} \, \rho^{ph}_{ai}(\Omega)  - \sum_{ijkbd} {\cal T}^{\dagger}_{ib}(\tau,\Omega) \, {\cal T}^{\dagger}_{kd}(\tau,\Omega) \, \rho^{ph}_{bj}(\Omega) \, \bar{v}_{jdik} \nonumber \\
&& - \sum_{iklbcd} {\cal T}^{\dagger}_{ib}(\tau,\Omega) \, {\cal T}^{\dagger}_{kd}(\tau,\Omega) \, \rho^{ph}_{dl}(\Omega) \, \bar{v}_{blic} \, \rho^{ph}_{ck}(\Omega) \nonumber \\
&& + \sum_{ikabcd} {\cal T}^{\dagger}_{ib}(\tau,\Omega) \, {\cal T}^{\dagger}_{kd}(\tau,\Omega) \, \bar{v}_{bdac} \, \rho^{ph}_{ai}(\Omega) \, \rho^{ph}_{ck}(\Omega) \nonumber \\
&& - \sum_{iklabd} {\cal T}^{\dagger}_{ib}(\tau,\Omega) \, {\cal T}^{\dagger}_{kd}(\tau,\Omega) \, \rho^{ph}_{dl}(\Omega) \, \bar{v}_{blak} \, \rho^{ph}_{ai}(\Omega) \nonumber \\
&& - \sum_{ijkbcd} {\cal T}^{\dagger}_{ib}(\tau,\Omega) \, {\cal T}^{\dagger}_{kd}(\tau,\Omega) \, \rho^{ph}_{bj}(\Omega) \, \bar{v}_{jdic} \, \rho^{ph}_{ck}(\Omega) \nonumber \\
&& + \sum_{ijklbd} {\cal T}^{\dagger}_{ib}(\tau,\Omega) \, {\cal T}^{\dagger}_{kd}(\tau,\Omega) \, \rho^{ph}_{bj}(\Omega) \, \rho^{ph}_{dl}(\Omega) \, \bar{v}_{jlik}  \nonumber \\
&& - \sum_{ijkabd} {\cal T}^{\dagger}_{ib}(\tau,\Omega) \, {\cal T}^{\dagger}_{kd}(\tau,\Omega) \, \rho^{ph}_{bj}(\Omega) \, \bar{v}_{jdak}  \, \rho^{ph}_{ai}(\Omega) \nonumber \\
&& - \sum_{iklabcd} {\cal T}^{\dagger}_{ib}(\tau,\Omega) \, {\cal T}^{\dagger}_{kd}(\tau,\Omega) \, \rho^{ph}_{dl}(\Omega) \, \bar{v}_{blac}  \, \rho^{ph}_{ai}(\Omega) \, \rho^{ph}_{ck}(\Omega) \nonumber \\
&& + \sum_{ijklbcd} {\cal T}^{\dagger}_{ib}(\tau,\Omega) \, {\cal T}^{\dagger}_{kd}(\tau,\Omega)\, \rho^{ph}_{bj}(\Omega) \, \rho^{ph}_{dl}(\Omega) \, \bar{v}_{jlic}   \, \rho^{ph}_{ck}(\Omega) \nonumber \\
&& - \sum_{ijkabcd} {\cal T}^{\dagger}_{ib}(\tau,\Omega) \, {\cal T}^{\dagger}_{kd}(\tau,\Omega) \, \rho^{ph}_{bj}(\Omega) \, \bar{v}_{jdac}  \, \rho^{ph}_{ai}(\Omega) \, \rho^{ph}_{ck}(\Omega) \nonumber \\
&& + \sum_{ijklabd} {\cal T}^{\dagger}_{ib}(\tau,\Omega) \, {\cal T}^{\dagger}_{kd}(\tau,\Omega)\, \rho^{ph}_{bj}(\Omega) \, \rho^{ph}_{dl}(\Omega) \, \bar{v}_{jlak}   \, \rho^{ph}_{ai}(\Omega) \nonumber \\
&& + \sum_{ijklabcd} {\cal T}^{\dagger}_{ib}(\tau,\Omega) \, {\cal T}^{\dagger}_{kd}(\tau,\Omega)\, \rho^{ph}_{bj}(\Omega) \, \rho^{ph}_{dl}(\Omega) \, \bar{v}_{jlac}   \, \rho^{ph}_{ai}(\Omega)  \, \rho^{ph}_{ck}(\Omega) \, .
\end{eqnarray}
\end{subequations}
\end{widetext}

\subsubsection{Compact algebraic expressions}
\label{reducedalgebraicexpressions}

Algebraic expressions~\ref{expressionTC} lead to two observations. First, they solely invoke hole-particle matrix elements of  ${\cal T}^{\dagger}_{1}(\tau,\Omega)$ and ${\cal T}^{\dagger}_{2}(\tau,\Omega)$. This is a consequence of the connected character of the matrix elements in Eqs.~\ref{CCenergyequation} and~\ref{clusterexpansion}, which itself derives from the necessity to forbid any double counting of perturbation theory diagrams. Second, they are lengthy and seem hardly amenable to an efficient implementation in a CC code. However, they can be compacted efficiently by working with convenient left and right bi-orthogonal single-particle bases that we now introduce.

The right basis $\{| \tilde{\alpha} \rangle\}$ is obtained by applying the non-unitary transformation
\begin{eqnarray}
B(\Omega) &\equiv& \bbone + \rho^{ph}(\Omega) \, , \label{matrix}
\end{eqnarray}
onto the original basis $\{| \alpha \rangle\}$. Omitting for notational simplicity the explicit $\Omega$ dependence of the basis states thus obtained and separating original particle and hole states provides 
\begin{subequations}
\label{rightbasis}
\begin{eqnarray}
| \tilde{\imath} \rangle &=&  | i \rangle + \sum_{kc} | c \rangle \,  R_{ck}(\Omega) M^{-1}_{ki}(\Omega) \, , \label{rightbasis1} \\
| \tilde{a} \rangle &=&  | a \rangle \, . \label{rightbasis2}
\end{eqnarray}
\end{subequations}
Particle kets are thus left unchanged. The left basis  $\{\langle \tilde{\alpha} | \}$ is similarly obtained by applying the transformation
\begin{eqnarray}
B^{-1}(\Omega) &\equiv& \bbone - \rho^{ph}(\Omega) \, , \label{inversematrix}
\end{eqnarray}
onto the original basis $\{\langle \alpha |\}$ such that
\begin{subequations}
\label{leftbasis}
\begin{eqnarray}
\langle \tilde{\jmath} | &=&  \langle j | \, , \label{leftbasis1} \\
\langle \tilde{b} | &=&  \langle b | - \sum_{kl} R_{bk}(\Omega) M^{-1}_{kl}(\Omega) \, \langle l |  \, . \label{leftbasis2}
\end{eqnarray}
\end{subequations}
Hole bras are thus left unchanged. Although we use for simplicity the same notation to characterize left- and right-basis states, the tilde is meant to underline their bi-orthogonal character. The latter, indicated by $\langle \tilde{\alpha} | \tilde{\beta} \rangle = \delta_{\alpha\beta}$, derives from   $B^{-1}(\Omega)B(\Omega)=\bbone$, which can be easily verified. 

With these bi-orthogonal bases at hand, it is tedious but straightforward to demonstrate that the contributions to $t(\tau,\Omega)$ and $v(\tau,\Omega)$ in Eq.~\ref{expressionTC} can be rewritten as
\begin{subequations}
\label{expressionTandVC}
\begin{eqnarray}
\frac{\langle \Phi | T  |\Phi (\Omega) \rangle}{\langle \Phi |  \Phi (\Omega) \rangle} &=& \sum_{i} t_{\tilde{\imath}\tilde{\imath}}(\Omega) \nonumber \\
&\equiv& \langle \Phi | \tilde{T}(\Omega)  |\Phi \rangle \, ,  \\
\frac{\langle \Phi | {\cal T}^{\dagger}_{1}(\tau,\Omega)\, T  |\Phi (\Omega) \rangle_{c}}{\langle \Phi |  \Phi (\Omega) \rangle} &=&  \sum_{ia}  {\cal T}^{\dagger}_{ia}(\tau,\Omega) \, t_{\tilde{a}\tilde{\imath}}(\Omega) \nonumber \\ 
&\equiv& \langle \Phi | {\cal T}^{\dagger}_{1}(\tau,\Omega)\, \tilde{T}(\Omega)  |\Phi \rangle_{c}   \, ,   \\
\frac{\langle \Phi | V  |\Phi (\Omega) \rangle}{\langle \Phi |  \Phi (\Omega) \rangle} &=& \frac{1}{2}\sum_{ij} \bar{v}_{\tilde{\imath}\tilde{\jmath}\tilde{\imath}\tilde{\jmath}}(\Omega) \nonumber \\
&\equiv& \langle \Phi | \tilde{V} (\Omega)  |\Phi \rangle   \, ,  \\
\frac{\langle \Phi | {\cal T}^{\dagger}_{1}(\tau,\Omega)\, V  |\Phi (\Omega) \rangle_{c}}{\langle \Phi |  \Phi (\Omega) \rangle} &=& \sum_{ija}  {\cal T}^{\dagger}_{ia}(\tau,\Omega) \, \bar{v}_{\tilde{a}\tilde{\jmath}\tilde{\imath}\tilde{\jmath}}(\Omega) \nonumber \\ 
&\equiv& \langle \Phi | {\cal T}^{\dagger}_{1}(\tau,\Omega)\, \tilde{V} (\Omega)  |\Phi \rangle_{c}  \, ,  \\
\frac{\langle \Phi | {\cal T}^{\dagger}_{2}(\tau,\Omega)\, V  |\Phi (\Omega) \rangle_{c}}{\langle \Phi |  \Phi (\Omega) \rangle} &=& \frac{1}{4} \sum_{ijab}  {\cal T}^{\dagger}_{ijab}(\tau,\Omega) \, \bar{v}_{\tilde{a}\tilde{b}\tilde{\imath}\tilde{\jmath}}(\Omega) \nonumber \\
&\equiv& \langle \Phi | {\cal T}^{\dagger}_{2}(\tau,\Omega)\, \tilde{V} (\Omega)  |\Phi \rangle_{c}  \, ,  \\
\frac{\langle \Phi | {\cal T}^{\dagger\, 2}_{1}(\tau,\Omega)\, V  |\Phi (\Omega) \rangle_{c}}{\langle \Phi |  \Phi (\Omega) \rangle} &=& \sum_{ijab} {\cal T}^{\dagger}_{ia}(\tau,\Omega) \, {\cal T}^{\dagger}_{jb}(\tau,\Omega) \, \bar{v}_{\tilde{a}\tilde{b}\tilde{\imath}\tilde{\jmath}}(\Omega) \nonumber \\ 
&\equiv& \langle \Phi | {\cal T}^{\dagger\, 2}_{1}(\tau,\Omega)\, \tilde{V} (\Omega)  |\Phi \rangle_{c}   \, .  
\end{eqnarray}
\end{subequations}
where an argument $\Omega$ has been added to the matrix elements at play to underline their dependence on the rotational angle through the bi-orthogonal basis states. Correspondingly, we have introduced the transformed operator $\tilde{O}(\Omega)$ of a n-body operator $O$ through
\begin{eqnarray}
\tilde{O}(\Omega) &\equiv& \left(\frac{1}{n!}\right)^2 \!\!\!\!\!\! \sum_{\alpha \ldots \beta\gamma\ldots\delta}\!\!\!\!\!\! O_{\tilde{\alpha}\ldots\tilde{\beta} \tilde{\gamma}\ldots\tilde{\delta}}(\Omega) \, a^{\dagger}_{\alpha} \ldots a^{\dagger}_{\beta} \, a_{\delta} \ldots a_{\gamma} \, , \label{deftransformedOp}
\end{eqnarray}
where creation and annihilation operators refer to the original eigenbasis $\{| \alpha \rangle\}$ of $H_0$ while left and right indices of the matrix elements refer to the associated bi-orthogonal system introduced above. 

The result obtained in Eq.~\ref{expressionTandVC} is remarkable. The {\it off-diagonal} linked/connected kinetic- and potential-energy kernels, originally expanded on the basis of the {\it off-diagonal} Wick theorem, are equal to the corresponding {\it diagonal} matrix elements of {\it transformed} kinetic- and potential-energy operators and expanded on the basis of the {\it standard}, i.e. diagonal, Wick theorem~\cite{wick50a}. This key result can be summarized as
\begin{subequations}
\label{wickvswick}
\begin{eqnarray}
h(\tau,\Omega) &=& \frac{\langle \Phi | e^{{\cal T}^{\dagger}(\tau,\Omega)} H  |\Phi (\Omega) \rangle_{c}}{\langle \Phi |  \Phi (\Omega) \rangle} \label{wickvswick1} \\
&=&  \langle \Phi | e^{{\cal T}^{\dagger}(\tau,\Omega)} \tilde{H}(\Omega)  |\Phi \rangle_{c} \, . \label{wickvswick2}
\end{eqnarray}
\end{subequations}
Correspondingly, the algebraic expressions in Eq.~\ref{expressionTandVC} are formally {\it identical} to  standard CC equations~\cite{shavitt09a}, with the sole difference that one must insert matrix elements expressed in the bi-orthogonal system rather than in the original eigenbasis of $H_{0}$.  Consequently, the routines used to compute those expressions in a standard CC code can be utilized directly at the price of computing and storing matrix elements of $T$ and $V$ in the bi-orthogonal system, which can be achieved by matrix-matrix multiplication. Noticeably, the contributions to the energy at play in standard CC theory are recovered from  Eq.~\ref{expressionTandVC} for $\Omega=0$ since the bi-orthogonal bases reduce to the original one in this case, i.e. $\tilde{T}(0)=T$ and $\tilde{V}(0)=V$.

\subsubsection{First-order perturbation theory}
\label{firstorderToperators}

\begin{figure}[t!]
\begin{center}
\includegraphics[clip=,width=0.44\textwidth,angle=0]{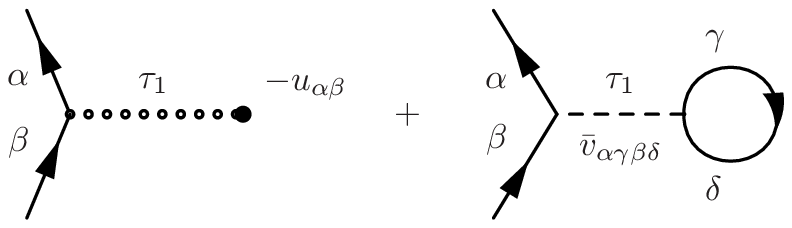}\\ \vspace{0.4cm}
\includegraphics[clip=,width=0.18\textwidth,angle=0]{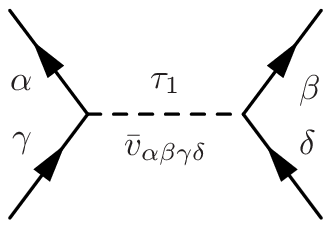}\\
\end{center}
\caption{
\label{1storderTamplitudes}
Feynman one-body (first line) and two-body (second line) cluster amplitudes at first order in perturbation theory.}
\end{figure}

Feynman diagrams contributing to one- and two-body cluster amplitudes at first order in perturbation theory are displayed in Fig.~\ref{1storderTamplitudes} and give
\begin{widetext}
\begin{subequations}
\label{1stordercluster}
\begin{eqnarray}
{\cal T}^{\dagger \, (1)}_{\alpha\beta}(\tau_1,\tau_2;\Omega) &=& \sum_{\gamma\delta} \bar{v}_{\alpha\gamma\beta\delta} \, G^{0}_{\delta\gamma}(\tau_1,\tau_1;\Omega) \, \delta(\tau_1-\tau_2) + u_{\alpha\beta} \, \delta(\tau_1-\tau_2) \, , \label{1storderclustera} \\
{\cal T}^{\dagger \, (1)}_{\alpha\beta\gamma\delta}(\tau_1,\tau_2,\tau_3,\tau_4;\Omega) &=& - \bar{v}_{\alpha\beta\gamma\delta} \, \delta(\tau_1-\tau_2) \, \delta(\tau_2-\tau_3) \, \delta(\tau_3-\tau_4) \, \, . \label{1storderclusterb}
\end{eqnarray}
\end{subequations}

Inserting these expressions into Eqs.~\ref{T1matrixelements} and~\ref{T2matrixelements}, one obtains associated Goldstone amplitudes
\begin{subequations}
\label{1storderclusterB}
\begin{eqnarray}
{\cal T}^{\dagger \, (1)}_{\alpha\beta}(\tau,\Omega) &=&  -\sum_{j} \frac{\bar{v}_{\alpha j\beta j}}{e_\beta-e_\alpha} \, \left[1 - e^{-\tau(e_\beta-e_\alpha)} \right] + \frac{u_{\alpha\beta}}{e_\beta-e_\alpha} \, \left[1 - e^{-\tau(e_\beta-e_\alpha)} \right] \nonumber \\
&& - \sum_{jb} \frac{\bar{v}_{\alpha j\beta b}}{e_\beta+e_b-e_\alpha-e_j} \, \rho^{ph}_{bj}(\Omega) \, \left[1 - e^{-\tau(e_\beta+e_b-e_\alpha-e_j)} \right] \, , \label{1storderclusterBa}  \\
{\cal T}^{\dagger \, (1)}_{\alpha\beta\gamma\delta}(\tau,\Omega) &=& - \frac{\bar{v}_{\alpha\beta\gamma\delta}}{e_\gamma+e_\delta-e_\alpha-e_\beta} \, \left[1 - e^{-\tau(e_\gamma+e_\delta-e_\alpha-e_\beta)} \right] \, \, , \label{1storderclusterBb}
\end{eqnarray}
\end{subequations}
\end{widetext}
such that ${\cal T}^{\dagger \, (1)}_{2}(\tau,\Omega)$ does not depend on $\Omega$. One can check that ${\cal T}^{\dagger \, (1)}_{1}(0,\Omega)={\cal T}^{\dagger \, (1)}_{2}(0,\Omega)=0$ as it should be. We observe in Eq.~\ref{1storderclusterB} that hole-particle cluster amplitudes, the only ones to effectively appear in the theory, are well defined and display a finite limit when $\tau$ goes to infinity.

With those expressions at hand, perturbative contributions to $h(\tau,\Omega)$ can be given in a compact form following Eq.~\ref{expressionTandVC}, i.e. lengthy zero and first-order expressions provided in App.~\ref{diagramsE} reduce to
\begin{subequations}
\label{expressionTandVCMBPT}
\begin{eqnarray}
t^{(0)}(\tau,\Omega) &=& \sum_{i} t_{\tilde{\imath}\tilde{\imath}}(\Omega) \, ,  \\
v^{(0)}(\tau,\Omega) &=& \frac{1}{2}\sum_{ij} \bar{v}_{\tilde{\imath}\tilde{\jmath}\tilde{\imath}\tilde{\jmath}}(\Omega)  \, ,  \\
t^{(1)}(\tau,\Omega) &=& \sum_{ia}  {\cal T}^{\dagger (1)}_{ia}(\tau,\Omega) \, t_{\tilde{a}\tilde{\imath}}(\Omega)   \, ,   \\
v^{(1)}(\tau,\Omega) &=& \sum_{ija}  {\cal T}^{\dagger (1)}_{ia}(\tau,\Omega) \, \bar{v}_{\tilde{a}\tilde{\jmath}\tilde{\imath}\tilde{\jmath}}(\Omega)  \nonumber \\
&+& \frac{1}{4} \sum_{ijab}  {\cal T}^{\dagger (1)}_{ijab}(\tau,\Omega) \, \bar{v}_{\tilde{a}\tilde{b}\tilde{\imath}\tilde{\jmath}}(\Omega)   \, ,
\end{eqnarray}
\end{subequations}
with the expressions of first-order one- and two-body cluster amplitudes provided by Eq.~\ref{1storderclusterB}.

\begin{figure}[t!]
\begin{center}
\includegraphics[clip=,width=0.14\textwidth,angle=0]{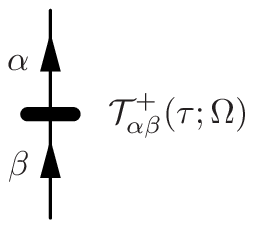} \\ \vspace{0.4cm}
\includegraphics[clip=,width=0.32\textwidth,angle=0]{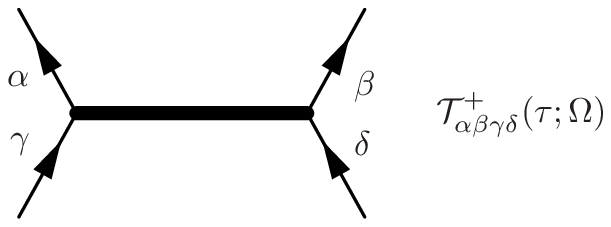}\\
\end{center}
\caption{
\label{diagbetadependentclusters}
Goldstone diagrams representing  one- (first line) and two-body (second line) cluster amplitudes.}
\end{figure}

\subsubsection{Diagrammatic}
\label{Diagrammatictransformed}

\begin{figure}[t!]
\begin{center}
\includegraphics[clip=,width=0.20\textwidth,angle=0]{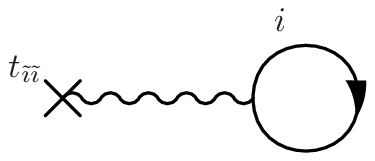}\\ \vspace{0.4cm}
\includegraphics[clip=,width=0.18\textwidth,angle=0]{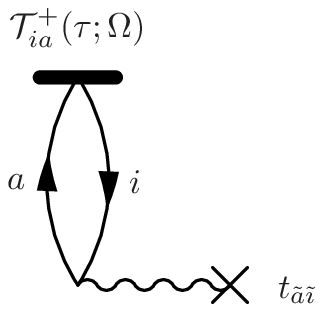}\\
\end{center}
\caption{
\label{diagramst}
Goldstone coupled-cluster diagrams contributing to $t(\tau,\Omega)$.}
\end{figure}

\begin{figure}[t!]
\begin{center}
\includegraphics[clip=,width=0.21\textwidth,angle=0]{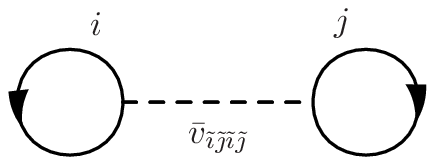}\\ \vspace{0.4cm}
\includegraphics[clip=,width=0.18\textwidth,angle=0]{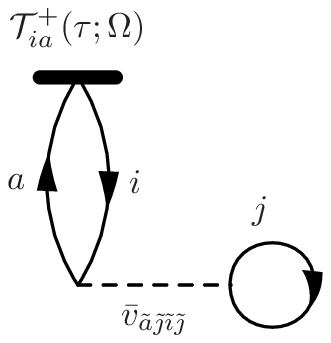}\\ \vspace{0.4cm}
\includegraphics[clip=,width=0.19\textwidth,angle=0]{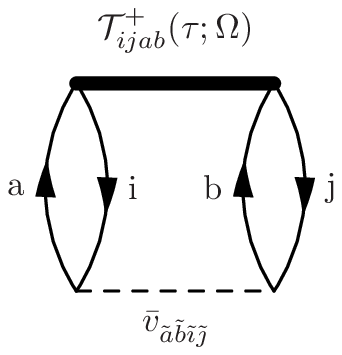}\\ \vspace{0.4cm}
\includegraphics[clip=,width=0.22\textwidth,angle=0]{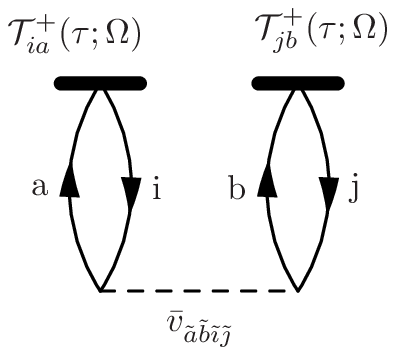}\\
\end{center}
\caption{
\label{diagramsv}
Goldstone coupled-cluster diagrams contributing to $v(\tau,\Omega)$.}
\end{figure}

Goldstone amplitudes of ${\cal T}^{\dagger}_{1}(\tau,\Omega)$ and ${\cal T}^{\dagger}_{2}(\tau,\Omega)$ are represented diagrammatically in Fig.~\ref{diagbetadependentclusters}. Algebraic expressions in Eq.~\ref{expressionTandVC} correspond to the diagrammatic representation given in Figs.~\ref{diagramst} and~\ref{diagramsv}, where the rules are now the same as in standard  CC theory~\cite{shavitt09a}, except that transformed interaction vertices are to be used. Of course, due to the formal similarity of the algebraic expressions, the diagrammatic representation given in Figs.~\ref{diagramst} and~\ref{diagramsv} mirrors exactly the one at play in standard CC theory.

\subsubsection{Amplitude equations}
\label{dynamicalequations}

In order to effectively compute the various contributions to $h(\tau,\Omega)$ in Eq.~\ref{expressionTandVC}, one must have at hand hole-particle matrix elements of the cluster operators. Thus, and as for any CC-based approach, we must identify the equations of motion that determine those $\Omega$-dependent amplitudes.

To make the expressions more compact, we now introduce n-tuply excited off-diagonal energy and norm kernels through
\begin{subequations}
\label{amplitudekernels}
\begin{eqnarray}
N^{ab\ldots}_{ij\ldots}(\tau,\Omega) &\equiv& \langle \Psi (\tau) | \bbone | \Phi^{ab\ldots}_{ij\ldots}(\Omega) \rangle   \,\, , \label{amplitudekernels1} \\
H^{ab\ldots}_{ij\ldots}(\tau,\Omega) &\equiv& \langle \Psi (\tau) | H | \Phi^{ab\ldots}_{ij\ldots}(\Omega) \rangle   \,\, , \label{amplitudekernels2}
\end{eqnarray}
\end{subequations}
where $| \Phi^{ab\ldots}_{ij\ldots}(\Omega) \rangle \equiv A^{ab\ldots}_{ij\ldots} | \Phi(\Omega) \rangle$, with the operator $A^{ab\ldots}_{ij\ldots}$ defined in Eq.~\ref{phexcitation}. From Eq.~\ref{schroedinger}, one obtains that
\begin{equation}
H^{ab\ldots}_{ij\ldots}(\tau,\Omega) = -\partial_{\tau} N^{ab\ldots}_{ij\ldots}(\tau,\Omega)  \, . \label{dynamicalkernels}
\end{equation}
In App.~\ref{amplitudeequations}, we demonstrate in detail how Eq.~\ref{dynamicalkernels} eventually provides the equations of motion satisfied by the n-tuply excited ($\tau$- and $\Omega$-dependent) hole-particle cluster amplitudes under the form
\begin{equation}
h^{ab\ldots}_{ij\ldots}(\tau,\Omega) = -\partial_{\tau} {\cal T}^{\dagger}_{ij\ldots ab\ldots}(\tau,\Omega)  \,  \, , \label{reduceddynamicalkernels}
\end{equation}
where the n-tuply excited {\it linked/connected} energy kernel is
\begin{equation}
h^{ab\ldots}_{ij\ldots}(\tau,\Omega) \equiv \frac{\langle \Phi | e^{{\cal T}^{\dagger}(\tau,\Omega)} H  | \Phi^{ab\ldots}_{ij\ldots}(\Omega) \rangle_{c}}{\langle \Phi |  \Phi(\Omega) \rangle} \, , \label{CCamplitudekernels}
\end{equation}
whose connected character denotes that (i) cluster operators are all connected to $H$ and that (ii) no contraction is to be considered among cluster operators or within any given cluster operator, such that the expansion naturally terminates. One should note that no contraction occurs within the operator $A^{ab\ldots}_{ij\ldots}$ given that any such contraction is, as in standard CC, zero by virtue of Eq.~\ref{contractionsrho2}.

Equations~\ref{reduceddynamicalkernels} and~\ref{CCamplitudekernels} generalize in a transparent way time-dependent CC equations~\cite{Kvaal:2012ys,Pigg:2012sw} to {\it off-diagonal}, i.e. $\Omega$-dependent, cluster amplitudes. Due to the termination of the exponential, singly- and doubly-excited off-diagonal linked/connected energy kernels read as
\begin{widetext}
\begin{subequations}
\label{termination}
\begin{eqnarray}
h^{a}_{i}(\tau,\Omega) &=& \langle \Phi |  \Big[\bbone + {\cal T}^{\dagger}_{1}(\tau,\Omega)  + {\cal T}^{\dagger}_{2}(\tau,\Omega) + {\cal T}^{\dagger}_{3}(\tau,\Omega) \nonumber \\
&& \hspace{0.6cm}  + {\cal T}^{\dagger}_{2}(\tau,\Omega) \, {\cal T}^{\dagger}_{1}(\tau,\Omega)  +  \frac{1}{2} {\cal T}^{\dagger\, 2}_{1}(\tau,\Omega) +  \frac{1}{3!} {\cal T}^{\dagger\, 3}_{1}(\tau,\Omega) \Big] H   |\Phi^{a}_{i} (\Omega) \rangle_{c}\langle \Phi |  \Phi(\Omega) \rangle^{-1} \, , \label{termination1} \\ 
h^{ab}_{ij}(\tau,\Omega) &=& \langle \Phi |  \Big[\bbone + {\cal T}^{\dagger}_{1}(\tau,\Omega)  + {\cal T}^{\dagger}_{2}(\tau,\Omega)   + {\cal T}^{\dagger}_{3}(\tau,\Omega)   + {\cal T}^{\dagger}_{4}(\tau,\Omega) +  \frac{1}{2} {\cal T}^{\dagger\, 2}_{1}(\tau,\Omega) + {\cal T}^{\dagger}_{2}(\tau,\Omega) \, {\cal T}^{\dagger}_{1}(\tau,\Omega) + {\cal T}^{\dagger}_{3}(\tau,\Omega) \, {\cal T}^{\dagger}_{1}(\tau,\Omega)  \nonumber \\
&& \hspace{0.6cm}  +  \frac{1}{2} {\cal T}^{\dagger\, 2}_{2}(\tau,\Omega) +  \frac{1}{3!} {\cal T}^{\dagger\, 3}_{1}(\tau,\Omega)  +  \frac{1}{2} {\cal T}^{\dagger}_{2}(\tau,\Omega) \, {\cal T}^{\dagger\, 2}_{1}(\tau,\Omega)  +  \frac{1}{4!} {\cal T}^{\dagger\, 4}_{1}(\tau,\Omega) \Big] H   |\Phi^{ab}_{ij} (\Omega) \rangle_{c}\langle \Phi |  \Phi(\Omega) \rangle^{-1} \, , \label{termination2}
\end{eqnarray}
\end{subequations}
\end{widetext}
respectively, and are thus formally identical to the usual, i.e. diagonal ($\Omega=0$), CC expressions~\cite{shavitt09a}. This is true for any n-tuply excited energy kernel. It can be demonstrated, similarly to Eq.~\ref{expressionTandVC}, that the expanded expressions of $h^{a}_{i}(\tau,\Omega)$ and $h^{ab}_{ij}(\tau,\Omega)$ obtained through the application of the off-diagonal Wick theorem are formally identical to the usual, i.e. diagonal ($\Omega=0$), CC formulae~\cite{shavitt09a} as long as one uses {\it transformed} kinetic- and potential-energy operators in place of the original ones, i.e.
\begin{equation}
h^{ab\ldots}_{ij\ldots}(\tau,\Omega) = \langle \Phi | e^{{\cal T}^{\dagger}(\tau,\Omega)} \tilde{H}(\Omega) | \Phi^{ab\ldots}_{ij\ldots} \rangle_{c} \, . \label{CCamplitudekernelsbis}
\end{equation}
It is unnecessary to reproduce here these well-known expressions and the associated diagrams~\cite{shavitt09a}.

At a given $\tau$, one can initialize the amplitude equations with the cluster amplitudes obtained at first order in MBPT, as provided in Eq.~\ref{1storderclusterB} for ${\cal T}^{\dagger}_{1}$ and ${\cal T}^{\dagger}_{2}$. Once $\Omega$- and $\tau$-dependent cluster amplitudes have been obtained by solving the set of coupled time-dependent equations~\ref{reduceddynamicalkernels}, the connected part of the energy kernel $h(\tau,\Omega)$ can be computed through Eq.~\ref{expressionTandVC}.  

As a matter of fact, one is eventually only interested in the infinite imaginary time limit. In this limit, the scheme becomes stationary such that the static amplitude equations are obtained  by setting the right-hand side of Eq.~\ref{reduceddynamicalkernels} to zero 
\begin{equation}
h^{ab\ldots}_{ij\ldots}(\Omega) = 0  \, , \label{staticamplitudeequations}
\end{equation}
which naturally extends standard CC amplitude equations.


\subsection{Norm kernel}
\label{Secnormkernel}

\subsubsection{Position of the problem}

Standard many-body theories typically compute the ground-state energy from the reduced {\it diagonal} energy kernel according to
\begin{eqnarray}
E^{J}_{0} &=& \lim\limits_{\tau \to \infty} {\cal H}(\tau,0) =  h(0) \, , \label{usualenergy}
\end{eqnarray}
rather than from Eq.~\ref{projected_energy}. In this situation, one must thus solely compute the linked/connected energy kernel at $\Omega=0$ without worrying about the norm kernel. For practical purposes, one can indeed choose to work with intermediate normalization from the outset, i.e. with ${\cal N}(\tau,0)=1$ for all $\tau$.

A key difficulty encountered presently relates to the necessity to capture the {\it change} of the reduced norm kernel ${\cal N}(\tau,\Omega)$ with $\Omega$ as it constitutes a key ingredient entering Eq.~\ref{projected_energy} and Eqs.~\ref{integratedkernels}-\ref{yrast_projected_energy}. The necessity to access ${\cal N}(\tau,\Omega)$ as a function of $\Omega$ is a significant formal and technical complication compared to standard CC theory. Indeed, finding an expansion of a norm kernel based on a coupled cluster ansatz that naturally terminates is a challenge. In particular, starting from the perturbative expansion $N(\tau,\Omega)=e^{-\tau\varepsilon_0 + n(\tau,\Omega)} \, \langle \Phi | \Phi(\Omega) \rangle$ developed in Sec.~\ref{normkernel2} is not immediately helpful as the diagrams making up $n(\tau,\Omega)$ are not linked to an operator at a fixed time. As a result, one cannot trivially rewrite  $n(\tau,\Omega)$ as a naturally terminating cluster expansion. We now explain how this apparent difficulty can be overcome. 

\subsubsection{Key property}

In the case of an {\it exact} many-body calculation, Eqs.~\ref{expandedkernels1} and~\ref{expandedkernels3} trivially lead, for any $\tau$, to
\begin{subequations}
\label{restoreLieoperators}
\begin{eqnarray}
\frac{\int_{D_{SU(2)}} \! d\Omega \, D^{J \, \ast}_{MK}(\Omega) \, {\cal J}_z(\tau,\Omega)}{\int_{D_{SU(2)}} \! d\Omega \, D^{J \, \ast}_{MK}(\Omega) \, {\cal N}(\tau,\Omega)} &=&  M\hbar \, , \label{restoreLieoperators1} 
\end{eqnarray}
and
\begin{eqnarray}
\frac{\int_{D_{SU(2)}} \! d\Omega \, D^{J \, \ast}_{MK}(\Omega) \, {\cal J}^2(\tau,\Omega)}{\int_{D_{SU(2)}} \! d\Omega \, D^{J \, \ast}_{MK}(\Omega) \, {\cal N}(\tau,\Omega)} &=& J(J+1)\hbar^2  \, , \label{restoreLieoperators2} 
\end{eqnarray}
\end{subequations}
independently of $K$ and $(M,K)$, respectively. In particular, Eq.~\ref{restoreLieoperators2} states that, in coherence with the lowest energy associated with angular momentum $J$ (Eq.~\ref{yrast_projected_energy} or Eq.~\ref{projected_energy_MBPT} below), one has that
\begin{widetext}
\begin{eqnarray}
\lim\limits_{\tau \to \infty} \frac{\sum_{MK} f^{J\ast}_M \, f^{J}_K  \int_{D_{SU(2)}} \! d\Omega  \,\, D^{J \, \ast}_{MK}(\Omega) \, \, {\cal J}^2(\tau,\Omega)}{\sum_{MK} f^{J\ast}_M \, f^{J}_K  \int_{D_{SU(2)}} \! d\Omega  \,\, D^{J \, \ast}_{MK}(\Omega) \,\, {\cal N}(\tau,\Omega)} = J(J+1)\hbar^2 , \label{projected_angular_momentum}
\end{eqnarray}
which testifies that the corresponding state is indeed an eigenstate of the total angular momentum with eigenvalue $J(J+1)\hbar^2$. Employing the Baker-Campbell-Hausdorff identity to commute $J_z$ with $R(\Omega)$
\begin{equation}
R(\Omega) J_z = \left[\sin \beta \, \cos \alpha \, J_x + \sin \beta \, \sin \alpha \, J_y + \cos \beta \, J_z\right] R(\Omega)
\end{equation}
one obtains the third remarkable identity
\begin{equation}
\frac{\int_{D_{SU(2)}} \!  \, D^{J \, \ast}_{MK}(\Omega) \, \Big[\sin \beta \, \cos \alpha \, {\cal J}_x(\tau,\Omega) + \sin \beta \, \sin \alpha \, {\cal J}_y(\tau,\Omega) + \cos \beta \, {\cal J}_z(\tau,\Omega) \Big]}{\int_{D_{SU(2)}} \!  \, D^{J \, \ast}_{MK}(\Omega) \, {\cal N}(\tau,\Omega)} =  K\hbar \, , \label{restoreLieoperators3} 
\end{equation}
\end{widetext}
Eventually, Eqs.~\ref{restoreLieoperators} and~\ref{restoreLieoperators3} stress the fact that we {\it know} a priori the results that must be obtained  from the integral over the domain of the group  of off-diagonal kernels associated with operators of the Lie algebra (contrarily to the energy). Consequently, the key question is: what happens to Eqs.~\ref{restoreLieoperators} and~\ref{restoreLieoperators3} when ${\cal J}_i(\tau,\Omega)$, ${\cal J}^2(\tau,\Omega)$ and ${\cal N}(\tau,\Omega)$ are approximated? Or to rephrase the question more appropriately: what constraint(s) does restoring the symmetry {\it exactly}, i.e. fulfilling  Eqs.~\ref{restoreLieoperators} and~\ref{restoreLieoperators3}, impose on the truncation scheme used to approximate the kernels? Addressing this question below delivers the proper approach to the reduced norm kernel. 

\subsubsection{Coupled differential equations}

We derive three coupled first-order ordinary differential equations (ODE) fulfilled, at each imaginary time $\tau$, by ${\cal N}(\tau,\Omega)$. To do so, we perform the explicit derivative of the rotation operator $R(\Omega)$ with respect to the Euler angles inside the Wigner D-functions entering ${\cal N}(\tau,\Omega)$. We then employ the Baker-Campbell-Hausdorff identity repeatedly to bring the Lie algebra operators $J_i$ to the left of  $R(\Omega)$ in the resulting kernels. Further exploiting that the reduced kernel of an operator $O$ can be factorized according to ${\cal O}(\tau,\Omega)=o(\tau,\Omega) {\cal N}(\tau,\Omega)$, where $o(\tau,\Omega)$ denotes the corresponding linked/connected kernel, we finally arrive at
\begin{widetext}
\begin{subequations}
\label{NormkernelODE}
\begin{eqnarray}
\frac{\partial}{\partial \alpha} \, {\cal N}(\tau,\Omega)  + \frac{i}{\hbar} \, j_z(\tau,\Omega) \, {\cal N}(\tau,\Omega) &=& 0 \, ,\label{NormkernelODE1} \\
\frac{\partial}{\partial \beta} \, {\cal N}(\tau,\Omega)  - \frac{i}{\hbar} \Big[\sin \alpha \, j_x(\tau,\Omega) - \cos \alpha \, j_y(\tau,\Omega)\Big] {\cal N}(\tau,\Omega) &=& 0  \, ,\label{NormkernelODE2} \\
\frac{\partial}{\partial \gamma} \, {\cal N}(\tau,\Omega)  + \frac{i}{\hbar} \Big[\sin \beta \, \cos \alpha \, j_x(\tau,\Omega) + \sin \beta \, \sin \alpha \, j_y(\tau,\Omega) + \cos \beta \, j_z(\tau,\Omega) \Big] {\cal N}(\tau,\Omega) &=& 0 \, ,\label{NormkernelODE3} 
\end{eqnarray}
\end{subequations}
\end{widetext}
with the initial condition ${\cal N}(\tau,0)=1$. Equation~\ref{NormkernelODE} demonstrates that, while the off-diagonal norm kernel does not itself possess a naturally terminating CC expansion, it can be related to the linked/connected kernels of the operators making up the Lie algebra, which themselves possess naturally terminating CC expansions. Computing the latter, which can be done by substituting the kinetic operator with $J_i$ in Eq.~\ref{CCenergyequation1}, ${\cal N}(\tau,\Omega)$ is accessed by integrating Eqs.~\ref{NormkernelODE1}, \ref{NormkernelODE2} and~\ref{NormkernelODE3} numerically. It is to be noted that, as for the one-body kinetic energy operator, the terminating expansion of $j_i(\tau,\Omega)$ is formally exact at and beyond the single level. To obtain the corresponding algebraic expressions one must compute matrix elements of $J_i$ in the bi-orthogonal bases.

\begin{figure}[t!]
\begin{center}
\includegraphics[clip=,width=0.21\textwidth,angle=0]{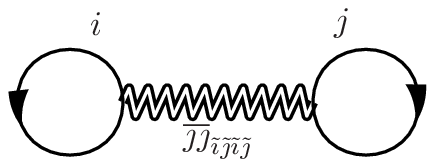}\\ \vspace{0.4cm}
\includegraphics[clip=,width=0.18\textwidth,angle=0]{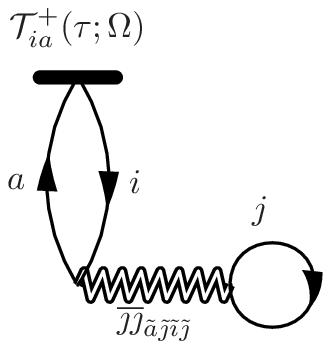}\\ \vspace{0.4cm}
\includegraphics[clip=,width=0.19\textwidth,angle=0]{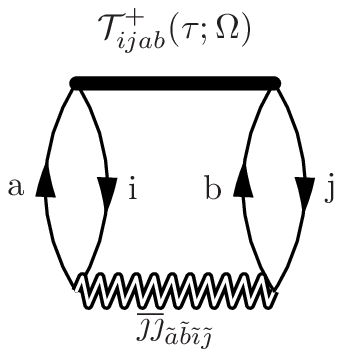}\\ \vspace{0.4cm}
\includegraphics[clip=,width=0.22\textwidth,angle=0]{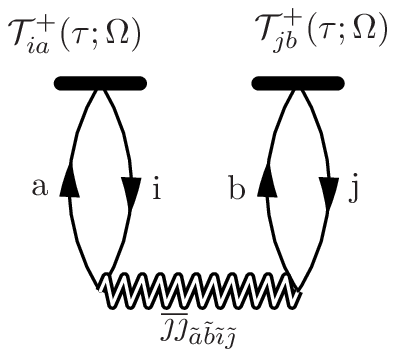}\\
\end{center}
\caption{
\label{diagramsjj}
Goldstone coupled-cluster diagrams contributing to $j^2_{(2)}(\tau,\Omega)$.}
\end{figure}

In addition to authorizing the computation of ${\cal N}(\tau,\Omega)$ from kernels displaying naturally terminating CC expansions, the scheme proposed above ensures that the angular momentum is indeed exactly restored at any truncation order in the proposed many-body method. This scheme is thus not only satisfying from the computational standpoint but also from the physical one. For instance, assuming that $j_z(\tau,\Omega)$ and ${\cal N}(\tau,\Omega)$ are indeed related through Eq.~\ref{NormkernelODE1}, we have that
\begin{widetext}
\begin{eqnarray}
\int_{D_{SU(2)}} \! d\Omega \, D^{J \, \ast}_{MK}(\Omega) \, {\cal J}_z(\tau,\Omega) &=& +i\hbar \int_{D_{SU(2)}} \! d\Omega \, D^{J \, \ast}_{MK}(\Omega) \, \frac{\partial}{\partial\alpha}  \, {\cal N}(\tau,\Omega) \nonumber \\
&=& -i\hbar \int_{D_{SU(2)}} \! d\Omega \,  \frac{\partial}{\partial\alpha} \, D^{J \, \ast}_{MK}(\Omega) \, {\cal N}(\tau,\Omega) \nonumber \\
&=& M\hbar \int_{D_{SU(2)}} \! d\Omega \, D^{J \, \ast}_{MK}(\Omega) \, {\cal N}(\tau,\Omega) \, ,
\end{eqnarray}
\end{widetext}
where an integration by part with respect to the angle $\alpha$ was performed to go from the first to the second line before invoking Eq.~\ref{wignerODE1} to obtain the final result. This demonstrates that, if the reduced norm kernel satisfies Eq.~\ref{NormkernelODE1}, then Eq.~\ref{restoreLieoperators1} is exactly fulfilled {\it independently of the approximation made on} $j_z(\tau,\Omega)$, i.e. independently of the order at which the CC scheme is truncated. Similarly, Eq.~\ref{NormkernelODE3} ensures that identity~\ref{restoreLieoperators3} is always fulfilled. Last but not least, the three coupled ODEs~\ref{NormkernelODE} can be combined to demonstrate that 
\begin{widetext}
\begin{eqnarray}
-\left[\frac{\partial}{\partial \beta}\left(\sin \beta \frac{\partial}{\partial \beta}\right) + \frac{1}{\sin^2\beta} \left(\frac{\partial^2}{\partial \alpha^2} - 2\cos \beta\frac{\partial^2}{\partial \alpha\partial \gamma} +  \frac{\partial^2}{\partial \gamma^2}\right)\right] {\cal N}(\tau,\Omega) &=&   \frac{1}{\hbar^2}{\cal J}^2(\tau,\Omega) \label{checkJ2} \\ 
&=& \frac{1}{\hbar^2}j^2(\tau,\Omega) {\cal N}(\tau,\Omega)\, , \nonumber
\end{eqnarray}
\end{widetext}
\begin{figure}[t!]
\begin{center}
\includegraphics[clip=,width=0.20\textwidth,angle=0]{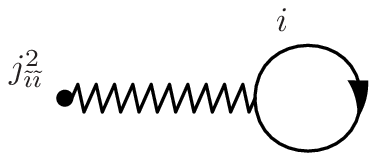}\\ \vspace{0.4cm}
\includegraphics[clip=,width=0.18\textwidth,angle=0]{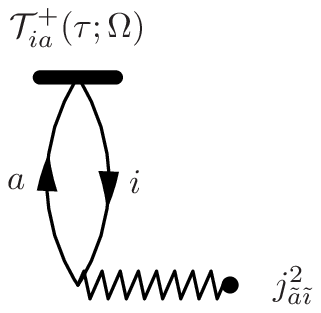}\\
\end{center}
\caption{
\label{diagramsj2}
Goldstone coupled-cluster diagrams contributing to $j^2_{(1)}(\tau,\Omega)$.}
\end{figure}
is satisfied. Via two consecutive integrations by parts, Eq.~\ref{checkJ2} can be shown to provide Eq.~\ref{restoreLieoperators2}, and thus Eq.~\ref{projected_angular_momentum}. To check that the eigenvalue $J(J+1)\hbar^2$ is indeed obtained in a practical calculation when ${\cal N}(\tau,\Omega)$ is obtained by integrating Eq.~\ref{NormkernelODE} at a given order in the CC expansion, the left-hand side of Eq.\ref{restoreLieoperators2} can be computed explicitly from $j^2(\tau,\Omega)$ at the same order in the expansion. The expression of $j^2(\tau,\Omega)$ is obtained by substituting $T$ ($V$) with $J^2_{(1)}$ ($J^2_{(2)}$) in Eq.~\ref{CCenergyequation1} (Eq.~\ref{CCenergyequation2}). The corresponding algebraic expressions requires the computation of the matrix elements of $J^2_{(1)}$ and $J^2_{(2)}$ in the bi-orthogonal bases.
The corresponding diagrams are displayed in Fig.~\ref{diagramsj2} and~\ref{diagramsjj} for illustration. Those diagrams obviously mirror those displayed in Fig.~\ref{diagramst} and~\ref{diagramsv} for the kinetic and potential energy kernels, respectively.

Eventually, the fact that ${\cal N}(\tau,\Omega)$ is determined from the group structure, i.e. from kernels associated with members of the Lie algebra, is  very natural in the present context. Once extracted at a given CC order through Eq.~\ref{NormkernelODE}, the norm kernel can be consistently used in the computation of the energy as is discussed below in Sec.~\ref{symrestE}.

\subsubsection{Lowest order}

When reducing the CC calculation to lowest order, i.e. taking ${\cal T}^{\dagger}_{1}(\tau,\Omega)={\cal T}^{\dagger}_{2}(\tau,\Omega)=0$ such that  
\begin{subequations}
\begin{eqnarray}
j^{(0)}_i(\tau,\Omega) &=& j_i(0,\Omega) = \langle \Phi | J_i | \Phi(\Omega) \rangle \, , \\
j^{2(0)}(\tau,\Omega) &=& j^2(0,\Omega) =  \langle \Phi | J^2 | \Phi(\Omega) \rangle \, ,
\end{eqnarray}
\end{subequations}
Equation~\ref{NormkernelODE} reduces to the ODEs satisfied by the norm kernel ${\cal N}^{(0)}(\tau,\Omega)=\langle \Phi | \Phi(\Omega) \rangle$ (Eq.~\ref{kernel}) at play in angular-momentum projected HF theory, e.g. see Refs.~\cite{Har82a,Ena99a}. In other words, the present scheme consistently extends the computation of the norm kernel at play in projected HF theory to any order in the angular-momentum restored CC theory.
 
\subsubsection{Time dependence}

As pointed out in Sec.~\ref{structure1} and demonstrated in App.~\ref{structure2}, the perturbative expansion of the (reduced) norm kernel provides the characteristic time dependence of the connected kernel $\aleph(\tau,\Omega)=\ln {\cal N}(\tau,\Omega)-\ln  \langle \Phi | \Phi(\Omega) \rangle$, in particular in the large $\tau$ limit (Eq.\ref{betalim2}) where it goes to a constant value. In the context of the present section, it is possible to extract this characteristic time dependence on the basis of, e.g., the first order ODE~\ref{NormkernelODE1} that can be rewritten as
\begin{eqnarray}
\frac{\partial}{\partial \alpha} \, \ln {\cal N}(\tau,\Omega) &=& \frac{\partial}{\partial \alpha} \, \aleph(\tau,\Omega) + \frac{\partial}{\partial \alpha} \, \ln  \langle \Phi | \Phi(\Omega) \rangle \nonumber \\
&=&  - \frac{i}{\hbar} \, j_z(\tau,\Omega)  \, ,\label{timedependencenormkernel} 
\end{eqnarray}
such that one is left analyzing the characteristic time dependence of the linked/connected kernel associated with the one-body operator $J_z$. Such an analysis is proposed in App.~\ref{structure3} for the kinetic energy kernel $t(\tau,\Omega)$ and is equally valid for $j_z(\tau,\Omega)$, which permits to recover Eq.\ref{betalim2}. Indeed, the generic time structure of $j_z(\tau,\Omega)$ remains untouched by the integral over $\alpha$ that must be performed to access ${\cal N}(\tau,\Omega)$ through Eq.~\ref{timedependencenormkernel}.

\subsection{Angular-momentum-restored energy}
\label{symrestE}

\subsubsection{Energy of the yrast states}

The symmetry-restored energy associated with the lowest state carrying a given angular momentum is computed through Eq.~\ref{yrast_projected_energy2nd}, which now reads as
\begin{widetext}
\begin{eqnarray}
E^{J}_{0} &=& \frac{\sum_{MK} f^{J\ast}_M \, f^{J}_K  \int_{D_{SU(2)}} \! d\Omega  \,\, D^{J \, \ast}_{MK}(\Omega) \, \, h(\Omega) \,\, {\cal N}(\Omega)}{\sum_{MK} f^{J\ast}_M \, f^{J}_K  \int_{D_{SU(2)}} \! d\Omega  \,\, D^{J \, \ast}_{MK}(\Omega) \,\, {\cal N}(\Omega)} . \label{projected_energy_MBPT}
\end{eqnarray}
\end{widetext}
where $h(\Omega)$ is evaluated from Eq.~\ref{wickvswick2} at $\tau=+\infty$ at the same order in cluster operators, i.e. including singles, doubles, triples\ldots, as $j_i(\Omega)$ and $j^2(\Omega)$ entering the computation of ${\cal N}(\Omega)$ through the integration of Eq.~\ref{NormkernelODE}. 

Expressing the energy in terms of the {\it reduced} norm kernel in Eq.~\ref{projected_energy_MBPT} is essential. Indeed, the fact that ${\cal N}(\tau,\Omega)$ goes to a finite number in the large $\tau$ limit, contrarily to $N(\tau,\Omega)$ that goes exponentially to zero, is mandatory to make the ratio in Eq.~\ref{projected_energy_MBPT} well defined and numerically controllable. Eventually, the consistency of the approach relies on the fact that all linked/connected kernels at play are truncated at the same order in the cluster operators. 

For a system that spontaneously breaks rotational symmetry at the mean-field level, i.e. whose reference state $| \Phi \rangle$ is spontaneously deformed, the set of yrast states typically provides the rotational band built on top of the ground state. As members of the rotational band are accessed beyond the lowest order in the CC expansion (see Sec.~\ref{PHFsection} below), their energy include dynamical correlations, i.e. the consistent mixing of collective and individual dynamics.

\subsubsection{Angular-momentum projected HF theory}
\label{PHFsection}

Applying the proposed scheme at lowest order in cluster operators, i.e. taking ${\cal T}^{\dagger}_{n}(\tau,\Omega)=0$ for all $n$, one recovers the angular-momentum projected Hartree-Fock (PHF) theory~\cite{ring80a,blaizot86} (assuming that the reference state $| \Phi \rangle$ is obtained from a deformed HF calculation). The associated kernels read as 
\begin{subequations}
\label{PHFkernels}
\begin{eqnarray}
h^{(0)}(\tau,\Omega) &=& h(0,\Omega) = \frac{\langle \Phi | H | \Phi(\Omega) \rangle}{\langle \Phi  | \Phi(\Omega) \rangle} \, ,  \label{PHFkernelsH} \\
{\cal N}^{(0)}(\tau,\Omega) &=& {\cal N}(0,\Omega) = \langle \Phi  | \Phi(\Omega) \rangle \label{PHFkernelsN} \, ,
\end{eqnarray}
\end{subequations}
for all $\tau$, such that the symmetry-restored energy becomes
\begin{subequations}
\label{projected_HF}
\begin{eqnarray*}
E^{J (0)}_{0} &=& \frac{\sum_{MK} f^{J\ast}_M \, f^{J}_K  \int_{D_{SU(2)}} \! d\Omega  \,\, D^{J \, \ast}_{MK}(\Omega) \, \, \langle \Phi | H | \Phi(\Omega) \rangle}{\sum_{MK} f^{J\ast}_M \, f^{J}_K  \int_{D_{SU(2)}} \! d\Omega  \,\, D^{J \, \ast}_{MK}(\Omega) \,\, \langle \Phi  | \Phi(\Omega) \rangle} \label{projected_HF1} \\
&=& \frac{\langle \Phi | H | \Phi^{JM}_{0} \rangle}{\langle \Phi | \Phi^{JM}_{0} \rangle} \label{projected_HF2} \\
&=& \frac{\langle \Phi^{JM}_{0} | H | \Phi^{JM}_{0} \rangle}{\langle \Phi^{JM}_{0} | \Phi^{JM}_{0} \rangle} \, ,
\end{eqnarray*}
\end{subequations}
where the (un-normalized) angular-momentum projected HF wave-function $| \Phi^{JM}_{0} \rangle \equiv \sum_{K} f^{J}_K  P^{J}_{MK} | \Phi \rangle$ is defined from the transfer operator
\begin{equation}
P^{J}_{MK} \equiv  \frac{2J+1}{16\pi^{2}}\int_{D_{SU(2)}} \!\!\!d\Omega \, D_{MK}^{J\ast}(\Omega) \, R(\Omega) \, ,
\end{equation}
satisfying 
\begin{subequations}
\label{projector}
\begin{eqnarray}
P^{J}_{MK}\,P^{J'}_{M'K'} &=& \delta_{JJ'}\delta_{KM'}P^{J}_{MK'} \, , \label{projector1} \\
P^{J \dagger}_{MK} &=& P^{J}_{KM} \, , \label{projector2} \\ 
\left[H,P^{J}_{MK}\right] &=& 0 \, .
\end{eqnarray}
\end{subequations}

\subsubsection{Ground-state energy}

We now focus on the ground state. Its energy is the lowest obtained from Eq.~\ref{projected_energy_MBPT}, which define the corresponding angular momentum $J_0$.If one were to sum all diagrams in the computation of $h(\Omega)$ and ${\cal N}(\Omega)$, the symmetry restoration would become dispensable by definition. Indeed, the complete sum of symmetry-breaking diagrams is known to fulfill the symmetry. This can be recovered by using property~\ref{deriveeOmega} stating that $h(\Omega)$ becomes independent of $\Omega$ in the exact limit. In this case, the linked/connected energy kernel comes out of the integral and one obtains
\begin{eqnarray}
E^{J_0}_0 &=&  h(0) \, . \label{standard_MBPT}
\end{eqnarray}
The benefit of the method arises as soon as the expansion is truncated. Indeed, $h(\Omega)$ acquires a dependence on $\Omega$ that signals the breaking of the symmetry generated by the truncation. As such, the method authorizes the summation of standard sets of diagrams (i.e. dealing with so-called {\it dynamical} correlations) while leaving the non-perturbative symmetry-restoration process (i.e. dealing with so-called {\it static} correlations) to be achieved at each truncation order through the integration over the domain of the group. As a matter of fact, one can rewrite the expression for the ground-state energy as
\begin{widetext}
\begin{eqnarray}
E^{J_0}_0 &=&  h(0) + \frac{\sum_{MK} f^{J_0\ast}_M \, f^{J_0}_K  \int_{D_{SU(2)}} \! d\Omega \,\, D^{J_0 \, \ast}_{MK}(\Omega) \,\, \left[h(\Omega)-h(0) \right] \,\, {\cal N}(\Omega)}{\sum_{MK} f^{J_0\ast}_M \, f^{J_0}_K \int_{D_{SU(2)}} \! d\Omega \,\, D^{J_0 \, \ast}_{MK}(\Omega) \,\,  {\cal N}(\Omega)} \, , \label{rewritingresult}
\end{eqnarray}
\end{widetext}
such that the effect of the angular-momentum restoration itself can be viewed, at any truncation order, as a {\it correction} to standard symmetry-unrestricted CC results provided by the approximation to $h(0)$. Of course, the second term in the right-hand side of Eq.~\ref{rewritingresult} is zero if (i) the unperturbed state $| \Phi \rangle$ does not break the symmetry and/or (ii) one sums all diagrams in $h(\Omega)$. 

It will be of interest to investigate schemes that reduce the computational effort significantly, e.g. one could compute $h(0)$ through symmetry-unrestricted CC theory while evaluating the kernels in the symmetry restoration term perturbatively.



\subsection{Discussion}

Let us now make one last set of comments
\begin{itemize}
\item The symmetry-restored CC theory solely invokes {\it hole-particle} cluster amplitudes ${\cal T}^{\dagger}_{ij\ldots ab\ldots}(\tau,\Omega)$. Although the complete set of matrix elements was a priori considered in the definition of the cluster operators in Eq.~\ref{clusteroperators}, particle-hole, particle-particle and hole-hole matrix elements remain inactive and can be set to zero as in standard CC theory. This is due to (i) the linked/connected character of the operator kernels at play, (ii) the ability to invoke the {\it reduced} norm kernel that provides intermediate normalization at $\Omega=0$ and (iii) the fact that this reduced norm kernel is itself entirely determined from linked/connected kernels. Thus, one can eventually rewrite, e.g., single and double cluster operators as
\begin{subequations}
\label{ToperatorsB}
\begin{eqnarray}
{\cal T}_{1}(\tau,\Omega) &\equiv& \frac{1}{(1!)^2} \sum_{ia} {\cal T}_{ai}(\tau,\Omega)  \, a^{\dagger}_{a} \, a_{i}  \, , \label{T1B} \\ 
{\cal T}_{2}(\tau,\Omega) &\equiv& \frac{1}{(2!)^2} \sum_{ijab} {\cal T}_{abij}(\tau,\Omega) \, a^{\dagger}_{a} \, a^{\dagger}_{b} \, a_{j} \, a_{i}  \label{T2B} \, ,
\end{eqnarray}
\end{subequations}
which naturally extends cluster operators at play in standard CC theory, i.e. at $\Omega$.
\item For each value of $\Omega$, matrix elements of $T$, $V$, $J_x$, $J_y$, $J_z$, $J^2_{(1)}$ and $J^2_{(2)}$ must be computed and stored in the bi-orthogonal bases. With those of $T$, $V$ at hand, the imaginary-time amplitude equations to be solved read as standard SR-CC equations. With the cluster amplitudes at hand, all the linked/connected kernels of interest can be evaluated and the ODEs fulfilled by the reduced norm kernel can eventually be solved.
\item Eventually, the CC scheme only needs to be solved at $\tau=+\infty$, i.e. the imaginary-time formulation becomes superfluous and one is left with the static version of the many-body formalism.
\item The present scheme is of multi-reference character but reduces in practice to a set of $N_{\text{sym}}$ single-reference-like CC calculations. Typically, $N_{\text{sym}}\sim 10$ per rotation angle. The factor of $10$ is an estimation based on the discretization of the integrals over Euler angles in Eq.~\ref{projected_energy_MBPT} typically needed to achieve convergence in the computation of low angular-momentum states of even-even nuclei at the PHF level~\cite{bender03b}. The full-fledged approach involving a three-dimensional integral thus leads to a significant numerical cost, i.e. of the order of $10^3$ deformed SR-CC calculations. Fortunately, the very large majority of doubly open shell even-even nuclei are likely to be well described on the basis of an axially-symmetric reference state $| \Phi \rangle$, i.e. a Slater determinant fulfilling $J_z| \Phi \rangle =0$. In such a case, the formalism is simplified and the integration over the domain of $SU(2)$ reduces to a single integral over the Euler angle $\beta$. It is thus recommended to first implement this version of the theory. The corresponding set of simplified equations is provided for reference in App.~\ref{axialsymmetry}.
\item It could be of interest to design an approximation of the presently proposed many-body formalism based on a Kamlah expansion~\cite{kamlah68a} in the future.
\end{itemize}

\subsection{Implementation algorithm}

Let us eventually synthesize the steps the owner of a single-reference coupled-cluster code allowing for the breaking of rotational symmetry must follow to implement the symmetry restoration procedure\footnote{In order to actually fit with an existing SR-CC code, one must eventually proceed to the hermitian conjugation of the quantities and equations that are referred to.}.
\begin{enumerate}
\item Solve, e.g., symmetry-unrestricted Hartree-Fock equations in the basis of interest to obtain the (deformed) reference state $| \Phi \rangle$ (see Eq.~\ref{slater}). We denote by $N_b=N_h+N_p$ the dimension of the one-body Hilbert space, where $N_h$ denotes the number of occupied states of $| \Phi \rangle$ and $N_p$ the number of unoccupied states.
\item Discretize the intervals of integration over the three Euler angles $\Omega\equiv(\alpha,\beta,\gamma)$. 
\item For each combination of Euler angles
\begin{enumerate}
\item Compute the $N_b \times N_b$ matrix $R_{\alpha\beta}(\Omega)\equiv \langle \alpha | R(\Omega) | \beta \rangle$ in the resulting, e.g., Hartree-Fock single-particle basis. Build the matrix $M_{ij}(\Omega)$ as the $N_h \times N_h$ reduction of $R_{\alpha\beta}(\Omega)$ to the subspace of hole states of $| \Phi \rangle$ and compute its inverse $M^{-1}(\Omega)$.
\item Built the $N_p \times N_h$ rectangular matrix $\rho^{ph}_{ai}(\Omega)\equiv \sum_{i=1}^{N_h}R_{aj}(\Omega)M^{-1}_{ji}(\Omega)$ (see Eq.~\ref{transdens}).
\item Build the bi-orthogonal bases according to Eqs.~\ref{rightbasis} and~\ref{leftbasis}. 
\item Transform the matrix elements of $T$ and $V$ into the bi-orthogonal system to generate the matrix elements of $\tilde{T}(\Omega)$ and $\tilde{V}(\Omega)$, respectively.
\item Initialize the coupled-cluster amplitudes through first-order perturbation theory; e.g. at the singles and doubles level, apply Eq.~\ref{1storderclusterB} for $\tau\rightarrow+\infty$ to obtain ${\cal T}^{\dagger (1)}_{ia}(\Omega)$ and ${\cal T}^{\dagger (1)}_{ijab}(\Omega)$.
\item Run the single-reference coupled-cluster code using the matrix elements of $\tilde{T}(\Omega)$ and $\tilde{V}(\Omega)$ along with the zeroth-iteration amplitudes, e.g.,  ${\cal T}^{\dagger (1)}_{ia}(\Omega)$ and ${\cal T}^{\dagger (1)}_{ijab}(\Omega)$.
\item Using the converged amplitudes, e.g. ${\cal T}^{\dagger}_{ia}(\Omega)$ and ${\cal T}^{\dagger}_{ijab}(\Omega)$, compute and store the linked/connected kernels $h(\Omega)=t(\Omega)+v(\Omega)$, $j_x(\Omega)$, $j_y(\Omega)$, $j_z(\Omega)$ and $j^2(\Omega)=j^2_{(1)}(\Omega)+j^2_{(2)}(\Omega)$. These kernels invoke one or two-body operators and can thus all be computed on the basis of the explicit algebraic expressions given in Eq.~\ref{expressionTandVC}, i.e. the kernels associated with the angular momentum operators can be computed within the single-reference coupled-cluster code by adapting the calculation of the energy.
\end{enumerate}
\item Using $j_x(\Omega)$, $j_y(\Omega)$ and $j_z(\Omega)$ for the discretized values of the Euler angles, along with the initial condition ${\cal N}(0)=1$, integrate the three coupled ODEs\footnote{Given that one is solely interested in the $\tau\rightarrow+\infty$ limit, the argument can be simply ignored in Eq.~\ref{NormkernelODE}.} given in Eq.~\ref{NormkernelODE} to obtain ${\cal N}(\Omega)$ for each combination of Euler angles. 
\item Solve the Hill-Wheeler-Griffin equation to obtain the weights $f^{J}_K$ (Eq.~\ref{HWG}).
\item Calculate the energy of the yrast states according to Eq.~\ref{projected_energy_MBPT}. Check that the angular momentum is indeed exactly restored by computing Eqs.~\ref{restoreLieoperators1}, \ref{restoreLieoperators1} and~\ref{projected_angular_momentum} (ignoring the time argument).
\end{enumerate}

A few remarks are in order
\begin{itemize}
\item Steps 2 and 3 can be carried out independently for each combination of the Euler angles and is thus amenable to a trivial parallelization. Eventually one solely needs to retrieve and store the value of $h(\Omega)=t(\Omega)+v(\Omega)$, $j_x(\Omega)$, $j_y(\Omega)$, $j_z(\Omega)$ and $j^2(\Omega)=j^2_{(1)}(\Omega)+j^2_{(2)}(\Omega)$.
\item The integrals over $\alpha$ and $\gamma$ can be discretized with a trapezoidal rule, while a Gauss-Legendre quadrature can be used to integrate the integral over $\beta$ (see, e.g., Ref.~\cite{Bender:2008zv}).
\item In practice, the domain of integration in step 6 can often be drastically reduced thanks to the remaining symmetries of the reference state $| \Phi \rangle$\footnote{For a detailed discussion on this question including the advanced case of odd nuclei, see Ref.~\cite{ballythesis}.}
\begin{enumerate}
\item If interested in even-even systems, i.e. in integer values of $J$, the domain of integration over $\alpha$ can first be reduced to $[0,2\pi]$. This corresponds to using only 1/2 of the full $16\pi^2$ integration volume necessary to treat systems with half-integer spin.
\item Time-reversal, parity and signature symmetries allow to further reduce the integration intervals to $\alpha \in [0,2\pi]$, $\beta \in [0,\pi/2]$ and $\gamma \in [0,\pi]$. This corresponds to using only 1/16 of the full $8\pi^2$ integration volume for systems with integer spin. For reference, the number of points in these intervals necessary to compute $^{24}$Mg at the projected mean-field level with high numerical precision is of the order of 6 for $\alpha$, 18 for $\beta$, and 12 for $\gamma$, which corresponds to 24, 36, and 24 points in the full $8\pi^2$ integration volume necessary for integer J values~\cite{Bender:2008zv}.
\item Axial symmetry leaves the sole dependence on the Euler angle $\beta$ over the interval $\beta \in [0,\pi/2]$. This corresponds to a drastic reduction of the numerical cost of the method, and yet allows the treatment of a very large number of doubly open-shell even-even nuclei. Additionally, using 10 points over $[0,\pi/2]$ is typically sufficient to obtain states with $J=0,2,4$ with good precision. One then needs to increase the number discretized values to reach states with larger angular momentum with good precision. Additionally, it can of interest to investigate the validity of the topological gaussian overlap approximation~\cite{onishi66,Hagino:2003dn} in the present context. At the projected mean-field level, a precision better than 200 keV was systematically obtained for the binding energy of $J=0$ even-even ground states with only two non-zero $\beta$ values over the interval $[0,\pi/2]$~\cite{Bender:2005ri}.
\end{enumerate}
\item When considering the reduced form of the formalism based on an axially symmetric reference state, the solving of the Hill-Wheeler-Griffin equation is no longer necessary (see App.~\ref{axialsymmetry}).
\end{itemize}

\section{Conclusions}

The present work addresses a long-term challenge of ab-initio many-body theory, i.e. it extends symmetry-unrestricted Rayleigh-Schroedinger many-body perturbation theory and coupled-cluster theory in such a way that the broken symmetry is {\it exactly} restored at any truncation order. The newly proposed symmetry-restored CC formalism authorizes the computation of connected diagrams while consistently incorporating static correlations through the non-perturbative restoration of the broken symmetry.  The approach is meant to be valid for any symmetry that can be (spontaneously) broken by the reference state and to be applicable to any system independently of its closed-shell, near degenerate or open-shell character.  In the present work we focus on the breaking and the restoration of $SU(2)$ rotational symmetry associated with angular momentum conservation. The scheme  accesses the complete yrast spectroscopy, i.e. the lowest energy for each angular momentum. Standard symmetry-restricted and symmetry-unrestricted MBPT and CC theories, along with angular-momentum-projected Hartree-Fock theory, are recovered as particular cases of the many-body formalism developed in the present work.

The goal being to resolve the near-degenerate nature of the ground state, the proposed extension is necessarily of multi-reference character. However, the multi-reference nature is different from any of the multi-reference coupled-cluster methods developed in quantum chemistry~\cite{bartlett07a}, i.e. reference states are not obtained from one another via particle-hole excitations but via highly non-perturbative symmetry transformations. Most importantly, the presently proposed method leads to solving a set of $N_{\text{sym}}$ single-reference-like coupled-cluster calculations, where $N_{\text{sym}}$ is typically of the order of $10$ per Euler angle parametrizing $SU(2)$. When considering an axially symmetric reference state, i.e. a Slater determinant that is an eigenstate of $J_z$ with the eigenvalue $M=0$, the scheme reduces to a single integration over the Euler angle $\beta$.

The symmetry-restored CC formalism offers a wealth of potential applications and further extensions appropriate to the {\it ab initio} description of open-shell atomic nuclei. Indeed, mid-mass open-shell nuclei are currently being addressed through symmetry-unrestricted CC methods. First, the recently proposed Bogoliubov CC theory that breaks global gauge symmetry associated with particle-number conservation permits the natural treatment of singly open-shell nuclei~\cite{signoracci13a}. Doubly open-shell nuclei can already be tackled by breaking rotational symmetry associated with angular-momentum conservation. Eventually, an even better description of doubly open-shell nuclei can be achieved by breaking both $U(1)$ and $SU(2)$ symmetries at the same time. Although such symmetry-unrestricted CC calculations efficiently access open-shell systems, the results are contaminated by the breaking of the symmetry. The presently proposed extension overcomes this limitation by restoring, in a consistent fashion, good angular momentum $(J,M)$. The next step is to implement the formalism in view of those applications. It will be of interest to see if such a route constitutes an efficient alternative to multi-reference or symmetry-adapted single-reference CC theories employed in quantum chemistry to address open-shell molecules and bond breaking. In the near future, the present work will be further extended to the $U(1)$ Lie group associated with particle number conservation. Following the same steps as for $SU(2)$, the necessary use of Bogoliubov algebra~\cite{signoracci13a} simply complicates the computation of the algebraic expressions.  Eventually, both symmetry restorations can be combined to best describe doubly open-shell nuclei.

Mid-mass singly open-shell nuclei have also been recently addressed through symmetry-unrestricted Green's function calculations under the form of self-consistent Gorkov Green's function theory~\cite{soma11a,Soma:2012zd,Barbieri:2012rd,Soma:2013vca}. It is thus of interest to develop the equivalent to the presently-proposed symmetry-restored CC formalism within the frame of self-consistent Green's function techniques.  

Last but not least, symmetry-restored MBPT and CC theories provide well-founded, formally exact, references for the so-far empirical multi-reference nuclear energy density functional (EDF) method. Multi-reference EDF calculations are known to be compromised with serious pathologies when the off-diagonal EDF kernel is not strictly computed as the matrix element of an {\it effective} Hamilton operator between a product state and its rotated partner~\cite{dobaczewski07,Lacroix:2008rj,Bender:2008rn,Duguet:2008rr,Duguet:2010cv,Duguet:2013dga}, i.e. when it does not take the strict form of an (effective) projected Hartree-Fock theory. Starting from the newly proposed many-body formalism, one could derive {\it safe} parametrizations of the off-diagonal EDF kernel that go beyond projected Hartree-Fock, most probably under the form of orbital- and symmetry-angle-dependent energy functionals. This remains to be investigated in the future.

\begin{acknowledgments}
The author wishes to thank G. Ripka very deeply for enlightening discussions that were instrumental to make this work possible and for the careful proofreading of the manuscript. The author wishes to thank R. J. Bartlett, M. Bender, H. Hergert, T. Lesinski, P. Piecuch, G. Scuseria, A. Signoracci and V. Som\`a for interesting comments as well as V. Som\`a for generating the set of diagrams used in the manuscript.
\end{acknowledgments}

\begin{appendix}

\section{Perturbative expansion of ${\cal U}(\tau)$}
\label{perturbativeannexe}

The imaginary-time evolution operator ${\cal U}(\tau)$ can be expanded in powers of $H_{1}$. Taking $\tau$ real, one writes
\begin{equation}
{\cal U}(\tau)  \equiv e^{-\tau H_{0}} \, {\cal U}_1(\tau) \label{evoloperatorApp1}
\end{equation}
where
\begin{eqnarray}
{\cal U}_1(\tau)  &=& e^{\tau H_{0}} \, e^{-\tau(H_{0}+H_{1})} \, , 
\end{eqnarray}
and thus
\begin{eqnarray}
\partial_{\tau} {\cal U}_1(\tau) &=&-e^{\tau H_{0}} \, H_{1} \, e^{-\tau H_{0}} \, {\cal U}_1(\tau) \, \, .
\end{eqnarray}
The formal solution to the latter equation reads
\begin{eqnarray}
{\cal U}_1(\tau)  &=& \textmd{T}e^{-\int_{0}^{\tau}dt H_{1}(t)} \, , \label{reducedevolutionop1} 
\end{eqnarray}
where $\textmd{T}$ is a time-ordering operator and where $H_{1}(\tau)$ defines the perturbation in the interaction representation
\begin{eqnarray}
H_{1}(\tau)  &\equiv& e^{\tau H_{0}} \, H_{1} \, e^{-\tau H_{0}} \, . \label{reducedevolutionop2}
\end{eqnarray}
Eventually, the full solution reads
\begin{equation}
{\cal U}(\tau)=e^{-\tau H_{0}} \, \textmd{T}e^{-\int_{0}^{\tau}d\tau H_{1}(\tau)
}  \, . \label{exp1}
\end{equation}

\section{Perturbative expansion of $N(\tau,\Omega)$}
\label{diagrams}

The contributions to $N(\tau,\Omega)$ can be represented by vacuum-to-vacuum Feynman diagrams. Thanks to the exponentiation property of Eq.~\ref{wexp3}, only connected diagrams $n(\tau,\Omega)$ need to be computed. The rules to compute the latter are given below and follow closely Chap.~5 of Ref.~\cite{blaizot86}. First- and second-order diagrams are then explicitly computed making use of the identities provided in App.~\ref{usefulID}. Last but not least, the characteristic dependence of the diagrams contributing to $n(\tau,\Omega)$ on $\tau$ and $\Omega$ is analyzed, in particular in the large $\tau$ limit.

\subsection{Labeled diagrams}

In the faithful representation, there is a one to one correspondence between a diagram and a system of contractions. Direct-product matrix elements of $V$ in the interaction representation are symbolized by the vertices
\begin{equation}%
v_{\alpha\beta\gamma\delta}  \,  \raisebox{-0.85cm}{\includegraphics[
height=0.7167in,
width=1.3in
]%
{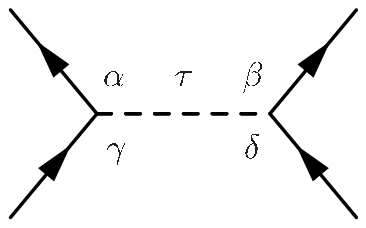}%
}%
\label{vijkltau}%
\end{equation}
while those of $-U$ are represented by
\begin{equation}%
 \raisebox{-0.85cm}{\includegraphics[
height=0.7167in,
width=1.2in
]%
{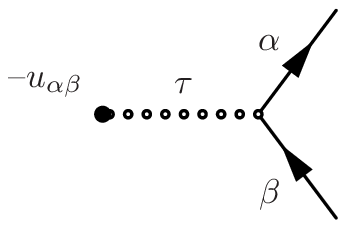}%
}%
\label{fijtau}%
\end{equation}
Off-diagonal contractions are represented by oriented lines denoting off-diagonal one-body propagators
\begin{equation}%
G^{0}_{\alpha\beta}(\tau_1,\tau_2 ; \Omega)   \,  \raisebox{-1.1cm}{\includegraphics[
height=1.0in,
width=0.3in
]%
{figures/MBPTdiag04_v2.eps}%
}%
\label{gitau12}%
\end{equation}
where the dependence on $\Omega$ is left implicit. Equation \ref{gitau12} associates the same graphical element independently of the actual time ordering, i.e. to two contractions that differ by a sign. Such a sign ambiguity can be lifted by adding an overall sign rule~\cite{blaizot86}. A one to one correspondence between a diagram and a system of contractions is achieved by multiplying the contribution of a diagram by $(-1)^{n_{c}}$, where $n_{c}$ is the number of closed loops formed by the fermion propagators in that diagram.


A mere displacement, i.e. a translation, of \emph{labeled} vertices yields a topologically identical diagram that corresponds to the {\it same} system of contractions. Two apparently different labeled diagrams that are topologically identical are just one diagram.

\subsection{Rules for labeled diagrams}
\label{sec:labdiag}

A diagram of order $n$ consists of closed loops formed by $2n$ oriented fermion propagators and $n$ interaction vertices attached to the fermion propagators. Given a diagram, assign a time label $\tau$ to each vertex as in Eqs.~\ref{vijkltau} and~\ref{fijtau} and assign single-particle indices to each propagator as in Eq.~\ref{gitau12}, along with time labels corresponding to the vertices its end points are attached to. Discard all topologically identical diagrams. The contribution of the diagram is obtained as follows
\begin{enumerate}
\item  Each vertex contributes a factor given by Eq.~\ref{vijkltau} or
~\ref{fijtau}.

\item  Each propagator contributes a factor~\ref{gitau12} whose actual expression is given by Eq.~\ref{decompoprop} or Eq.~\ref{decompopropsametime}.

\item  Sum over all single-particle labels and integrate all time variables from $0$ to $\tau$.

\item  Multiply the result by $(-1)^{n_{c}+n}/n!$, where $n=n_{v}+n_{u}$ denotes the number of $v$ and $u$ vertices, and where $n_{c}$ is the
number of closed loops formed by the fermion propagators.

\item  Multiply the result by $2^{-n_{v}}$.
\end{enumerate}

\subsection{Unlabeled diagrams and symmetry factors}

In a labeled diagram, each vertex bears a time variable $\tau$. However all the time variables are integrated over the same time interval, namely from $0$ to $\tau$. Therefore the time labels of a diagram can be exchanged without modifying the contribution of the diagram and $n!$ diagrams are generated by the exchange of time labels. They may all be represented by a single \emph{unlabeled diagram}, that is, by a diagram in which the vertices bear no time label. However, not all permutations of the time labels give in fact rise to topologically distinct labeled diagrams such that multiplying the result obtained for the one representative unlabeled diagram (which is eventually computed by re-assigning it some time labels) by $n!$ actually over counts the number of original topologically distinct labeled diagrams. Consequently, if $S$ is the number of permutations of the time labels that lead to a topologically identical diagram, the contribution of the representative unlabeled diagram calculated according to the rules stated in Sec.~\ref{sec:labdiag}, should eventually be multiplied by a factor $n!/S$.

Furthermore, matrix elements~\ref{vijkltau} display the left-right symmetry $v_{\alpha\beta\gamma\delta} = v_{\beta\alpha\delta\gamma}$ due to the invariance of the two-body interaction under the exchange of the two identical fermions. Diagrammatically, such a left-right transformation corresponds to permuting the extremities of the vertex. Consequently, diagrams that correspond to one-another by a given number of such permutations actually give the same results such that one shall only compute one representative diagram of this kind. Given a diagram containing $n_v$ vertices, one naively obtains $2^{n_v}$ different diagrams by permuting their extremities. However, it may happen that some of those permutations actually transform a diagram into a (topologically) identical diagram. This leads us to extend the definition of $S$ so as to include the number of permutations of both the time labels and the extremities of vertices~\ref{vijkltau} that transform a diagram into a topologically identical diagram. Then the contribution of an unlabeled representative diagram, calculated according to the rules stated in Sec.~\ref{sec:labdiag}, should be multiplied by $2^{n_{v}}n!/S$. This effectively modifies rules 4 and 5 of Sec.~\ref{sec:labdiag}. The factor $S$ is called the \emph{symmetry factor} of the diagram.

\subsection{Rules for unlabeled diagrams}
\label{sec:unlabdiag}

A diagram of order $n$ consists of closed loops formed by $2n$ oriented fermion propagators and $n$ interaction vertices attached to the fermion propagators. Choose a representative unlabeled diagram as defined above and assign a time label $\tau$ to each vertex as in Eqs.~\ref{vijkltau} and~\ref{fijtau}. Assign then single-particle indices to each propagator as in Eq.~\ref{gitau12} together with time labels corresponding to the vertices its end points are attached to. The contribution of the diagram is obtained as follows
\begin{enumerate}
\item Each vertex contributes a factor given either by Eq.~\ref{vijkltau} or~\ref{fijtau}.

\item  Each propagator contributes a factor~\ref{gitau12} whose actual expression is given by Eq.~\ref{decompoprop} or ~\ref{decompopropsametime}.

\item  Sum over all single-particle labels and integrate all time variables from $0$ to $\tau$.

\item  Multiply the result by $(-1)^{n_{c}+n}/{S}$, where $n=n_{v}+n_{u}$ is the number of vertices $v$ and $u$, where $n_{c}$ is the
number of closed loops formed by the fermion propagators, and where $S$ is the symmetry factor of the diagram. The symmetry factor is obtained by considering the group of both permutations of the time labels and of the extremities of the vertices (\ref{vijkltau}) which do not change the contribution of a diagram. The symmetry factor is then equal to the number of elements of the subgroup which transform the diagram into an (topologically) identical diagram.
\end{enumerate}

\subsection{Hugenholtz diagrams}

We eventually use antisymmetrized matrix elements such that a single "expanded" diagram has to be considered for each Hugenholtz diagram to determine its appropriate sign. Furthermore, we eventually rephrase the content of the symmetry factor by stating that it is obtained by considering the group of both permutations of the time labels and the factor $2^{n_{\text{eq}}}$, where $n_{\text{eq}}$ denotes the number of pairs of equivalent lines, i.e. pairs of lines which both leave from an initial vertex and end in the same final vertex.

\subsection{Diagrams contributing to $n\left(\tau,\Omega\right)$}
\label{combinatorial}

Three comments are in order prior to computing diagrams explicitly
\begin{enumerate}
\item  Any connected vacuum-to-vacuum diagram $n^{(n)}\left(\tau,\Omega\right)$ of order $n$ contains $2n$ propagator lines $G^{0}(\Omega)$. Splitting each propagator line into its two contributions $G^{0}$ and $G^{0,ph}(\Omega)$ (Eqs.~\ref{decompoprop}-\ref{twocontribprop}) generates $C^{p}_{2n}$ terms with $p\leq 2n$ propagator lines equal to $G^{0,ph}(\Omega)$. Overall, there are $\sum_{p=0}^{2n} C^{p}_{2n} = 2^{2n}$ contributions to the diagram. The $\Omega$-independent part $n^{(n)}\left(\tau\right)$ is equal to the contribution $(p=0)$ in which all propagator lines are equal to $G^{0}$. The $2^{2n}\!-\!1$ other terms make up $\aleph^{(n)}\left(\tau,\Omega\right)$.
\item One is eventually interested in the expressions of the diagrams in the limit $\tau\rightarrow\infty$. Those expressions are easily obtained from the formulae provided below that are valid for any finite value of $\tau \in [0,\infty[$. 
\item It may seem at first that various contributions to $\aleph(\tau,\Omega)$ contain dangerous denominators that can be zero for certain combinations of particle and hole indices. One can however check that each such contribution is in fact well behaved in the limit where the denominator in question goes to zero.
\end{enumerate}

\subsubsection{First order}

The first order diagram involving $V$ is the off-diagonal Hartree-Fock diagram. The corresponding diagram reads
\begin{equation}%
\raisebox{-0.8cm}{\includegraphics[
height=0.7in,
width=1.7in
]%
{figures/MBPTdiag05_v2.eps}%
}%
\label{hartree}%
\end{equation}
It has a symmetry factor $S=2$, contains two loops and one interaction vertex. Its expression is
\begin{eqnarray}
n^{(1)}_V(\tau,\Omega)&=&-\frac{1}{2} \sum_{\alpha\beta\gamma\delta}\int_{0}^{\tau}d\tau_1 \, \bar{v}_{\alpha\beta\gamma\delta} \, G^{0}_{\gamma\alpha}(\tau_1,\tau_1 ; \Omega) \, G^{0}_{\delta\beta}(\tau_1,\tau_1 ; \Omega) \nonumber \, .
\end{eqnarray}
Expressing each equal-time propagator according to Eqs.~\ref{decompopropsametime}-\ref{twocontribpropequaltime}, one eventually obtains
\begin{widetext}
\begin{subequations}
\begin{eqnarray}
n^{(1)}_V(\tau)&=& -\frac{\tau}{2} \sum_{ij} \bar{v}_{ijij} \label{gamhar2} \, , \\
\aleph^{(1)}_V(\tau,\Omega) &=&-\sum_{ija}\frac{\bar{v}_{ijaj}}{e_{a}-e_{i}} \, \rho^{ph}_{ai}(\Omega) \, \Big(1-e^{-\tau\left(
e_{a}-e_{i}\right)}\Big)  \label{hartree1omega} \\
&&-\frac{1}{2}\sum_{ijab}\frac{\bar{v}_{ijab}}{e_{a}+e_{b}-e_{i}-e_{j}} \,  \Big( 1-e^{-\tau\left(e_{a}+e_{b}-e_{i}-e_{j}\right)}\Big) \, \rho^{ph}_{ai}(\Omega) \, \rho^{ph}_{bj}(\Omega) \, . \nonumber
\end{eqnarray}
\end{subequations}
\end{widetext}

The first-order contribution deriving from the one-body potential $U$ is 
\begin{equation}%
\raisebox{-0.8cm}{\includegraphics[
height=0.7in,
width=1.7in
]%
{figures/MBPTdiag06_v2.eps}%
}%
\label{1storderU}%
\end{equation}
It has a symmetry factor $S=1$ and contains one loop along with one interaction vertex. Its contribution reads 
\begin{equation}
n^{(1)}_U(\tau,\Omega)= \sum_{\alpha\beta}\int_{0}^{\tau}d\tau_1 \, (-u_{\alpha\beta}) \, G^{0}_{\beta\alpha}(\tau_1,\tau_1 ; \Omega) \label{Ufirstorder} \, .
\end{equation}
One eventually obtains
\begin{subequations}
\begin{eqnarray}
n^{(1)}_U(\tau) &=& \tau \sum_{i} u_{ii} \label{Ufirstorder0} \, , \\
\aleph^{(1)}_U(\tau,\Omega) &=&  \sum_{ia} \frac{u_{ia}}{e_{a} -e_{i}} \, \Big(1-e^{-\tau (e_{a} -e_{i})}\Big) \, \rho^{ph}_{ai}(\Omega) \, . \label{Ufirstorder1}
\end{eqnarray}
\end{subequations}

\subsubsection{Second order}

The first second-order connected vacuum-to-vacuum diagram involving twice the two-body interaction is
\begin{equation}%
\raisebox{-0cm}{\includegraphics[
height=1.2in,
width=1.3in
]%
{figures/MBPTdiag07_v2.eps}%
}%
\label{2p2hunlab}%
\end{equation}
It contains two loops and two interaction vertices. Its symmetry factor is $S=8$ and its contribution reads
\begin{widetext}
\begin{eqnarray}
n^{(2)}_{V1}(\tau,\Omega) &=& \frac{1}{8}\sum_{\alpha\beta\gamma\delta}\sum_{\epsilon\zeta\eta\theta}\int_{0}^{\tau}d\tau_{1}%
d\tau_{2}\; \bar{v}_{\alpha\beta\gamma\delta} \, \bar{v}_{\epsilon\zeta\eta\theta} \, G^{0}_{\gamma\epsilon}(\tau_1,\tau_2 ; \Omega) \, G^{0}_{\eta\alpha}(\tau_2,\tau_1 ; \Omega) \, G^{0}_{\delta\zeta}(\tau_1,\tau_2 ; \Omega) \, G^{0}_{\theta\beta}(\tau_2,\tau_1 ; \Omega) \, . \label{total2ndorderdiag}
\end{eqnarray}
One eventually obtains
\begin{subequations}
\begin{eqnarray}
n^{(2)}_{V1}(\tau) &=& +\frac{1}{4}\sum_{ijab} \frac{\bar{v}_{ijab}\bar{v}_{abij}}{e_{a}+e_{b}-e_{i}-e_{j}} \, \left\{\tau+\frac
{e^{-\tau\left(e_{a}+e_{b}-e_{i}-e_{j}\right)  }-1}{e_{a}+e_{b}-e_{i}-e_{j}} \right\} \, , \\
\aleph^{(2)}_{V1}(\tau,\Omega) &=& + \frac{1}{4} \sum_{ijabc} \frac{\bar{v}_{ijab} \, \bar{v}_{abcj}}{e_{a}+e_{b}-e_{j}-e_{c}} \left\{ \frac{e^{-\tau(e_{c}-e_{i})}-1}{e_{i}-e_{c}} - \frac{e^{-\tau(e_{a}+e_{b}-e_{i}-e_{j})}-1}{e_{i}+e_{j}-e_{a}-e_{b}}\right\} \, \rho^{ph}_{ci}(\Omega) \\
&& - \frac{1}{4} \sum_{ijkab} \frac{\bar{v}_{iajk} \, \bar{v}_{jkba}}{e_{a}+e_{i}-e_{j}-e_{k}} \left\{ \frac{e^{-\tau(e_{b}-e_{i})}-1}{e_{i}-e_{b}} - \frac{e^{-\tau(e_{a}+e_{b}-e_{j}-e_{k})}-1}{e_{j}+e_{k}-e_{a}-e_{b}}\right\}\, \rho^{ph}_{bi}(\Omega)  \nonumber \\
&& - \frac{1}{4} \sum_{ijkab} \frac{\bar{v}_{ijab} \, \bar{v}_{kbij}}{e_{b}+e_{k}-e_{i}-e_{j}} \left\{ \frac{e^{-\tau(e_{a}-e_{k})}-1}{e_{k}-e_{a}} - \frac{e^{-\tau(e_{a}+e_{b}-e_{i}-e_{j})}-1}{e_{i}+e_{j}-e_{a}-e_{b}}\right\} \, \rho^{ph}_{ak}(\Omega) \nonumber \\
&& + \frac{1}{4} \sum_{ijabc} \frac{\bar{v}_{abci} \, \bar{v}_{jiab}}{e_{a}+e_{b}-e_{c}-e_{i}} \left\{ \frac{e^{-\tau(e_{c}-e_{j})}-1}{e_{j}-e_{c}} - \frac{e^{-\tau(e_{a}+e_{b}-e_{i}-e_{j})}-1}{e_{i}+e_{j}-e_{a}-e_{b}}\right\}\, \rho^{ph}_{cj}(\Omega) \nonumber \\
&& + \frac{1}{4} \sum_{ijklab} \frac{\bar{v}_{ijab} \, \bar{v}_{klij}}{e_{k}+e_{l}-e_{i}-e_{j}} \left\{ \frac{e^{-\tau(e_{a}+e_{b}-e_{k}-e_{l})}-1}{e_{k}+e_{l}-e_{a}-e_{b}} - \frac{e^{-\tau(e_{a}+e_{b}-e_{i}-e_{j})}-1}{e_{i}+e_{j}-e_{a}-e_{b}}\right\} \, \rho^{ph}_{ak}(\Omega) \, \rho^{ph}_{bl}(\Omega) \nonumber \\
&& + \frac{1}{4} \sum_{ijabcd} \frac{\bar{v}_{abcd} \, \bar{v}_{ijab}}{e_{a}+e_{b}-e_{c}-e_{d}} \left\{ \frac{e^{-\tau(e_{c}+e_{d}-e_{i}-e_{j})}-1}{e_{i}+e_{j}-e_{c}-e_{d}} - \frac{e^{-\tau(e_{a}+e_{b}-e_{i}-e_{j})}-1}{e_{i}+e_{j}-e_{a}-e_{b}}\right\} \, \rho^{ph}_{ci}(\Omega) \, \rho^{ph}_{dj}(\Omega)  \nonumber \\
&& + \sum_{ijkabc} \frac{\bar{v}_{ijab} \, \bar{v}_{kbcj}}{e_{b}+e_{k}-e_{j}-e_{c}} \left\{ \frac{e^{-\tau(e_{a}+e_{c}-e_{i}-e_{k})}-1}{e_{i}+e_{k}-e_{a}-e_{c}} - \frac{e^{-\tau(e_{a}+e_{b}-e_{i}-e_{j})}-1}{e_{i}+e_{j}-e_{a}-e_{b}}\right\} \, \rho^{ph}_{ak}(\Omega) \, \rho^{ph}_{ci}(\Omega)  \nonumber \\
&& - \frac{1}{2}\sum_{ijkabcd} \frac{\bar{v}_{ijab} \, \bar{v}_{akcd}}{e_{a}+e_{k}-e_{c}-e_{d}} \left\{ \frac{e^{-\tau(e_{b}+e_{c}+e_{d}-e_{i}-e_{j}-e_{k})}-1}{e_{i}+e_{j}+e_{k}-e_{b}-e_{c}-e_{d}} - \frac{e^{-\tau(e_{a}+e_{b}-e_{i}-e_{j})}-1}{e_{i}+e_{j}-e_{a}-e_{b}}\right\} \, \rho^{ph}_{ci}(\Omega) \, \rho^{ph}_{bk}(\Omega) \, \rho^{ph}_{dj}(\Omega) \nonumber \\
&& +\frac{1}{2}\sum_{ijklabc} \frac{\bar{v}_{ijla} \, \bar{v}_{lkbc}}{e_{i}+e_{j}-e_{l}-e_{a}} \left\{ \frac{e^{-\tau(e_{a}+e_{b}+e_{c}-e_{i}-e_{j}-e_{k})}-1}{e_{i}+e_{j}+e_{k}-e_{a}-e_{b}-e_{c}} - \frac{e^{-\tau(e_{b}+e_{c}-e_{l}-e_{k})}-1}{e_{l}+e_{k}-e_{b}-e_{c}}\right\} \, \rho^{ph}_{bi}(\Omega) \, \rho^{ph}_{ak}(\Omega) \, \rho^{ph}_{cj}(\Omega)   \nonumber \\
&& - \frac{1}{4} \sum_{ijklabcd} \frac{\bar{v}_{jlac} \, \bar{v}_{ikbd}}{e_j+e_l-e_a-e_c}  \left\{ \frac{e^{-\tau(e_a+e_b+e_c+e_d-e_i-e_j-e_k-e_l)}-1}{e_a+e_b+e_c+e_d-e_i-e_j-e_k-e_l} \right. \nonumber \\
&& \hspace{7.5cm} \left. - \frac{e^{-\tau(e_{b}+e_{d}-e_{i}-e_{k})}-1}{e_{b}+e_{d}-e_{i}-e_{k}}\right\} \, \rho^{ph}_{ai}(\Omega) \, \rho^{ph}_{bj}(\Omega) \, \rho^{ph}_{ck}(\Omega) \, \rho^{ph}_{dl}(\Omega)   \,\,\, . \nonumber
\end{eqnarray}
\end{subequations}
\end{widetext}

The other second-order connected vacuum-to-vacuum diagram involving twice the two-body interaction is
\begin{equation}%
\raisebox{-0cm}{\includegraphics[
height=1.2in,
width=1.3in
]%
{figures/MBPTdiag08_v2.eps}%
}%
\label{2p2hunlab2}%
\end{equation}
It contains three loops and two interaction vertices. Its symmetry factor is $S=2$ and its contribution reads
\begin{widetext}
\begin{eqnarray}
n^{(2)}_{V2}(\tau,\Omega) &=& -\frac{1}{2}\sum_{\alpha\beta\gamma\delta}\sum_{\epsilon\zeta\eta\theta}\int_{0}^{\tau}d\tau_{1}%
d\tau_{2}\; \bar{v}_{\alpha\beta\gamma\delta} \, \bar{v}_{\epsilon\zeta\eta\theta} \, G^{0}_{\eta\epsilon}(\tau_2,\tau_2 ; \Omega) \, G^{0}_{\gamma\alpha}(\tau_1,\tau_1 ; \Omega) \, G^{0}_{\delta\zeta}(\tau_1,\tau_2 ; \Omega) \, G^{0}_{\theta\beta}(\tau_2,\tau_1 ; \Omega) \, , 
\end{eqnarray}
which eventually gives
\begin{subequations}
\begin{eqnarray}
n^{(2)}_{V2}(\tau) &=& +\sum_{ijka} \frac{\bar{v}_{iaik} \, \bar{v}_{jkja}}{e_a-e_k} \left\{ \tau + \frac{e^{-\tau(e_a-e_k)}-1}{e_a-e_k}  \right\} \, , \\
\aleph^{(2)}_{V2}(\tau,\Omega) &=& + \sum_{ijkla} \frac{\bar{v}_{ikil} \, \bar{v}_{jlja}}{e_k-e_l} \left\{\frac{e^{-\tau(e_a-e_k)}-1}{e_a-e_k} -\frac{e^{-\tau(e_a-e_l)}-1}{e_a-e_l}  \right\} \, \rho^{ph}_{ak}(\Omega)  \\
&&  - \sum_{ijkab} \frac{\bar{v}_{ibia} \, \bar{v}_{jkjb}}{e_b-e_a} \left\{\frac{e^{-\tau(e_a-e_k)}-1}{e_a-e_k} -\frac{e^{-\tau(e_b-e_k)}-1}{e_b-e_k}  \right\} \, \rho^{ph}_{ak}(\Omega)  \nonumber \\
&&  - \sum_{ijkab} \frac{\bar{v}_{ibik} \, \bar{v}_{jkab}}{e_b-e_k} \left\{\frac{e^{-\tau(e_a-e_j)}-1}{e_a-e_j} -\frac{e^{-\tau(e_a+e_b-e_j-e_k)}-1}{e_a+e_b-e_j-e_k}  \right\} \, \rho^{ph}_{aj}(\Omega)  \nonumber \\
&&  - \sum_{ijkab} \frac{\bar{v}_{ibak} \, \bar{v}_{jkjb}}{e_i+e_b-e_a-e_k} \left\{\frac{e^{-\tau(e_a-e_i)}-1}{e_a-e_i} -\frac{e^{-\tau(e_b-e_k)}-1}{e_b-e_k}  \right\} \, \rho^{ph}_{ai}(\Omega)  \nonumber \\
&&  + \sum_{ijklab} \frac{\bar{v}_{ilia} \, \bar{v}_{jkjb}}{e_l-e_a} \left\{\frac{e^{-\tau(e_a+e_b-e_k-e_l)}-1}{e_a+e_b-e_k-e_l} -\frac{e^{-\tau(e_b-e_k)}-1}{e_b-e_k}  \right\} \, \rho^{ph}_{ak}(\Omega) \, \rho^{ph}_{bl}(\Omega)  \nonumber \\
&&  + \sum_{ijklab} \frac{\bar{v}_{ilik} \, \bar{v}_{jkab}}{e_l-e_k} \left\{\frac{e^{-\tau(e_a+e_b-e_j-e_l)}-1}{e_a+e_b-e_j-e_l} -\frac{e^{-\tau(e_a+e_b-e_j-e_k)}-1}{e_a+e_b-e_j-e_k}  \right\} \, \rho^{ph}_{aj}(\Omega) \, \rho^{ph}_{bl}(\Omega)  \nonumber \\
&&  - \sum_{ijkabc} \frac{\bar{v}_{icib} \, \bar{v}_{jkac}}{e_c-e_b} \left\{\frac{e^{-\tau(e_a+e_b-e_j-e_k)}-1}{e_a+e_b-e_j-e_k} -\frac{e^{-\tau(e_a+e_c-e_j-e_k)}-1}{e_a+e_c-e_j-e_k}  \right\} \, \rho^{ph}_{aj}(\Omega) \, \rho^{ph}_{bk}(\Omega)  \nonumber \\
&&  + \sum_{ijklab} \frac{\bar{v}_{ilak} \, \bar{v}_{jkjb}}{e_i+e_l-e_a-e_k} \left\{\frac{e^{-\tau(e_a+e_b-e_i-e_l)}-1}{e_a+e_b-e_i-e_l} -\frac{e^{-\tau(e_b-e_k)}-1}{e_b-e_k}  \right\} \, \rho^{ph}_{ai}(\Omega) \, \rho^{ph}_{bl}(\Omega)  \nonumber \\
&&  - \sum_{ijkabc} \frac{\bar{v}_{icab} \, \bar{v}_{jkjc}}{e_i+e_c-e_a-e_b} \left\{\frac{e^{-\tau(e_a+e_b-e_i-e_k)}-1}{e_a+e_b-e_i-e_k} -\frac{e^{-\tau(e_c-e_k)}-1}{e_c-e_k}  \right\} \, \rho^{ph}_{ai}(\Omega) \, \rho^{ph}_{bk}(\Omega)  \nonumber \\
&&  - \sum_{ijkabc} \frac{\bar{v}_{icak} \, \bar{v}_{jkbc}}{e_i+e_c-e_a-e_k} \left\{\frac{e^{-\tau(e_a+e_b-e_i-e_j)}-1}{e_a+e_b-e_i-e_j} -\frac{e^{-\tau(e_b+e_c-e_j-e_k)}-1}{e_b+e_c-e_j-e_k}  \right\} \, \rho^{ph}_{ai}(\Omega) \, \rho^{ph}_{bj}(\Omega)  \nonumber \\
&&  + \sum_{ijklabc} \frac{\bar{v}_{ilib} \, \bar{v}_{jkac}}{e_l-e_b} \left\{\frac{e^{-\tau(e_a+e_b+e_c-e_j-e_k-e_l)}-1}{e_a+e_b+e_c-e_j-e_k-e_l} -\frac{e^{-\tau(e_a+e_c-e_j-e_k)}-1}{e_a+e_c-e_j-e_k}  \right\} \, \rho^{ph}_{aj}(\Omega) \, \rho^{ph}_{bk}(\Omega) \, \rho^{ph}_{cl}(\Omega)  \nonumber \\
&&  + \sum_{ijklabc} \frac{\bar{v}_{ilab} \, \bar{v}_{jkjc}}{e_i+e_l-e_a-e_b} \left\{\frac{e^{-\tau(e_a+e_b+e_c-e_i-e_k-e_l)}-1}{e_a+e_b+e_c-e_i-e_k-e_l} -\frac{e^{-\tau(e_c-e_k)}-1}{e_c-e_k}  \right\} \, \rho^{ph}_{ai}(\Omega) \, \rho^{ph}_{bk}(\Omega) \, \rho^{ph}_{cl}(\Omega)  \nonumber \\
&&  + \sum_{ijklabc} \frac{\bar{v}_{ilak} \, \bar{v}_{jkbc}}{e_i+e_l-e_a-e_k} \left\{\frac{e^{-\tau(e_a+e_b+e_c-e_i-e_j-e_l)}-1}{e_a+e_b+e_c-e_i-e_j-e_l} -\frac{e^{-\tau(e_b+e_c-e_j-e_k)}-1}{e_b+e_c-e_j-e_k}  \right\} \, \rho^{ph}_{ai}(\Omega) \, \rho^{ph}_{bj}(\Omega) \, \rho^{ph}_{cl}(\Omega)  \nonumber \\
&&  - \sum_{ijkabcd} \frac{\bar{v}_{idac} \, \bar{v}_{jkbd}}{e_i+e_d-e_a-e_c} \left\{\frac{e^{-\tau(e_a+e_b+e_c-e_i-e_j-e_k)}-1}{e_a+e_b+e_c-e_i-e_j-e_k} -\frac{e^{-\tau(e_b+e_d-e_j-e_k)}-1}{e_b+e_d-e_j-e_k}  \right\} \, \rho^{ph}_{ai}(\Omega) \, \rho^{ph}_{bj}(\Omega) \, \rho^{ph}_{ck}(\Omega)  \nonumber \\
&&  + \sum_{ijklabcd} \frac{\bar{v}_{ilac} \, \bar{v}_{jkbd}}{e_i+e_l-e_a-e_c} \left\{\frac{e^{-\tau(e_a+e_b+e_c+e_d-e_i-e_j-e_k-e_l)}-1}{e_a+e_b+e_c+e_d-e_i-e_j-e_k-e_l} \right. \nonumber \\
&& \hspace{7.5cm} \left.  -\frac{e^{-\tau(e_b+e_d-e_j-e_k)}-1}{e_b+e_d-e_j-e_k}  \right\} \, \rho^{ph}_{ai}(\Omega) \, \rho^{ph}_{bj}(\Omega) \, \rho^{ph}_{ck}(\Omega) \, \rho^{ph}_{dl}(\Omega)  \nonumber \, \, .
\end{eqnarray}
\end{subequations}
\end{widetext}

The second-order connected vacuum-to-vacuum diagram involving the two-body interaction and the one-body potential is
\begin{equation}%
\raisebox{-0cm}{\includegraphics[
height=1.2in,
width=1.3in
]%
{figures/MBPTdiag09_v2.eps}%
}%
\label{2p2hunlab3}%
\end{equation}
It contains two loops and two interaction vertices. Its symmetry factor is $S=1$ and its contribution reads
\begin{widetext}
\begin{eqnarray}
n^{(2)}_{UV}(\tau,\Omega) &=& +\sum_{\alpha\beta\gamma\delta}\sum_{\epsilon\zeta}\int_{0}^{\tau}d\tau_{1}%
d\tau_{2}\; \bar{v}_{\alpha\beta\gamma\delta} \, (-u_{\epsilon\zeta}) \,  G^{0}_{\gamma\alpha}(\tau_1,\tau_1 ; \Omega) \,  G^{0}_{\delta\epsilon}(\tau_1,\tau_2 ; \Omega) \, G^{0}_{\zeta\beta}(\tau_2,\tau_1 ; \Omega) \, , 
\end{eqnarray}
which eventually gives
\begin{subequations}
\begin{eqnarray}
n^{(2)}_{UV}(\tau) &=& -\sum_{ija} \frac{u_{ja} \, \bar{v}_{iaij}}{e_a-e_j} \left\{ \tau + \frac{e^{-\tau(e_a-e_j)}-1}{e_a-e_j}  \right\} -\sum_{ija} \frac{u_{aj} \, \bar{v}_{ijia}}{e_a-e_j} \left\{ \tau + \frac{e^{-\tau(e_a-e_j)}-1}{e_a-e_j}  \right\} \, , \\
\aleph^{(2)}_{UV}(\tau,\Omega) &=& - \sum_{ijka} \frac{u_{ja} \, \bar{v}_{ikij}}{e_k-e_j} \left\{\frac{e^{-\tau(e_a-e_k)}-1}{e_a-e_k} -\frac{e^{-\tau(e_a-e_j)}-1}{e_a-e_j}  \right\} \, \rho^{ph}_{ak}(\Omega)  \\
&&  - \sum_{ijka} \frac{u_{jk} \, \bar{v}_{ikia}}{e_j-e_k} \left\{\frac{e^{-\tau(e_a-e_j)}-1}{e_a-e_j} -\frac{e^{-\tau(e_a-e_k)}-1}{e_a-e_k}  \right\} \, \rho^{ph}_{aj}(\Omega)   \nonumber \\
&&  + \sum_{ijab} \frac{u_{jb} \, \bar{v}_{ibia}}{e_b-e_a} \left\{\frac{e^{-\tau(e_a-e_j)}-1}{e_a-e_j} -\frac{e^{-\tau(e_b-e_j)}-1}{e_b-e_j}  \right\} \, \rho^{ph}_{aj}(\Omega)   \nonumber \\
&&  + \sum_{ijab} \frac{u_{ab} \, \bar{v}_{ijia}}{e_a-e_b} \left\{\frac{e^{-\tau(e_b-e_j)}-1}{e_b-e_j} -\frac{e^{-\tau(e_a-e_j)}-1}{e_a-e_j}  \right\} \, \rho^{ph}_{bj}(\Omega)   \nonumber \\
&&  + \sum_{ijab} \frac{u_{bj} \, \bar{v}_{ijab}}{e_b-e_j} \left\{\frac{e^{-\tau(e_a-e_i)}-1}{e_a-e_i} -\frac{e^{-\tau(e_a+e_b-e_i-e_j)}-1}{e_a+e_b-e_i-e_j}  \right\} \, \rho^{ph}_{ai}(\Omega)   \nonumber \\
&&  + \sum_{ijab} \frac{u_{jb} \, \bar{v}_{ibaj}}{e_i+e_b-e_a-e_j} \left\{\frac{e^{-\tau(e_a-e_i)}-1}{e_a-e_i} -\frac{e^{-\tau(e_b-e_j)}-1}{e_b-e_j}  \right\} \, \rho^{ph}_{ai}(\Omega)   \nonumber \\
&&  + \sum_{ijkab} \frac{u_{jb} \, \bar{v}_{ikia}}{e_a-e_k} \left\{\frac{e^{-\tau(e_a+e_b-e_j-e_k)}-1}{e_a+e_b-e_j-e_k} -\frac{e^{-\tau(e_b-e_j)}-1}{e_b-e_j}  \right\} \, \rho^{ph}_{aj}(\Omega) \, \rho^{ph}_{bk}(\Omega)   \nonumber \\
&&  - \sum_{ijkab} \frac{u_{jk} \, \bar{v}_{ikab}}{e_j-e_k} \left\{\frac{e^{-\tau(e_a+e_b-e_i-e_j)}-1}{e_a+e_b-e_i-e_j} -\frac{e^{-\tau(e_a+e_b-e_i-e_k)}-1}{e_a+e_b-e_i-e_k}  \right\} \, \rho^{ph}_{ai}(\Omega) \, \rho^{ph}_{bj}(\Omega)   \nonumber \\
&&  - \sum_{ijkab} \frac{u_{jb} \, \bar{v}_{ikia}}{e_b-e_j} \left\{\frac{e^{-\tau(e_a-e_k)}-1}{e_a-e_k} -\frac{e^{-\tau(e_a+e_b-e_j-e_k)}-1}{e_a+e_b-e_j-e_k}  \right\} \, \rho^{ph}_{aj}(\Omega) \, \rho^{ph}_{bk}(\Omega)   \nonumber \\
&&  + \sum_{ijabc} \frac{u_{bc} \, \bar{v}_{ijab}}{e_b-e_c} \left\{\frac{e^{-\tau(e_a+e_c-e_i-e_j)}-1}{e_a+e_c-e_i-e_j} -\frac{e^{-\tau(e_a+e_b-e_i-e_j)}-1}{e_a+e_b-e_i-e_j}  \right\} \, \rho^{ph}_{ai}(\Omega) \, \rho^{ph}_{cj}(\Omega)   \nonumber \\
&&  + \sum_{ijkab} \frac{u_{jb} \, \bar{v}_{ikaj}}{e_i+e_k-e_a-e_j} \left\{\frac{e^{-\tau(e_b-e_j)}-1}{e_b-e_j} -\frac{e^{-\tau(e_a+e_b-e_i-e_k)}-1}{e_a+e_b-e_i-e_k}  \right\} \, \rho^{ph}_{ai}(\Omega) \, \rho^{ph}_{bk}(\Omega)   \nonumber \\
&&  - \sum_{ijabc} \frac{u_{jc} \, \bar{v}_{icab}}{e_i+e_c-e_a-e_b} \left\{\frac{e^{-\tau(e_c-e_j)}-1}{e_c-e_j} -\frac{e^{-\tau(e_a+e_b-e_i-e_j)}-1}{e_a+e_b-e_i-e_j}  \right\} \, \rho^{ph}_{ai}(\Omega) \, \rho^{ph}_{bj}(\Omega)   \nonumber \\
&&  - \sum_{ijkabc} \frac{u_{jc} \, \bar{v}_{ikab}}{e_a+e_b-e_i-e_k} \left\{\frac{e^{-\tau(e_c-e_j)}-1}{e_c-e_j} -\frac{e^{-\tau(e_a+e_b+e_c-e_i-e_j-e_k)}-1}{e_a+e_b+e_c-e_i-e_j-e_k}  \right\} \, \rho^{ph}_{ai}(\Omega) \, \rho^{ph}_{bj}(\Omega) \, \rho^{ph}_{ck}(\Omega)   \nonumber \\
&&  - \sum_{ijkabc} \frac{u_{jc} \, \bar{v}_{ikab}}{e_c-e_j} \left\{\frac{e^{-\tau(e_a+e_b-e_i-e_k)}-1}{e_a+e_b-e_i-e_k} -\frac{e^{-\tau(e_a+e_b+e_c-e_i-e_j-e_k)}-1}{e_a+e_b+e_c-e_i-e_j-e_k}  \right\} \, \rho^{ph}_{ai}(\Omega) \, \rho^{ph}_{bj}(\Omega) \, \rho^{ph}_{ck}(\Omega)   \nonumber \, .
\end{eqnarray}
\end{subequations}
\end{widetext}

The second-order connected vacuum-to-vacuum diagram involving twice the one-body potential is
\begin{equation}%
\raisebox{-0cm}{\includegraphics[
height=1.2in,
width=1.3in
]%
{figures/MBPTdiag10_v2.eps}%
}%
\label{2p2hunlab4}%
\end{equation}
It contains one loop and two interaction vertices. Its symmetry factor is $S=2$ and its contribution reads
\begin{widetext}
\begin{eqnarray}
n^{(2)}_{UU}(\tau,\Omega) &=& -\frac{1}{2}\sum_{\alpha\beta}\sum_{\gamma\delta}\int_{0}^{\tau}d\tau_{1}%
d\tau_{2}\; (-u_{\alpha\beta}) \, (-u_{\gamma\delta}) \,  G^{0}_{\delta\alpha}(\tau_2,\tau_1 ; \Omega) \, G^{0}_{\beta\gamma}(\tau_1,\tau_2 ; \Omega) \, , 
\end{eqnarray}
which eventually gives
\begin{subequations}
\begin{eqnarray}
n^{(2)}_{UU}(\tau) &=& +\sum_{ia} \frac{u_{ia} \, u_{ai}}{e_a-e_i} \left\{ \tau + \frac{e^{-\tau(e_a-e_i)}-1}{e_a-e_i}  \right\} \, , \\
\aleph^{(2)}_{UU}(\tau,\Omega) &=& - \sum_{iab} \frac{u_{ia} \, u_{ab}}{e_a-e_b} \left\{\frac{e^{-\tau(e_b-e_i)}-1}{e_b-e_i} -\frac{e^{-\tau(e_a-e_i)}-1}{e_a-e_i}  \right\} \, \rho^{ph}_{bi}(\Omega)  \\
&&  + \sum_{ija} \frac{u_{ja} \, u_{ij}}{e_i-e_j} \left\{\frac{e^{-\tau(e_a-e_i)}-1}{e_a-e_i} -\frac{e^{-\tau(e_a-e_j)}-1}{e_a-e_j}  \right\} \, \rho^{ph}_{ai}(\Omega)   \nonumber \\
&&  - \sum_{ijab} \frac{u_{ja} \, u_{ib}}{e_b-e_i} \left\{\frac{e^{-\tau(e_a-e_j)}-1}{e_a-e_j} -\frac{e^{-\tau(e_a+e_b-e_i-e_j)}-1}{e_a+e_b-e_i-e_j}  \right\} \, \rho^{ph}_{ai}(\Omega)  \, \rho^{ph}_{bj}(\Omega)   \nonumber \, .
\end{eqnarray}
\end{subequations}
\end{widetext}

\subsection{Structure of $n(\tau,\Omega)$ in $\tau$ and $\Omega$}
\label{structure2}

We now provide a proof of the characteristic dependence of $n(\tau,\Omega)$ on $\tau$ and $\Omega$ (eventually in the limit of large $\tau$). As such, we demonstrate that the $\Omega$-{\it independent} part $n(\tau)= n(\tau,0)$ and the $\Omega$-{\it dependent} part $\aleph\left(\tau,\Omega\right)=n(\tau,\Omega)-n(\tau,0)$ display different dependencies on $\tau$.  

As the proof relies on an analysis of the generic topology of the diagrams contributing to $n(\tau,\Omega)$, let us start from the perturbative expansion
\begin{widetext}
\begin{eqnarray}
n(\tau,\Omega) &=& \langle \Phi | \textmd{T}e^{-\int_{0}^{\tau}dt H_{1}\left(t\right)} | \Phi(\Omega) \rangle_{c} \nonumber  \\
&=& \langle \Phi | 1- \int_{0}^{\tau
}d\tau_1 H_{1}\left(  \tau_1\right)+\frac{1}{2!}\int_{0}^{\tau}d\tau_{2}d\tau
_{1}\textmd{T}\left[H_{1}\left(  \tau_{2}\right)  H_{1}\left(  \tau_{1}\right)  \right]+... |  \Phi(\Omega) \rangle_{c} \nonumber\\
&=& \Big\{  \sum_{n=0}^{\infty}\frac{\left(  -\right)^{n}}{n!}\int_{0}^{\tau}d\tau_{n}\ldots d\tau_{1} \frac{1}{2}\!\!\!\!\sum_{\alpha_n\beta_n\gamma_n\delta_n}v_{\alpha_n\beta_n\gamma_n\delta_n} \ldots \frac{1}{2}\!\!\!\!\sum_{\alpha_1\beta_1\gamma_1\delta_1}v_{\alpha_1\beta_1\gamma_1\delta_1}  \nonumber\\
&& \hspace{2.5cm} \times \langle \Phi |  \textmd{T}\left[a_{\alpha_n}^{\dagger}\left(  \tau_{n}\right)
a_{\beta_n}^{\dagger}\left(  \tau_{n}\right)  
a_{\delta_n}\left(  \tau_{n}\right)
a_{\gamma_n}\left(  \tau_{n}\right)
 \ldots
a_{\alpha_1}^{\dagger}\left(  \tau_{1}\right)
a_{\beta_1}^{\dagger}\left(  \tau_{1}\right)  
a_{\delta_1}\left(  \tau_{1}\right)
a_{\gamma_1}\left(  \tau_{1}\right) 
\right]  |  \Phi(\Omega) \rangle_{c} \nonumber \\
&& + \sum_{n=0}^{\infty}\frac{\left(  -\right)^{n}}{n!}\int_{0}^{\tau}d\tau_{n}\ldots d\tau_{1}\sum_{\alpha_n\beta_n}(-u_{\alpha_n\beta_n}) \ldots \sum_{\alpha_1\beta_1}(-u_{\alpha_1\beta_1})  \nonumber \\
&& \hspace{6.9cm} \times \langle \Phi |  \textmd{T}\left[
a_{\alpha_n}^{\dagger}\left(  \tau_{n}\right)
a_{\beta_n}\left(  \tau_{n}\right)\ldots
a_{\alpha_1}^{\dagger}\left(  \tau_{1}\right)
a_{\beta_1}\left(  \tau_{1}\right)   
\right]  |  \Phi(\Omega) \rangle_{c} \nonumber\\
&& + \, \text{all cross terms involving both} \, \, V(\tau_p) \, \, \text{and} \, \, -\!U(\tau_p) \Big\} \,, \label{expansionenergykernelbeta}
\end{eqnarray}
\end{widetext}
where one notices the connected character of the considered strings of contractions.

A term of order $n$ in Eq.~\ref{expansionenergykernelbeta} contains $k$ operators $-U(\tau_p)$ and $n-k$ operators $V(\tau_p)$ whose time labels are integrated over. For simplicity, and without any lack of generality, we focus on the case with $k=n$. The proof can be extended to $k=0,\ldots,n-1$ without any difficulty. This contribution to $n^{(n)}(\tau,\Omega)$ is a sum (over single-particle indices and all possible time orderings) of terms proportional to
\begin{widetext}
\begin{eqnarray}
\int_{0}^{\tau}\!\!d\tau_{n}\!\!\int_{0}^{\tau_n}\!\!d\tau_{n-1}\ldots\!\! \int_{0}^{\tau_{2}}\!\!d\tau_{1} \, e^{\tau_n(e_{\alpha_n}-e_{\beta_n})}  \ldots e^{\tau_1(e_{\alpha_1}-e_{\beta_1})}\, \langle \Phi |a_{\alpha_n}^{\dagger} a_{\beta_n}  \ldots a_{\alpha_1}^{\dagger}
a_{\beta_1} |  \Phi(\Omega) \rangle_{c} \, , \label{typicaldiagram1}
\end{eqnarray}
\end{widetext}
where the time dependencies have been extracted from the creation and annihilation operators (see Eq.~\ref{aalphatau}). The particular time ordering considered in Eq.~\ref{typicaldiagram1} does not limit the generality of the analysis given that any contribution to $n^{(n)}(\tau,\Omega)$ can be written in this form. For compactness, Eq.~\ref{typicaldiagram1} can be rewritten as
\begin{widetext}
\begin{eqnarray}
\int_{0}^{\tau}\!\!d\tau_{n}\, A_n(\tau_n; b_1,\ldots b_n) \, \langle \Phi |  a_{\alpha_n}^{\dagger} a_{\beta_n}  \ldots a_{\alpha_1}^{\dagger}
a_{\beta_1} |  \Phi(\Omega) \rangle_{c}\, , \label{typicaldiagram2}
\end{eqnarray}
where $b_p\equiv e_{\alpha_p}-e_{\beta_p}$ captures the coefficient in front of $\tau_p$ in the original time integral. At first order, we have $A_1(\tau_1; b_1) = e^{\tau_1 b_1}$. For $n \geq 2$, one finds the recurrence relation
\begin{eqnarray}
A_{n}(\tau_n; b_1,\ldots, b_n) &=& e^{\tau_n b_n} \int_{0}^{\tau_n}\!\!d\tau_{n\!-\!1}  \, A_{n-1}(\tau_{n\!-\!1}; b_1,\ldots b_{n-1}) \, ,
\end{eqnarray}
which can be solved using identities provided in App.~\ref{usefulID}. Except for the first-order case, the connected character of the diagram implies that $b_p \neq 0$ as the creation and annihilation operators stemming from an operator $U(\tau_p)$ cannot be contracted together in the matrix element appearing in Eq.~\ref{typicaldiagram1}. This eventually leads to the general structure
\begin{eqnarray}
A_n(\tau_n; b_1,\ldots b_n) &\equiv&  c^{(n)}_n \, e^{\tau_n b_{n}} + c^{(n)}_{n-1} \, e^{\tau_n (b_{n}+b_{n-1})} +\ldots + c^{(n)}_{1} \, e^{\tau_n (b_{n}+b_{n-1}+\ldots +b_2+b_1)} \, , \label{timeintegral}
\end{eqnarray}
\end{widetext}
where each $c^{(n)}_p\,(p = 1,\ldots,n)$ is a function of the $b_p\,(p=1,\ldots,n-1)$.  Eventually, the dependence of $c^{(n)}_p$ and $d_p\equiv \sum_{k=p}^{n} b_k (p=1,\ldots,n)$ on single-particle energies $\{e_\alpha\}$ depend on the particular (i.e. differ for each) set of contractions arising from the matrix element in Eq.~\ref{typicaldiagram2}. To attain the desired result, we must thus consider two separate cases. 

We first focus on the $\Omega$-{\it independent} contribution, i.e. on the result obtained by setting $\Omega=0$ in Eq.~\ref{typicaldiagram1}. In this  case, all the propagators (i.e. contractions emerging from the matrix element in Eq.~\ref{typicaldiagram1}) are {\it diagonal} according to Eq.~\ref{truc}. As a result, the index of each creation operator is equated with the index of one annihilation operator, which leads to the remarkable result
\begin{equation}
b_{n}+b_{n-1}+\ldots +b_2+b_1 = 0 \, , \label{remarkableidentity}
\end{equation}
whereas all the other incomplete combinations of the $b_p$ appearing in Eq.~\ref{timeintegral} differ from $0$. After integrating over $\tau_n$ in Eq.~\ref{typicaldiagram2}, one finally obtains the characteristic structure
\begin{eqnarray}
\tau c_{1} + \sum_{p=2}^{n} \frac{c^{(n)}_p}{d_p} \,\left(e^{\tau d_{p}}-1\right)  \, , \nonumber
\end{eqnarray}
where the $d_p$ with $p=2,\ldots,n$ are all strictly negative. This leads to the conclusion that $n(\tau,0)$ is made of a term proportional to $\tau$ and a series of terms that are either independent of $\tau$ or decreasing exponentially with it. Taking the large $\tau$ limit, one eventually recovers property~\ref{betalim1}, first proven in Refs.~\cite{hugenholtz57a,bloch58a}, that characterizes the dependence of $n(\tau)$ in the large $\tau$ limit.

We are now in position to study off-diagonal matrix elements obtained for $\Omega\neq 0$. Contributions to $\aleph\left(\tau,\Omega\right)$ are obtained as soon as (at least) one contraction originating from the matrix element in Eq.~\ref{typicaldiagram1} is proportional to $\rho^{ph}(\Omega)$ (second term of Eq.~\ref{truc}). This means that at least one propagator is {\it not} diagonal and connects one particle index with one hole index. In turn, property~\ref{remarkableidentity} is not fulfilled any more and no exponent in Eq.~\ref{timeintegral} is equal to $0$. One thus obtains the characteristic structure of the term making up $\aleph\left(\tau,\Omega\right)$ under the form
\begin{eqnarray}
\sum_{p=1}^{n} \frac{c^{(n)}_p}{d_p} \,\left(e^{\tau d_{p}}-1\right) \, G^{(n)}_p[\rho^{ph}(\Omega)]  \,  , \label{finalstructure2}
\end{eqnarray}
where {\it all} the $d_p (p=1,\ldots,n)$ including $d_{1}$ are strictly negative and where $G^{(n)}_p[\rho^{ph}(\Omega)]$ is a polynomial in $\rho^{ph}(\Omega)$ whose constant term is zero and that may contain terms with up to power $n$. This leads to the conclusion that $\aleph\left(\tau,\Omega\right)$ is made of a sum of terms that are either independent of $\tau$ or decreasing exponentially with it. Contrarily to $n(\tau)$ that grows linearly when $\tau$ becomes infinitely large, $\aleph\left(\tau,\Omega\right)$ goes exponentially to a finite limit $\aleph\left(\Omega\right)$. This demonstrates the original result stated in Eq.~\ref{betalim2}.

\section{Perturbative expansion of $h(\tau,\Omega)$}
\label{diagramsE}

We compute zero- and first-order diagrams contributing to the off-diagonal linked/connected energy kernel $h(\tau,\Omega)$. After mentioning the slight change of diagrammatic rules due to the presence of the fixed-time operator at time $\tau=0$, we provide the expressions of zero- and first-order diagrams. 

\subsection{Diagrammatic rules}

One needs to slightly adapt the diagrammatic rules stated in Sec.~\ref{sec:unlabdiag} due to the presence of the fixed-time operator at time $\tau=0$. By convention, the order $n$ of a diagram is defined such that the fixed-time operator at $\tau=0$ is not taken into account. Since one does not integrate over the time label associated with the fixed-time operator at $\tau=0$, one must not consider this operator in the determination of the topologically identical diagrams obtained under the exchange of time labels when identifying the symmetry factor of the diagram. 

\subsection{Zero-order kinetic energy}

The zero-order labeled diagram is
\begin{equation}%
\raisebox{-0.8cm}{\includegraphics[
height=0.7in,
width=1.5in
]%
{figures/MBPTdiag12_v2.eps}%
}%
\label{hartreeT}%
\end{equation}
It contains one loop and one interaction vertex. It has a symmetry factor $S=1$ and reads
\begin{subequations}
\begin{eqnarray}
t^{(0)}(\tau,\Omega) &=&  -\sum_{\alpha\beta}  t_{\alpha\beta} \, G^{0}_{\beta\alpha}(0,0 ; \Omega)   \, \\
 &=& + \sum_{\alpha\beta}  t_{\alpha\beta} \, \rho_{\beta\alpha}(\Omega)  \nonumber \, \\
 &=& + \sum_{i} t_{ii} \label{0orderT} \\
&& +  \sum_{ia} t_{ia} \, \rho^{ph}_{ai}(\Omega) \, , \nonumber
\end{eqnarray}
\end{subequations}
i.e. it is independent of $\tau$.

\subsection{Zero-order potential energy}

The zero-order diagram reads
\begin{equation}%
\raisebox{-0.8cm}{\includegraphics[
height=0.7in,
width=1.55in
]%
{figures/MBPTdiag11_v2.eps}%
}%
\label{hartreey}%
\end{equation}
The diagram contains two loops and one interaction vertex. It has a symmetry factor $S=2$ and reads
\begin{subequations}
\begin{eqnarray}
v^{(0)}(\tau,\Omega) &=& \frac{1}{2} \sum_{\alpha\beta\gamma\delta}  \bar{v}_{\alpha\beta\gamma\delta} \, G^{0}_{\gamma\alpha}(0,0 ; \Omega) \, G^{0}_{\delta\beta}(0,0 ; \Omega)  \, \\
 &=& \frac{1}{2} \sum_{\alpha\beta\gamma\delta}  \bar{v}_{\alpha\beta\gamma\delta} \, \rho_{\gamma\alpha}(\Omega) \, \rho_{\delta\beta}(\Omega) \nonumber \\
 &=& + \frac{1}{2}\sum_{ij} \bar{v}_{ijij}  \label{0orderV} \\
&& + \frac{1}{2} \sum_{ijc} \bar{v}_{ijcj} \, \rho^{ph}_{ci}(\Omega) + \frac{1}{2} \sum_{ijd} \bar{v}_{ijid} \, \rho^{ph}_{dj}(\Omega)  \nonumber \\
&&+ \frac{1}{2} \sum_{ijab} \bar{v}_{ijab} \, \rho^{ph}_{ai}(\Omega) \, \rho^{ph}_{bj}(\Omega)  \, , \nonumber
\end{eqnarray}
\end{subequations}
i.e. it is independent of $\tau$.

\subsection{First-order kinetic energy}

The first-order diagram deriving from $V$ contains two loops and two interaction vertices. Its symmetry factor is $S=1$. The diagram is
\begin{equation}%
\raisebox{-0cm}{\includegraphics[
height=1.2in,
width=1.3in
]%
{figures/MBPTdiag14_v2.eps}%
}%
\label{2p2hlabx}%
\end{equation}
and its contribution reads
\begin{widetext}
\begin{subequations}
\begin{eqnarray}
t^{(1)}_{V}(\tau,\Omega) &=& -\sum_{\alpha\beta}\sum_{\epsilon\zeta\eta\theta}\int_{0}^{\tau}d\tau_{1}%
\; t_{\alpha\beta} \, \bar{v}_{\epsilon\zeta\eta\theta} \, G^{0}_{\eta\epsilon}(\tau_1,\tau_1 ; \Omega) \, G^{0}_{\theta\alpha}(\tau_1 ,0 ; \Omega)  \, G^{0}_{\beta\zeta}(0,\tau_1 ; \Omega)   \label{total2ndorderdiag3} \\
&=& -\sum_{ija} \frac{t_{ai} \, \bar{v}_{ijaj}}{e_{a}-e_{i}} \, \left[1- e^{-\tau \left(e_{a}-e_{i}\right)} \right] \\
&&- \sum_{ijab} \frac{t_{ai} \, \bar{v}_{ijab}}{e_{a}+e_{b}-e_{i}-e_{j}} \, \rho^{ph}_{bj}(\Omega) \, \left[1- e^{-\tau \left(e_{a}+e_{b}-e_{i}-e_{j}\right)} \right] \nonumber \\
&&+ \sum_{ijka} \frac{t_{ij} \, \bar{v}_{jkak}}{e_{a}-e_{j}} \, \rho^{ph}_{ai}(\Omega) \, \left[1- e^{-\tau \left(e_{a}-e_{j}\right)} \right] \nonumber \\
&&-\sum_{ijab} \frac{t_{ab} \, \bar{v}_{ijia}}{e_{a}-e_{j}} \, \rho^{ph}_{bj}(\Omega) \, \left[1- e^{-\tau \left(e_{a}-e_{j}\right)} \right] \nonumber \\
&&+ \sum_{ijkab} \frac{t_{jb} \, \bar{v}_{ikak}}{e_{a}-e_{i}} \, \rho^{ph}_{aj}(\Omega) \, \rho^{ph}_{bi}(\Omega) \, \left[1- e^{-\tau \left(e_{a}-e_{i}\right)} \right] \nonumber \\
&&- \sum_{ijabc} \frac{t_{ac} \, \bar{v}_{ijba}}{e_{a}+e_{b}-e_{i}-e_{j}} \, \rho^{ph}_{bi}(\Omega) \, \rho^{ph}_{cj}(\Omega) \, \left[1- e^{-\tau \left(e_{a}+e_{b}-e_{i}-e_{j}\right)} \right] \nonumber \\
&&+ \sum_{ijkab} \frac{t_{kj} \, \bar{v}_{ijab}}{e_{a}+e_{b}-e_{i}-e_{j}} \, \rho^{ph}_{ai}(\Omega) \, \rho^{ph}_{bk}(\Omega) \, \left[1- e^{-\tau \left(e_{a}+e_{b}-e_{i}-e_{j}\right)} \right] \nonumber \\
&&+ \sum_{ijkabc} \frac{t_{kc} \, \bar{v}_{ijab}}{e_{a}+e_{b}-e_{i}-e_{j}} \, \rho^{ph}_{bj}(\Omega) \, \rho^{ph}_{ak}(\Omega) \, \rho^{ph}_{ci}(\Omega) \, \left[1- e^{-\tau \left(e_{a}+e_{b}-e_{i}-e_{j}\right)} \right] \,  . \nonumber
\end{eqnarray}
\end{subequations}
\end{widetext}

The first-order diagram deriving from $U$ contains one loop and two interaction vertices. Its symmetry factor is $S=1$. The diagram is
\begin{equation}%
\raisebox{-0cm}{\includegraphics[
height=1.2in,
width=1.3in
]%
{figures/MBPTdiag13_v2.eps}%
}%
\label{2p2hlabz}%
\end{equation}
and its contribution reads
\begin{widetext}
\begin{subequations}
\begin{eqnarray}
t^{(1)}_{U}(\tau,\Omega) &=& \sum_{\alpha\beta}\sum_{\epsilon\zeta}\int_{0}^{\tau}d\tau_{1}%
\; t_{\alpha\beta} \, \left(-u_{\epsilon\zeta}\right) \, G^{0}_{\zeta\alpha}(\tau_1,0 ; \Omega) \, G^{0}_{\beta\epsilon}(0,\tau_1 ; \Omega)  \\
&=& + \sum_{ia} \frac{t_{ai} \, u_{ia}}{e_{a}-e_{i}} \, \left[1-e^{-\tau \left(e_{a}-e_{i} \right)}\right] \label{total2ndorderdiag4}  \\
&&- \sum_{ija} \frac{t_{ji} \, u_{ia}}{e_{a}-e_{i}} \, \rho^{ph}_{aj}(\Omega) \, \left[1-e^{-\tau \left(e_{a}-e_{i} \right)}\right] \nonumber \\
&& + \sum_{iab} \frac{t_{ab} \, u_{ia}}{e_{a}-e_{i}} \, \rho^{ph}_{bi}(\Omega) \, \left[1-e^{-\tau \left(e_{a}-e_{i} \right)}\right] \nonumber \\
&& - \sum_{ijab} \frac{t_{jb} \, u_{ia}}{e_{a}-e_{i}} \, \rho^{ph}_{aj}(\Omega) \, \rho^{ph}_{bi}(\Omega) \, \left[1-e^{-\tau \left(e_{a}-e_{i} \right)}\right] \, . \nonumber
\end{eqnarray}
\end{subequations}
\end{widetext}

\subsection{First-order potential energy}

The first diagram deriving from $V$ contains two loops and two interaction vertices. Its symmetry factor is $S=4$. The diagram is
\begin{equation}%
\raisebox{-0cm}{\includegraphics[
height=1.2in,
width=1.3in
]%
{figures/MBPTdiag15_v2.eps}%
}%
\label{2p2hlabt}%
\end{equation}
and its contribution reads
\begin{widetext}
\begin{subequations}
\begin{eqnarray}
v^{(1)}_{V1}(\tau,\Omega) &=& -\frac{1}{4}\sum_{\alpha\beta\gamma\delta}\sum_{\epsilon\zeta\eta\theta}\int_{0}^{\tau}d\tau_{1}%
\; \bar{v}_{\alpha\beta\gamma\delta} \, \bar{v}_{\epsilon\zeta\eta\theta} \, G^{0}_{\gamma\epsilon}(0,\tau_1 ; \Omega) \, G^{0}_{\eta\alpha}(\tau_1,0 ; \Omega) \, G^{0}_{\delta\zeta}(0,\tau_1 ; \Omega) \, G^{0}_{\theta\beta}(\tau_1,0 ; \Omega)   \\
&=& -\frac{1}{4}\sum_{ijab} \frac{|\bar{v}_{ijab}|^2}{e_{a}+e_{b}-e_{i}-e_{j}} \, \left[1- e^{-\tau \left(e_{a}+e_{b}-e_{i}-e_{j}\right)} \right] \label{total2ndorderdiag2} \\
&&-  \frac{1}{2}\sum_{ijabc} \frac{\bar{v}_{ijab}\bar{v}_{abcj}}{e_{a}+e_{b}-e_{i}-e_{j}} \, \rho^{ph}_{ci}(\Omega) \, \left[1- e^{-\tau \left(e_{a}+e_{b}-e_{i}-e_{j}\right)}\right] \nonumber \\
&&+ \frac{1}{2} \sum_{ijkab} \frac{\bar{v}_{ijab}\bar{v}_{kbij}}{e_{a}+e_{b}-e_{i}-e_{j}} \, \rho^{ph}_{ak}(\Omega) \, \left[1- e^{-\tau \left(e_{a}+e_{b}-e_{i}-e_{j}\right)}\right] \nonumber \\
&&-  \frac{1}{4}\sum_{ijklab} \frac{\bar{v}_{ijab}\bar{v}_{klij}}{e_{a}+e_{b}-e_{i}-e_{j}} \, \rho^{ph}_{ak}(\Omega)\, \rho^{ph}_{bl}(\Omega) \, \left[1- e^{-\tau \left(e_{a}+e_{b}-e_{i}-e_{j}\right)}\right] \nonumber \\
&&-  \frac{1}{4}\sum_{ijabcd} \frac{\bar{v}_{ijab}\bar{v}_{abcd}}{e_{a}+e_{b}-e_{i}-e_{j}} \, \rho^{ph}_{ci}(\Omega)\, \rho^{ph}_{dj}(\Omega) \, \left[1- e^{-\tau \left(e_{a}+e_{b}-e_{i}-e_{j}\right)}\right] \nonumber \\
&&+ \sum_{ijkabc} \frac{\bar{v}_{ijab}\bar{v}_{kbcj}}{e_{a}+e_{b}-e_{i}-e_{j}} \, \rho^{ph}_{ak}(\Omega)\, \rho^{ph}_{ci}(\Omega) \, \left[1- e^{-\tau \left(e_{a}+e_{b}-e_{i}-e_{j}\right)}\right]  \nonumber \\
&&+ \frac{1}{2} \sum_{ijkabcd} \frac{\bar{v}_{ijab}\bar{v}_{akcd}}{e_{a}+e_{b}-e_{i}-e_{j}} \, \rho^{ph}_{ci}(\Omega) \, \rho^{ph}_{bk}(\Omega)\, \rho^{ph}_{dj}(\Omega) \, \left[1- e^{-\tau \left(e_{a}+e_{b}-e_{i}-e_{j}\right)}\right] \nonumber \\
&&- \frac{1}{2} \sum_{ijklabc} \frac{\bar{v}_{ijab}\bar{v}_{klic}}{e_{a}+e_{b}-e_{i}-e_{j}} \, \rho^{ph}_{ak}(\Omega) \, \rho^{ph}_{bl}(\Omega)\, \rho^{ph}_{cj}(\Omega) \, \left[1- e^{-\tau \left(e_{a}+e_{b}-e_{i}-e_{j}\right)}\right] \nonumber \\
&&- \frac{1}{4}\sum_{ijklabcd} \frac{\bar{v}_{ijab}\bar{v}_{klcd}}{e_{a}+e_{b}-e_{i}-e_{j}} \, \rho^{ph}_{ak}(\Omega) \, \rho^{ph}_{ci}(\Omega) \, \rho^{ph}_{bl}(\Omega)\, \rho^{ph}_{dj}(\Omega) \, \left[1- e^{-\tau \left(e_{a}+e_{b}-e_{i}-e_{j}\right)}\right]  \,\,\, . \nonumber
\end{eqnarray}
\end{subequations}
\end{widetext}

The second diagram deriving from $V$ contains three loops and two interaction vertices. Its symmetry factor is $S=1$. The diagram is
\begin{equation}%
\raisebox{-0cm}{\includegraphics[
height=1.2in,
width=1.3in
]%
{figures/MBPTdiag16_v2.eps}%
}%
\label{2p2hlabm}%
\end{equation}
and its contribution reads
\begin{widetext}
\begin{subequations}
\begin{eqnarray}
v^{(1)}_{V2}(\tau,\Omega) &=& \sum_{\alpha\beta\gamma\delta}\sum_{\epsilon\zeta\eta\theta}\int_{0}^{\tau}d\tau_{1}%
\; \bar{v}_{\alpha\beta\gamma\delta} \, \bar{v}_{\epsilon\zeta\eta\theta} \, G^{0}_{\gamma\alpha}(0,0 ; \Omega) \, G^{0}_{\eta\epsilon}(\tau_1,\tau_1 ; \Omega) \, G^{0}_{\delta\zeta}(0,\tau_1 ; \Omega) \, G^{0}_{\theta\beta}(\tau_1, 0 ; \Omega)   \\
&=& -\sum_{ijka} \frac{\bar{v}_{iaik}\bar{v}_{jkja}}{e_{a}-e_{k}} \, \left[1- e^{-\tau \left(e_{a}-e_{k}\right)} \right] \label{total2ndorderdiagX} \\
&&-  \sum_{ijkab} \frac{\bar{v}_{ibak}\bar{v}_{jkjb}}{e_{b}-e_{k}} \, \rho^{ph}_{ai}(\Omega) \, \left[1- e^{-\tau \left(e_{b}-e_{k}\right)}\right] \nonumber \\
&&- \sum_{ijkab} \frac{\bar{v}_{ibik}\bar{v}_{jkab}}{e_{a}+e_{b}-e_{j}-e_{k}} \, \rho^{ph}_{aj}(\Omega) \, \left[1- e^{-\tau \left(e_{a}+e_{b}-e_{j}-e_{k}\right)}\right] \nonumber \\
&&- \sum_{ijkab} \frac{\bar{v}_{ibia}\bar{v}_{jkjb}}{e_{b}-e_{k}} \, \rho^{ph}_{ak}(\Omega) \, \left[1- e^{-\tau \left(e_{b}-e_{k}\right)}\right] \nonumber \\
&&+ \sum_{ijkla} \frac{\bar{v}_{ilik}\bar{v}_{jkja}}{e_{a}-e_{k}} \, \rho^{ph}_{al}(\Omega) \, \left[1- e^{-\tau \left(e_{a}-e_{k}\right)}\right] \nonumber \\
&&- \sum_{ijkabc} \frac{\bar{v}_{icak}\bar{v}_{jkbc}}{e_{b}+e_{c}-e_{j}-e_{k}} \, \rho^{ph}_{ai}(\Omega) \, \rho^{ph}_{bj}(\Omega) \, \left[1- e^{-\tau \left(e_{b}+e_{c}-e_{j}-e_{k}\right)}\right] \nonumber \\
&&- \sum_{ijkabc} \frac{\bar{v}_{icab}\bar{v}_{jkjc}}{e_{c}-e_{k}} \, \rho^{ph}_{ai}(\Omega) \, \rho^{ph}_{bk}(\Omega) \, \left[1- e^{-\tau \left(e_{c}-e_{k}\right)}\right] \nonumber \\
&&+ \sum_{ijklab} \frac{\bar{v}_{ilak}\bar{v}_{jkjb}}{e_{b}-e_{k}} \, \rho^{ph}_{ai}(\Omega) \, \rho^{ph}_{bl}(\Omega) \, \left[1- e^{-\tau \left(e_{b}-e_{k}\right)}\right] \nonumber \\
&&- \sum_{ijkabc} \frac{\bar{v}_{icib}\bar{v}_{jkac}}{e_{a}+e_{c}-e_{j}-e_{k}} \, \rho^{ph}_{aj}(\Omega) \, \rho^{ph}_{bk}(\Omega) \, \left[1- e^{-\tau \left(e_{a}+e_{c}-e_{j}-e_{k}\right)}\right] \nonumber \\
&&+ \sum_{ijklab} \frac{\bar{v}_{ilik}\bar{v}_{jkab}}{e_{a}+e_{b}-e_{j}-e_{k}} \, \rho^{ph}_{aj}(\Omega) \, \rho^{ph}_{bl}(\Omega) \, \left[1- e^{-\tau \left(e_{a}+e_{b}-e_{j}-e_{k}\right)}\right] \nonumber \\
&&+ \sum_{ijklab} \frac{\bar{v}_{ilia}\bar{v}_{jkjb}}{e_{b}-e_{k}} \, \rho^{ph}_{ak}(\Omega) \, \rho^{ph}_{bl}(\Omega) \, \left[1- e^{-\tau \left(e_{b}-e_{k}\right)}\right] \nonumber \\
&&+ \sum_{ijklabc} \frac{\bar{v}_{ilib}\bar{v}_{jkac}}{e_{a}+e_{c}-e_{j}-e_{k}} \, \rho^{ph}_{aj}(\Omega) \, \rho^{ph}_{bk}(\Omega) \, \rho^{ph}_{cl}(\Omega) \, \left[1- e^{-\tau \left(e_{a}+e_{c}-e_{j}-e_{k}\right)}\right] \nonumber \\
&&+ \sum_{ijklabc} \frac{\bar{v}_{ilab}\bar{v}_{jkjc}}{e_{c}-e_{k}} \, \rho^{ph}_{ai}(\Omega) \, \rho^{ph}_{bk}(\Omega) \, \rho^{ph}_{cl}(\Omega) \, \left[1- e^{-\tau \left(e_{c}-e_{k}\right)}\right] \nonumber \\
&&+ \sum_{ijklabc} \frac{\bar{v}_{ilak}\bar{v}_{jkbc}}{e_{b}+e_{c}-e_{j}-e_{k}} \, \rho^{ph}_{ai}(\Omega) \, \rho^{ph}_{bj}(\Omega) \, \rho^{ph}_{cl}(\Omega) \, \left[1- e^{-\tau \left(e_{b}+e_{c}-e_{j}-e_{k}\right)}\right] \nonumber \\
&&- \sum_{ijkabcd} \frac{\bar{v}_{idac}\bar{v}_{jkbd}}{e_{b}+e_{d}-e_{j}-e_{k}} \, \rho^{ph}_{ai}(\Omega) \, \rho^{ph}_{bj}(\Omega) \, \rho^{ph}_{ck}(\Omega) \, \left[1- e^{-\tau \left(e_{b}+e_{d}-e_{j}-e_{k}\right)}\right] \nonumber \\
&&+ \sum_{ijklabcd} \frac{\bar{v}_{ilac}\bar{v}_{jkbd}}{e_{b}+e_{d}-e_{j}-e_{k}} \, \rho^{ph}_{ai}(\Omega) \, \rho^{ph}_{bj}(\Omega) \, \rho^{ph}_{ck}(\Omega) \, \rho^{ph}_{dl}(\Omega) \, \left[1- e^{-\tau \left(e_{b}+e_{d}-e_{j}-e_{k}\right)}\right] \,\,\, .
\end{eqnarray}
\end{subequations}
\end{widetext}

The diagram deriving from $U$ contains two loops and two interaction vertices. Its symmetry factor is $S=1$. The diagram is
\begin{equation}%
\raisebox{-0cm}{\includegraphics[
height=1.2in,
width=1.3in
]%
{figures/MBPTdiag17_v2.eps}%
}%
\label{2p2hlabk}%
\end{equation}
and its contribution reads
\begin{widetext}
\begin{subequations}
\begin{eqnarray}
v^{(1)}_{U}(\tau,\Omega) &=& -\sum_{\alpha\beta\gamma\delta}\sum_{\epsilon\zeta}\int_{0}^{\tau}d\tau_{1}%
\; \bar{v}_{\alpha\beta\gamma\delta}  \, \left(-u_{\epsilon\zeta}\right) \, G^{0}_{\gamma\alpha}(0,0 ; \Omega) \, G^{0}_{\zeta\beta}(\tau_1,0 ; \Omega) \, G^{0}_{\delta\epsilon}(0,\tau_1 ; \Omega)  \label{total2ndorderdiag4x}  \\
&=& +\sum_{ija} \frac{\bar{v}_{iaij} \, u_{ja}}{e_{a}-e_{j}} \, \left[1-e^{-\tau \left(e_{a}-e_{j} \right)}\right]  \\
&&+ \sum_{iab} \frac{\bar{v}_{iabj} \, u_{ja}}{e_{a}-e_{j}} \, \rho^{ph}_{bi}(\Omega) \left[1-e^{-\tau \left(e_{a}-e_{j} \right)}\right] \nonumber \\
&&- \sum_{ijka} \frac{\bar{v}_{ijik} \, u_{ka}}{e_{a}-e_{k}} \, \rho^{ph}_{aj}(\Omega) \left[1-e^{-\tau \left(e_{a}-e_{k} \right)}\right] \nonumber \\
&&+ \sum_{ijab} \frac{\bar{v}_{iaib} \, u_{ja}}{e_{a}-e_{j}} \, \rho^{ph}_{bj}(\Omega) \left[1-e^{-\tau \left(e_{a}-e_{j} \right)}\right] \nonumber \\
&&- \sum_{ijkab} \frac{\bar{v}_{ijia} \, u_{kb}}{e_{b}-e_{k}} \, \rho^{ph}_{bj}(\Omega) \, \rho^{ph}_{ak}(\Omega) \left[1-e^{-\tau \left(e_{b}-e_{k} \right)}\right] \nonumber \\
&&+ \sum_{ijabc} \frac{\bar{v}_{iabc} \, u_{ja}}{e_{a}-e_{j}} \, \rho^{ph}_{bi}(\Omega) \, \rho^{ph}_{cj}(\Omega) \left[1-e^{-\tau \left(e_{a}-e_{j} \right)}\right] \nonumber \\
&&- \sum_{ijkab} \frac{\bar{v}_{ijak} \, u_{kb}}{e_{b}-e_{k}} \, \rho^{ph}_{ai}(\Omega) \, \rho^{ph}_{bj}(\Omega) \left[1-e^{-\tau \left(e_{b}-e_{k} \right)}\right] \nonumber \\
&&-  \sum_{ijkabc} \frac{\bar{v}_{ijab} \, u_{kc}}{e_{c}-e_{k}} \, \rho^{ph}_{ai}(\Omega) \, \rho^{ph}_{cj}(\Omega) \, \rho^{ph}_{bk}(\Omega) \left[1-e^{-\tau \left(e_{c}-e_{k} \right)}\right] \, . \nonumber
\end{eqnarray}
\end{subequations}
\end{widetext}

\subsection{Structure of $h(\tau,\Omega)$ in $\tau$ and $\Omega$}
\label{structure3}

We now provide a proof of the characteristic dependence of the linked/connected energy kernel $h(\tau,\Omega)$ on $\tau$ and $\Omega$. For simplicity, the derivation is provided for its kinetic energy part  $t(\tau,\Omega)$ but can be easily extended to its potential energy part $v(\tau,\Omega)$, or to the linked/connected kernel  $o(\tau,\Omega)$ of any operator $O$.

Starting from the linked/connected part of the perturbative expansion~\ref{expansionenergykernel}, we reason on a particular, though generic, diagram of arbitrary order $n$. A term of order $n$ in Eq.~\ref{expansionenergykernel} contains $k$ operators $-U(\tau_p)$ and $n-k$ operators $V(\tau_p)$ whose time labels are integrated over, plus the operator $T$ at time $0$. For simplicity, and without any lack of generality, we focus on the term with $k=n$. Such a contribution is a sum (over single-particle indices and all possible time orderings) of terms proportional to
\begin{widetext}
\begin{eqnarray}
\!\int_{0}^{\tau}\!\!d\tau_{n}\!\!\int_{0}^{\tau_n}\!\!d\tau_{n-1}\ldots\!\! \int_{0}^{\tau_{2}}\!\!d\tau_{1} \, e^{\tau_n(e_{\alpha_n}-e_{\beta_n})}  \ldots e^{\tau_1(e_{\alpha_1}-e_{\beta_1})}\, \langle \Phi |   a_{\alpha_n}^{\dagger} a_{\beta_n}  \ldots a_{\alpha_1}^{\dagger}
a_{\beta_1} a_{\epsilon}^{\dagger}  a_{\zeta} |  \Phi(\Omega) \rangle_{c} \, , \label{diagramlinkedtoT}
\end{eqnarray}
\end{widetext}
where the operators $ a_{\epsilon}^{\dagger}$ and $ a_{\zeta}$ comes from the kinetic energy operator $T$ and where the time dependence has been extracted from the creation and annihilation operators (see Eq.~\ref{aalphatau}). As $T$ is at time $0$, no time factor appears in connection with $ a_{\epsilon}^{\dagger}$ and $ a_{\zeta}$. The particular time ordering considered in Eq.~\ref{diagramlinkedtoT} does not limit the generality of the analysis given that any contribution to $t(\tau,\Omega)$ can be written under this form. In particular, it is clear that operators stemming from $T(0)$ always carry the smallest time and thus occupy the rightmost position. 

Equation~\ref{typicaldiagram1} can be written in the compact form
\begin{widetext}
\begin{eqnarray}
B_n(\tau; b_1,\ldots b_n) \, \langle \Phi |   a_{\alpha_n}^{\dagger} a_{\beta_n}  \ldots a_{\alpha_1}^{\dagger}
a_{\beta_1}a_{\epsilon}^{\dagger}  a_{\zeta} |  \Phi(\Omega) \rangle_{c}\, , \label{diagramlinkedtoT2}
\end{eqnarray}
where $b_p\equiv e_{\alpha_p}\!-\!e_{\beta_p}$ amounts to capturing the coefficient in front of $\tau_p$ in the original time integral and with the convention that $B_0(\tau)\equiv 1$. The explicit expressions of $B_1(\tau; b_1)$ and $B_2(\tau; b_1, b_2)$ can be easily extracted from Eqs.~\ref{integral1} and~\ref{integral3}, respectively. Further considering that
\begin{eqnarray}
B_{p}(\tau; b_1,\ldots, b_p) &=& \int_{0}^{\tau}\!\!d\tau_{p} \, e^{\tau_{p}b_{p}} \, B_{n}(\tau_{p\!-\!1}; b_1,\ldots b_{p\!-\!1}) \, ,
\end{eqnarray}
it is straightforward to prove by recurrence that the function $B_n(\tau; b_1,\ldots b_n)$ possesses the general structure
\begin{eqnarray}
B_n(\tau; b_1,\ldots b_n) &\equiv&  c^{(n)}_0 + c^{(n)}_1 \, e^{\tau d_{1}} + c^{(n)}_2 \, e^{\tau d_{2}} +\ldots + c^{(n)}_{n} \, e^{\tau d_{n}} \, , \label{diagramlinkedtoT3}
\end{eqnarray}
\end{widetext}
where $c^{(n)}_p\,(p = 0,\ldots,n)$ and $d_p\equiv\sum_{k=n-p+1}^{n} b_k (p = 1,\ldots,n)$ are functions of the $b_p\,(p=1,\ldots,n)$. Eventually, the actual dependence of $c^{(n)}_p$ and $d_p$ on the single-particle energies $\{e_\alpha\}$ depend on the particular (i.e. differ for each) set of contractions arising from the matrix element in Eq.~\ref{diagramlinkedtoT} and on whether $\Omega$ is zero or not. In all cases though, the key feature is that there exists one term with no dependence on  $\tau$ while all the others depend exponentially on it with a strictly negative coefficient. Focusing on a particular contraction, Eq.~\ref{diagramlinkedtoT} can be generically written as
 \begin{eqnarray}
c^{(n)}_0  F^{(n)}_0[\rho^{ph}(\Omega)] + \sum_{p=1}^{n} c^{(n)}_p \, e^{\tau d_{p}} \, F^{(n)}_p[\rho^{ph}(\Omega)]   \, , \label{diagramlinkedtoT4}
\end{eqnarray}
where $F^{(n)}_p[\rho^{ph}(\Omega)]$ is a polynomial in $\rho^{ph}(\Omega)$ whose constant term is always non zero and that may contain terms with up to power $n$. One can verify that zero- and first-order diagrams computed above do display such typical $\tau$ and $\Omega$ dependencies. This result demonstrates that $t(\tau,\Omega)$ goes in the large $\tau$ limit to a finite value $t(\Omega)$ that sums all the contributions embodied by the coefficient $c^{(n)}_0$.

\section{Cluster amplitude equations}
\label{amplitudeequations}

Starting from Eq.~\ref{dynamicalkernels}, we derive the equations of motion (Eqs.~\ref{reduceddynamicalkernels} and~\ref{CCamplitudekernels}) satisfied by hole-particle matrix elements of ${\cal T}^{\dagger}_{n}(\tau,\Omega)$. The derivations below are obtained by adapting to n-tuply excited energy and norm kernels the steps taken in Secs.~\ref{energykernel} and~\ref{energykernelMBPTtoCC} for the energy kernel. Being formally similar, those steps are not detailed here. 

\subsection{Energy equation}

With $A^{ab\ldots}_{ij\ldots}=\bbone$, Eq.~\ref{dynamicalkernels} provides 
\begin{equation}
H(\tau,\Omega) = -\partial_{\tau} N(\tau,\Omega) \, , \label{nonconnectedenergy}
\end{equation}
that expresses the energy kernel as the (imaginary-)time derivative of the norm kernel.

\subsection{Single amplitude equation}
\label{singles}

Considering the operator $A^{a}_{i}=a^{\dagger}_a a_i$ that creates a single excitation, Eq.~\ref{dynamicalkernels} provides 
\begin{equation}
H^{a}_{i}(\tau,\Omega) = -\partial_{\tau} N^{a}_{i}(\tau,\Omega) \, . \label{nonconnectedsingleamplitude}
\end{equation}

Let us start with $N^{a}_{i}(\tau,\Omega)$. Rewriting $| \Psi (\tau) \rangle$ in terms of ${\cal U}(\tau)$, expanding the latter through perturbation theory and applying the off-diagonal Wick theorem~\cite{balian69a}, one obtains the factorization of the singly-excited norm kernel as
\begin{equation}
N^{a}_{i}(\tau,\Omega) = n^{a}_{i}(\tau,\Omega) \, N(\tau,\Omega) \, , \label{factorizesinglenormkernel}
\end{equation}
where 
\begin{equation}
n^{a}_{i}(\tau,\Omega) \equiv \frac{\langle \Phi | {\cal U}(\tau)   A^{a}_{i} | \Phi(\Omega) \rangle_{c}}{\langle \Phi |  \Phi(\Omega) \rangle} \,  \label{factorizesinglenormkernela}
\end{equation}
contains the complete set of connected vacuum-to-vacuum diagrams {\it linked} to $A^{a}_{i}$. In the next step, this complete set of diagrams can be rewritten as
\begin{subequations}
\label{singleconnectednormkernel}
\begin{eqnarray}
n^{a}_{i}(\tau,\Omega) &=& \frac{\langle \Phi | {\cal T}^{\dagger}_{1}(\tau,\Omega)   A^{a}_{i} | \Phi(\Omega) \rangle_{c}}{\langle \Phi |  \Phi(\Omega) \rangle} \label{singleconnectednormkernel2} \\
&=& {\cal T}^{\dagger}_{ia}(\tau,\Omega) \label{singleconnectednormkernel1} \\
&=& \frac{\langle \Phi | e^{{\cal T}^{\dagger}(\tau,\Omega)}   A^{a}_{i} | \Phi(\Omega) \rangle_{c}}{\langle \Phi |  \Phi(\Omega) \rangle} \label{singleconnectednormkernel3} \, ,
\end{eqnarray}
\end{subequations}
where the rule is that no contraction is to be considered among cluster operators or within a cluster operator when expanding the exponential. Off-diagonal contractions within the operator $A^{ab\ldots}_{ij\ldots}$ are zero (Eq.~\ref{contractionsrho2}).  Expression~\ref{singleconnectednormkernel3} can thus be equated at no cost to Eq.~\ref{singleconnectednormkernel2} by virtue of the linked/connected character of the kernel.

Let us now come to $H^{a}_{i}(\tau,\Omega)$. Because of the presence of two fixed-time operators $A^{a}_{i}$ and $H$ in the matrix elements, perturbation theory leads to the typical structure
\begin{equation}
H^{a}_{i}(\tau,\Omega) = h^{a}_{i}(\tau,\Omega) \, N(\tau,\Omega)  + n^{a}_{i}(\tau,\Omega) \, H(\tau,\Omega) \, . \label{factorizesingleenergykernel}
\end{equation}
In Eq.~\ref{factorizesingleenergykernel} was introduced the kernel
\begin{subequations}
\label{singleconnectedenergykernel}
\begin{eqnarray}
h^{a}_{i}(\tau,\Omega) &\equiv& \frac{\langle \Phi | {\cal U}(\tau) H  A^{a}_{i} | \Phi(\Omega) \rangle_{c}}{\langle \Phi |  \Phi(\Omega) \rangle} \label{singleconnectedenergykernel1} \\
&=& \frac{\langle \Phi | e^{{\cal T}^{\dagger}(\tau,\Omega)} H  A^{a}_{i} | \Phi(\Omega) \rangle_{c}}{\langle \Phi |  \Phi(\Omega) \rangle} \label{singleconnectedenergykernel2} \,  ,
\end{eqnarray}
\end{subequations}
where operators in the matrix element are all connected together by strings of contractions. A crucial remark is here in order. In Eq.~\ref{singleconnectedenergykernel2}, the only cluster operator that could contract exclusively with the operators entering $A^{a}_{i}$ is ${\cal T}^{\dagger}_{1}(\tau,\Omega)$. However, if it were to happen, the product ${\cal T}^{\dagger}_{1}(\tau,\Omega)A^{a}_{i}$ would be disconnected from $H$ and  the other allowed ${\cal T}^{\dagger}_{n}(\tau,\Omega)$, which would contradict the fact that the matrix elements are connected, i.e. such contractions actually contribute to the second term in the right-hand side of Eq.~\ref{factorizesingleenergykernel}. Consequently, all allowed cluster operators are only partially contracted with $A^{a}_{i}$ and are thus necessarily contracted with $H$. Eventually, this is the actual meaning carried by the label $c$ in Eq.~\ref{singleconnectedenergykernel2}. This result allows us to recover the crucial termination of the expanded exponential at play in standard CC theory. As already discussed for the non-excited energy kernel, this termination and the connected nature of the resulting terms are usually obtained on the basis of the Baker-Campbell-Hausdorff identity and the standard Wick theorem. In the present case, such a property cannot be obtained directly and requires a long detour through perturbation theory applied to off-diagonal kernels. The actual expression of $h^{a}_{i}(\tau,\Omega)$ is given in Eq.~\ref{termination1}.

Inserting Eqs.~\ref{factorizesinglenormkernel} and~\ref{factorizesingleenergykernel} into Eq.~\ref{nonconnectedsingleamplitude}, utilizing Eq.~\ref{singleconnectednormkernel1} and combining the result with Eq.~\ref{nonconnectedenergy} eventually leads to the single amplitude equation under the practical form of Eq.~\ref{reduceddynamicalkernels}, i.e. 
\begin{equation}
h^{a}_{i}(\tau,\Omega) = -\partial_{\tau} {\cal T}^{\dagger}_{ia}(\tau,\Omega)  \, . \label{reducedsingleamplitudeequation}
\end{equation}

\subsection{Double amplitude equation}
\label{doubles}

Considering the operator $A^{ab}_{ij}=a^{\dagger}_a a_i a^{\dagger}_b a_j$ that creates a double excitation, Eq.~\ref{dynamicalkernels} provides 
\begin{equation}
H^{ab}_{ij}(\tau,\Omega) = -\partial_{\tau} N^{ab}_{ij}(\tau,\Omega) \, . \label{nonconnecteddoubleamplitude}
\end{equation}
Following the same steps as before, one first obtains
\begin{equation}
N^{ab}_{ij}(\tau,\Omega) = n^{ab}_{ij}(\tau,\Omega) \, N(\tau,\Omega) \, , \label{factorizedoublenormkernel}
\end{equation}
along with
\begin{widetext}
\begin{subequations}
\label{doubleconnectednormkernel}
\begin{eqnarray}
n^{ab}_{ij}(\tau,\Omega) &=& \frac{\langle \Phi | \Big[{\cal T}^{\dagger}_{2}(\tau,\Omega) +  \frac{1}{2} {\cal T}^{\dagger\, 2}_{1}(\tau,\Omega)\Big] A^{ab}_{ij} | \Phi(\Omega) \rangle_{c}}{\langle \Phi |  \Phi(\Omega) \rangle} \label{doubleconnectednormkernel1} \\
&=& {\cal T}^{\dagger}_{ijab}(\tau,\Omega) + {\cal T}^{\dagger}_{ia}(\tau,\Omega) \, {\cal T}^{\dagger}_{jb}(\tau,\Omega) - {\cal T}^{\dagger}_{ja}(\tau,\Omega) \, {\cal T}^{\dagger}_{ib}(\tau,\Omega) \label{doubleconnectednormkernel2} \\
&=& \frac{\langle \Phi | e^{{\cal T}^{\dagger}(\tau,\Omega)}   A^{ab}_{ij} | \Phi(\Omega) \rangle_{c}}{\langle \Phi |  \Phi(\Omega) \rangle} \label{doubleconnectednormkernel3} \, ,
\end{eqnarray}
\end{subequations}
where the same rules and explanations as before apply.
\end{widetext}

Coming to $H^{ab}_{ij}(\tau,\Omega)$, perturbation theory leads once again to the typical structure
\begin{eqnarray}
H^{ab}_{ij}(\tau,\Omega) &=& \frac{\langle \Phi | {\cal U}(\tau) H  A^{ab}_{ij} | \Phi(\Omega) \rangle_{c}}{\langle \Phi |  \Phi(\Omega) \rangle} \, N(\tau,\Omega)  \nonumber \\
&& + n^{ab}_{ij}(\tau,\Omega) \, H(\tau,\Omega) \, , \label{factorizedoubleenergykernel}
\end{eqnarray}
where operators in the first kernel on the right-hand side are all connected together by strings of contractions. Following the same steps as before, one proceeds to the identification of the cluster, which leads to
\begin{eqnarray}
\frac{\langle \Phi | {\cal U}(\tau) H  A^{ab}_{ij} | \Phi(\Omega) \rangle_{c}}{\langle \Phi |  \Phi(\Omega) \rangle} &=& \frac{\langle \Phi | e^{{\cal T}^{\dagger}(\tau,\Omega)} H  A^{ab}_{ij} | \Phi(\Omega) \rangle_{c}}{\langle \Phi |  \Phi(\Omega) \rangle} \label{doubleconnectedenergykernelintermediate} \,  .
\end{eqnarray}
Again, it is essential to detail the connected structure of this kernel. At this point, it can only be stated that the operators at play in Eq.~\ref{singleconnectedenergykernel} are all connected together through strings of contractions by virtue of the connected character of the associated diagrams. A priori, this leaves the possibility that a cluster operator is solely, and thus entirely, connected to $A^{ab}_{ij}$, i.e. that it is not connected to $H$. In the doubly-excited case, it can at most happen for ${\cal T}^{\dagger}_{1}(\tau,\Omega)$ or ${\cal T}^{\dagger}_{2}(\tau,\Omega)$. Contracting fully ${\cal T}^{\dagger}_{2}(\tau,\Omega)$ with $A^{ab}_{ij}$ leaves no possibility for the latter to further connect to $H$ and  contradicts the fact that all the operators are  connected together through strings of contractions, i.e. such a contribution is already included in the second term on the right-hand side of Eq.~\ref{factorizedoubleenergykernel}. As for ${\cal T}^{\dagger}_{1}(\tau,\Omega)$, the situation is more delicate. Let us thus consider contributions to Eq.~\ref{doubleconnectedenergykernelintermediate} where ${\cal T}^{\dagger}_{1}(\tau,\Omega)$ is fully contracted with $A^{ab}_{ij}$. This leaves one particle creation operator and one hole annihilation operator from $A^{ab}_{ij}$ to operate further contractions, i.e. a single-excitation operator $A^{c}_{k}$ with $c$ ($k$) equal to $a$ ($i$) or $b$ ($j$). For each term with $p\geq 1$ powers of ${\cal T}^{\dagger}_{1}(\tau,\Omega)$ in the exponential, i.e. terms proportional  to ${\cal T}^{\dagger \, p}_{1}(\tau,\Omega)/p!$, there are $p$ possibilities to fully contract a ${\cal T}^{\dagger}_{1}(\tau,\Omega)$ operator with $A^{ab}_{ij}$, which leaves ${\cal T}^{\dagger \, p\!-\!1}_{1}(\tau,\Omega)/(p\!-\!1)!$ for further contractions. Summing over all terms stemming from  the exponential, one can eventually re-factorize each time the full contribution of ${\cal T}^{\dagger}_{1}(\tau,\Omega)$ to the exponential. Performing the algebraic manipulations in details, one eventually arrives at
\begin{eqnarray}
\frac{\langle \Phi | e^{{\cal T}^{\dagger}(\tau,\Omega)} H  A^{ab}_{ij} | \Phi(\Omega) \rangle_{c}}{\langle \Phi |  \Phi(\Omega) \rangle} &=& h^{ab}_{ij}(\tau,\Omega) \label{doubleconnectedenergykernel}  \\
&& + h^{b}_{j}(\tau,\Omega) \, {\cal T}^{\dagger}_{ia}(\tau,\Omega) \nonumber \\
&& - h^{a}_{j}(\tau,\Omega) \, {\cal T}^{\dagger}_{ib}(\tau,\Omega) \nonumber \\
&& - h^{b}_{i}(\tau,\Omega) \, {\cal T}^{\dagger}_{ja}(\tau,\Omega) \nonumber \\
&& + h^{a}_{i}(\tau,\Omega) \, {\cal T}^{\dagger}_{jb}(\tau,\Omega) \nonumber \, , 
\end{eqnarray}
where $h^{ab}_{ij}(\tau,\Omega)$ denotes the contributions to the matrix elements where all cluster operators are {\it necessarily} contracted with $H$, which ultimately leads to the usual truncation of the exponential. The last four terms in Eq.~\ref{doubleconnectedenergykernel} gather all the contributions where a ${\cal T}^{\dagger}_{1}(\tau,\Omega)$ was fully contracted to $A^{ab}_{ij}$.

To eventually obtain the practical form of the double amplitude equation (Eq.~\ref{reduceddynamicalkernels}), one needs not only to insert  Eqs.~\ref{doubleconnectednormkernel2} and~\ref{doubleconnectedenergykernel} into Eq.~\ref{nonconnecteddoubleamplitude}, but one must also invoke the single amplitude equation (Eq.~\ref{reducedsingleamplitudeequation}) and the norm equation (Eq.~\ref{nonconnectedenergy}). In doing so, one finally obtains the equation of motion for hole-particle double amplitudes under the desired form
\begin{equation}
h^{ab}_{ij}(\tau,\Omega) = -\partial_{\tau} {\cal T}^{\dagger}_{ijab}(\tau,\Omega)  \, . \label{reduceddoubleamplitudeequation}
\end{equation}

\subsection{N-tuple amplitude equation}
\label{ntuple}

As for single and double amplitude equations, the derivation of the n-tuple amplitude equation invokes all the amplitude equations of lower rank. Reasoning by recurrence, one can prove Eq.~\ref{reduceddynamicalkernels} for any n-tuply excited amplitude.

\section{Axial symmetry}
\label{axialsymmetry}

We consider the angular-momentum restored CC formalism in the particular case where the reference state fulfills $J_z | \Phi \rangle =0$. In this case, the kernels depend on the sole Euler angle $\beta$. Given that $N^{J \mu}_{MK} = N^{J \mu}_{00} \delta_{M0} \delta_{K0}$, their expanded form (Eq.~\ref{expandedkernels}) reduces to
\begin{subequations}
\label{expandedkernelsaxial}
\begin{eqnarray}
N(\tau,\beta)  &=&  \sum_{\mu J} e^{-\tau E^{J}_{\mu}} \, N^{J \mu}_{00} \, d^{J}_{00}(\beta)  \, , \label{expandedkernelsaxial1} \\
H(\tau,\beta) &=&  \sum_{\mu J} e^{-\tau E^{J}_{\mu}} \, E^{J}_{\mu} \,  N^{J \mu}_{00} \, d^{J}_{00}(\beta) \, , \label{expandedkernelsaxial2} \\
J_z(\tau,\beta) &=&  0 \, , \label{expandedkernelsaxial3} \\
J^2(\tau,\beta) &=&  \sum_{\mu J} e^{-\tau E^{J}_{\mu}} \, J(J+1)\hbar^2  \, N^{J \mu}_{00} \, d^{J}_{00}(\beta) \, , \label{expandedkernelsaxial4}
\end{eqnarray}
\end{subequations}
where $d^{J}_{00}(\beta) = \langle \Psi^{J0}_{\mu} | R(0,\beta,0) | \Psi^{J0}_{\mu} \rangle$ and 
\begin{equation}
R(0,\beta,0)=e^{-\beta J_y} \, . 
\end{equation}
Consistently, the rotated reference state is obtained from a one-dimensional rotation $| \Phi(\beta) \rangle \equiv R(0,\beta,0)| \Phi \rangle$.

The connected/linked kernels $h^{ab\ldots}_{ij\ldots}(\tau,\beta)$, $j^2(\tau,\beta)$ and $j_i(\tau,\beta)$ at play in the theory display the same functional dependence on the cluster operators ${\cal T}^{\dagger}_n(\tau,\beta)$ as in the three dimensional case. They invoke matrix elements of the corresponding operators expressed in the bi-orthogonal system whose right (left) basis is obtained via the transformation $D(\beta)\equiv \bbone + \rho^{ph}(\beta)$ ($D^{-1}(\beta)\equiv \bbone - \rho^{ph}(\beta)$). Because of the axial symmetry, one notes that $j_z(\tau,\beta)=0$ and $j_x(\tau,\beta)=j_y(\tau,\beta)$. The amplitude equations naturally read
\begin{equation}
h^{ab\ldots}_{ij\ldots}(\tau,\beta) = -\partial_{\tau} {\cal T}^{\dagger}_{ij\ldots ab\ldots}(\tau,\beta) \,\, . \label{reduceddynamicalkernelsaxial}
\end{equation}

The reduced norm kernel now satisfies a single ODE. One can use either on the first-order ODE of Eq.~\ref{NormkernelODE2} or on the second-order one given by Eq.~\ref{checkJ2}. Focusing on the latter gives in the present case
\begin{eqnarray*}
-\frac{d^2}{d \beta^2}  \, {\cal N}(\tau,\beta)  -\cot \beta \, \frac{d}{d \beta}  \, {\cal N}(\tau,\beta) &=& \frac{j^2(\tau,\beta)}{\hbar^2} \, {\cal N}(\tau,\beta) \, , \label{NormkernelODEaxial3}
\end{eqnarray*}
which must be integrated under the initial conditions
\begin{subequations}
\begin{eqnarray}
{\cal N}(\tau,0)  &=& 1 \, , \\
\left. \frac{d}{d \beta} \, {\cal N}(\tau,\beta)\right|_{\beta=0} &=& -\frac{i}{\hbar} \, j_y(\tau,0)  \, . 
\end{eqnarray}
\end{subequations}

Eventually, the above set of equations only need to be solved at infinite imaginary time, i.e. in the stationary limit. Once this is done, the angular-momentum restored energy of the yrast states is obtained from a single integral over the $\beta$ angle according to
\begin{eqnarray}
E^{J}_{0} &=& \frac{\int_{0}^{\pi} \! d\beta \, \sin\beta \,\, d^{J \, \ast}_{00}(\beta) \, \, h(\beta) \,\, {\cal N}(\beta)}{\int_{0}^{\pi} \! d\beta \, \sin\beta  \,\, d^{J \, \ast}_{00}(\beta) \,\, {\cal N}(\beta)} \, , \label{projected_energy_axial}
\end{eqnarray}
whose lowest energy is nothing but the ground state energy $E^{J_0}_{0}$.

\section{Useful identities}
\label{usefulID}

\begin{widetext}
\begin{subequations}
\label{integral} 
\begin{eqnarray}
\int_{0}^{\tau}d\tau_1 \, e^{a\tau_1} &=& \frac{1}{a}\Big(e^{\tau a}-1\Big) \, , \label{integral1} \\
\int_{0}^{\tau}d\tau_{1}d\tau_{2}\,\theta\left(  \tau_{1}-\tau_{2}\right)
e^{a\left(  \tau_{1}-\tau_{2}\right)  } &=&\int_{0}^{\tau}d\tau_{1} \, e^{a\tau_{1}}\int_{0}^{\tau_{1}}d\tau
_{2} \, e^{-a\tau_{2}} = -\frac{\tau}{a}+\frac{1}{a^{2}%
}\Big(  e^{\tau a}-1\Big) \, , \label{integral2} \\
\int_{0}^{\tau}d\tau_{1}d\tau_{2}\,\theta\left(  \tau_{1}-\tau_{2}\right)
e^{a\tau_{1}+b\tau_{2}} &=& \int_{0}^{\tau}d\tau_{1} \, e^{a\tau_{1}}\int_{0}^{\tau_{1}}d\tau
_{2} \, e^{b\tau_{2}} = \frac{1}{b\left(  a+b\right)  } \Big(
e^{\tau\left(  a+b\right)  }-1\Big)   -\frac{1}{ab}\Big(  e^{\tau a}-1\Big) \,  . \label{integral3} 
\end{eqnarray}
\end{subequations}
\end{widetext}
Given that such integrals only appear in the theory with $a<0$ and $a+b<0$, one obtains 
\begin{subequations}
\label{limitintegral} 
\begin{eqnarray}
\lim\limits_{\tau \to \infty} \int_{0}^{\tau}d\tau \, e^{a\tau} &=& -\frac{1}{a} \, , \label{limitintegral1} \\
\lim\limits_{\tau \to \infty} \int_{0}^{\tau}d\tau_{1}d\tau_{2}\,\theta\left(  \tau_{1}-\tau_{2}\right)
e^{a\left(  \tau_{1}-\tau_{2}\right)  } &=& -\frac{\tau}{a}-\frac{1}{a^{2}}  \, , \label{limitintegral2} \\
\lim\limits_{\tau \to \infty}  \int_{0}^{\tau}d\tau_{1}d\tau_{2}\,\theta\left(  \tau_{1}-\tau_{2}\right)
e^{a\tau_{1}+b\tau_{2}} &=& \frac{1}{a(a+b)} \,  , \label{limitintegral3} 
\end{eqnarray}
\end{subequations}
where the first and third result are necessarily positive.

\end{appendix}

\bibliography{CC_rest_ang_mom}

\end{document}